\documentclass[12pt,preprint]{aastex}

\newcommand{\gps}{\ensuremath{g_{\rm P1}} }
\newcommand{\rps}{\ensuremath{r_{\rm P1}} }
\newcommand{\ips}{\ensuremath{i_{\rm P1}} }
\newcommand{\gpserr}{\ensuremath{g_{e,\rm P1}} }
\newcommand{\rpserr}{\ensuremath{r_{e,\rm P1}} }
\newcommand{\ipserr}{\ensuremath{i_{e,\rm P1}} }
\newcommand{\nrps}{\ensuremath{r_{{\rm P1},0}} }
\newcommand{\nips}{\ensuremath{i_{{\rm P1},0}} }
\newcommand{\ngps}{\ensuremath{g_{{\rm P1},0}} }

\makeatletter

\newcommand{\Rmnum}[1]{\expandafter\@slowromancap\romannumeral #1@}
\makeatother
    
\slugcomment{Accepted for publication in AJ}

\shorttitle{Cepheids in M31 -- The PAndromeda Cepheid sample}
\shortauthors{Kodric et al.}

\begin{document}

\title{Cepheids in M31 -- The PAndromeda Cepheid sample}

\author{Mihael Kodric\altaffilmark{1,2}, Arno Riffeser\altaffilmark{1,2}, Ulrich Hopp\altaffilmark{1,2}, Claus Goessl\altaffilmark{1,2}, Stella Seitz\altaffilmark{1,2}, Ralf Bender\altaffilmark{2,1}, Johannes Koppenhoefer\altaffilmark{2,1}, Christian Obermeier\altaffilmark{2,1}, Jan Snigula\altaffilmark{2,1}, Chien-Hsiu Lee\altaffilmark{1}, W. S. Burgett\altaffilmark{4}, P. W. Draper\altaffilmark{5}, K. W. Hodapp\altaffilmark{4}, N. Kaiser\altaffilmark{4}, R.-P. Kudritzki\altaffilmark{4}, N. Metcalfe\altaffilmark{5}, J. L. Tonry\altaffilmark{4}, R. J. Wainscoat\altaffilmark{4}}

\email{kodric@usm.lmu.de} 

\altaffiltext{1}{University Observatory Munich, Scheinerstrasse 1, 81679 Munich, Germany}
\altaffiltext{2}{Max Planck Institute for Extraterrestrial Physics, Giessenbachstrasse, 85748 Garching, Germany}
\altaffiltext{3}{Subaru Telescoope, NAOJ, 650 N Aohoku Pl., Hilo, HI 96720, USA}
\altaffiltext{4}{Institute for Astronomy, University of Hawaii at Manoa, Honolulu, HI 96822, USA}
\altaffiltext{5}{Department of Physics, Durham University, South Road, Durham DH1 3LE, UK}

\begin{abstract}
We present the largest Cepheid sample in M31 based on the complete \mbox{Pan-STARRS1} survey of Andromeda (PAndromeda) in the \rps, \ips and \gps bands. We find 2686 Cepheids with 1662 fundamental mode Cepheids, 307 first-overtone Cepheids, 278 type II Cepheids and 439 Cepheids with undetermined Cepheid type. Using the method developed by \citet{K13} we identify Cepheids by using a three dimensional parameter space of Fourier parameters of the Cepheid light curves combined with a color cut and other selection criteria. This is an unbiased approach to identify Cepheids and results in a homogeneous Cepheid sample. The Period-Luminosity relations obtained for our sample have smaller dispersions than in our previous work. We find a broken slope that we previously observed with HST data in \citet{K15}, albeit with a lower significance.
\end{abstract}


\section{Introduction\label{intro}}

Cepheids are standard candles and their Period-Luminosity relation (PLR), sometimes also referred to as Leavitt Law (\citet{1908AnHar..60...87L} and \citet{1912HarCi.173....1L}), can be used to determine extragalactic distances. In order to be used for distance determination the PLR needs to be calibrated by using Cepheids with already known distances. There are multiple galaxies (so called anchor galaxies) with well known distances that host the Cepheids used for this calibration. Typically one or a combination of these anchor galaxies are used for the calibration: the Milky Way, the Magellanic clouds, M106 and the Andromeda galaxy (M31). Cepheids in the Milky Way need very precise parallaxes. Until recently the number of Cepheids with accurate enough parallaxes was very limited. The Gaia mission \citep{2016A&A...595A...1G} did remedy this problem and will keep improving the parallaxes and therefore the precision of the calibration (see \citet{2018arXiv180410655R} for an application of the Gaia DR2 parallaxes on Milky Way Cepheids). The Magellanic clouds have been studied extensively (see e.g. \citet{2015AcA....65....1U} and \citet{2015AcA....65..297S} for the OGLE survey and \citet{2015ApJ...815...28G} for the Araucaria project) and there are a lot of known Cepheids. \citet{2013Natur.495...76P} use eclipsing binaries to determine the distance to the LMC with an uncertainty of only 2\%. M106 has a very accurate distance determined geometrically by a maser \citep{2013ApJ...775...13H} and is therefore also used as an anchor galaxy \citep{2015AJ....149..183H}. M31 is relatively nearby, the disk has a solar metallicity, and the full range of the PLR can be observed from ground and space based observatories in the optical and near-infrared. These advantages are partly compromised by the large extent on the sky and significant crowding caused by the large inclination are major issues for a M31 Cepheid survey. There have however been multiple surveys of M31 Cepheids and hence there are multiple Cepheid samples: The PAndromeda Cepheid sample (\citet{K13}; hereafter K13) with 2009 Cepheids, the \citet{2006A&A...459..321V} sample with 416 Cepheids, the 332 Cepheid sample from the DIRECT survey (\citet{1998AJ....115.1894S}, \citet{1998AJ....115.1016K}, \citet{1999AJ....117.2810S}, \citet{1999AJ....118..346K}, \citet{1999AJ....118.2211M}, \citet{2003AJ....126..175B}) and the WECAPP sample \citep{2006A&A...445..423F} with 126 Cepheids. There are also Cepheid samples making use of the PHAT HST survey of M31 \citep{2012ApJS..200...18D}: the \citet{2012ApJ...745..156R} sample with 68 Cepheids, the \citet{K15} (hereafter K15) sample with 371 Cepheids and the \citet{2015MNRAS.451..724W} sample with 175 Cepheids. See also \citet{2017arXiv170102507L} for a summary of time domain studies in M31. 

The PLR calibration is not the only goal of the PAndromeda project, but also to study systematic effects on the PLR. The effect of metallicity on the PLR needs further investigation (see e.g. \citet{2011ApJ...734...46F} and \citet{2011ApJ...741L..36M}). A metallicity effect on the PLR has to be accounted for when using the low metallicity Magellanic clouds PLR to determine the distance to a galaxy with a Milky way like metallicity like the Andromeda galaxy (M31). Another issue with the PLR is the possible existence of a broken slope \citep{2009A&A...493..471S}. The existence of the broken slope is heavily discussed in the literature. \citet{2008A&A...477..621N} find a broken slope, but linear relations in the Ks band and the Wesenheit PLR. \citet{2013ApJ...764...84I} however find linear Wesenheit PLRs, while \citet{2013MNRAS.431.2278G} find no linear relations but rather exponential ones. In K15 we find a significant broken slope in all relations, including the Wesenheit PLR. The Cepheid sample we obtain here is used in \citet{K18b} (herafter K18b) where we combine it with the PHAT data to obtain the largest HST Cepheid sample for M31 and to investigate the broken slope. Since Cepheids are part of the distance ladder used to determine the Hubble constant ($H_0$) the study of these systematic effects is not only important for the value and error budget of $H_0$, but also to ascertain how universal the PLR is in different galaxies. \citet{2016ApJ...826...56R} used Cepheids as part of their distance ladder in order to determine $H_0$  with an uncertainty level of only 2.4\%. \citet{2010ARA&A..48..673F} summarize previous projects focused on determining $H_0$ (for a more recent summary see \citet{2016ApJ...826...56R} and references therein). Another uncertainty in the determination of $H_0$ is the handling of outliers in the PLR, which is still discussed in literature (e.g. \citet{2014MNRAS.440.1138E} and \citet{2015arXiv150707523B}).

This paper is structured as follows: In section \ref{datareduction} we discuss the PAndromeda survey and the data reduction procedure used for this paper, which is different from K13. In section \ref{detection} we describe the Cepheid detection and classification. Section \ref{cat} describes the PAndromeda Cepheid catalog. The results are discussed in section \ref{results} followed by the conclusion in section \ref{conclusion}.

\section{PAndromeda data\label{datareduction}}

PAndromeda is a Pan-STARRS1 (\citet{2002SPIE.4836..154K}, \citet{2004AN....325..636H}, \citet{2009amos.confE..40T}, \citet{2012ApJ...750...99T} and \citet{2016arXiv161205560C}) survey dedicated to monitoring the Andromeda galaxy (M31). The original goal of the survey was microlensing \citep{shermanPAndromeda} but the focus has shifted towards cepheids (K13, \citet{2013ApJ...777...35L} and K15), eclipsing binaries \citep{2014ApJ...797...22L} and luminous blue variables \citep{2014ApJ...785...11L}. The 1.8 m Pan-STARRS1 (PS1) telescope has a very large field of view with $\sim$~7~deg$^2$, which makes it ideal for observing M31, since the complete disk can be observed in one exposure.

\subsection{PAndromeda survey\label{survey}}

\begin{figure}
\centering
\includegraphics[width=\linewidth]{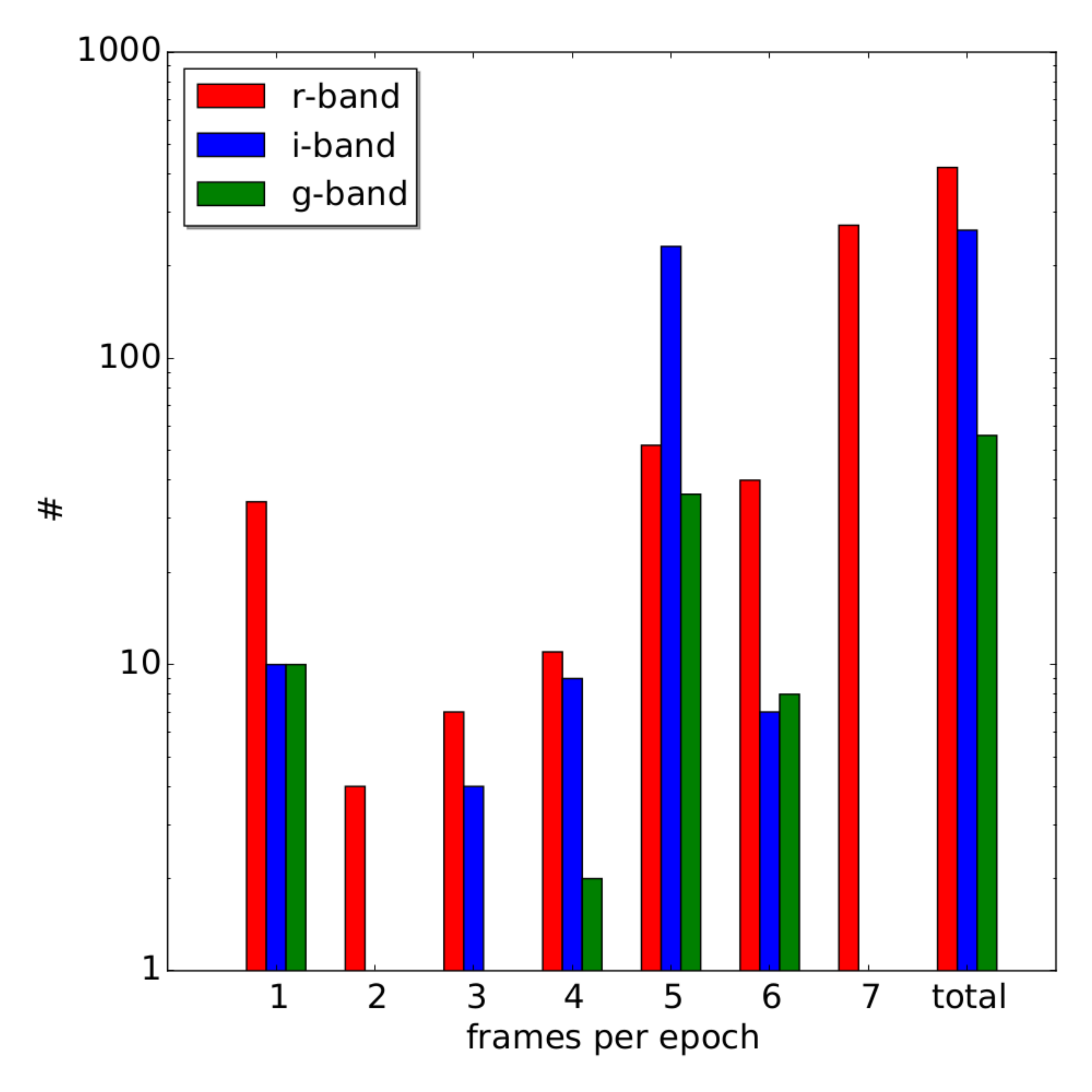}
\caption{Number of frames contributing to each epoch. For Cepheid light curves we require at least three frames per epoch. In total the PAndromeda data consists of up to 420 epochs in the $\rps$-band, up to 262 epochs in the $\ips$-band and up to 56 epochs in the $\gps$-band.
\label{frames-per-epoch}}
\end{figure}

M31 was observed from 2009-08-31 to 2010-01-21, 2010-07-23 to 2010-12-27, 2011-07-25 to 2011-11-22 and 2012-07-28 to 2012-10-31 with usually two visits per night in the \gps, \rps and \ips bands (see \citet{2012ApJ...750...99T} for information on the PS1 filter system). The general exposure time is 60 seconds per frame and each epoch (visit) is a stack of typically 7 frames in the \rps band and 5 frames in the \gps and \ips bands. The creation of this visit stack is described in the next section. Figure \ref{frames-per-epoch} shows the distribution of the number of frames making up one epoch. In total there are 420 epochs in the $\rps$-band, 262 epochs in the $\ips$-band and 56 epochs in the $\gps$-band. Due to masking (which will be discussed in the following section) the light curves usually have less epochs, so these are the maximum number of epochs. In addition, we disregard epochs that are based on less than three frames. In total PAndromeda has taken 2511 frames in the \rps band with a combined exposure time of 157904 seconds ($\approx$ 43.86h), 1260 frames in the \ips band with a total of 75840 seconds ($\approx$ 21.07h) and 246 frames in the \gps band with a combined exposure time of 15048 seconds ($\approx$ 4.18h).

The giga pixel camera (GPC) consists of 60 orthogonal transfer arrays (OTAs), where each OTA is made up of an array of 8 $\times$ 8 cells. One OTA has a field of view (FOV) of 21\arcmin~$\times$ 21\arcmin. The pixel scale is 0.25\arcsec~per pixel. The Pan-STARRS1 image processing pipeline (IPP; \citet{2006amos.confE..50M}) provides us with astrometrically aligned frames (the frames are also bias and flatfield corrected). During the alignment the exposures are remapped to a common grid, the so called skycells. Only the first data reduction steps are done by the IPP, the other steps done by ourselves are discussed in the next section. Compared to K13 we changed the layout and the size of the skycells so that they now cover 6647 px~$\times$ 6647 px (i.e. 27.7\arcmin~$\times$ 27.7\arcmin) and use the intrinsic pixel scale of 0.25\arcsec. Also we defined the skycells such that they overlap (by 6\arcmin), in order to have a simple way to check our data reduction for consistency. The covered area is $\sim$7~deg$^2$ in the \rps and \ips bands and $\sim$6.8~deg$^2$ in the \gps band. The central six skycells were split during our data reduction into four smaller skycells with 3524 px~$\times$ 3524 px (i.e. 14.7\arcmin~$\times$ 14.7\arcmin), for reasons discussed in the next section. The area covered in the \gps band is smaller since skycells 003, 004, 005, 006, 012 and 028 are not used due to the lack of data in their skycells. The area in K13 (with $\sim$2.6~deg$^2$) is much smaller than the area analyzed here. The skycell layout for this work (hereafter K18a) as well as for K13 is shown in the appendix.

\subsection{PAndromeda data reduction}

The data reduction is based on our previous pipeline used in K13. The overall approach of using difference imaging \citep{1998ApJ...503..325A} is the same. Apart from changing certain details the biggest change is to mask a large fraction of the images in order to minimize systematic errors that are due the poor PS1 image quality. The trade off for less systematics is a larger statistical error caused by the rigorous masking. This was mainly done to help with the microlensing search \citep{shermanPAndromeda} which was originally the main scientific goal of PAndromeda. As far as Cepheid light curves are concerned the fact of having more statistical noise is usually negated by the large increase in epochs. But some area is lost due to the masking or has substantially less epochs due to the masking. Therefore we do not recover all Cepheids found in K13 and all eclipsing binaries found in \citet{2014ApJ...797...22L}.

As mentioned in the previous section our data reduction starts with astrometrically aligned frames provided by the IPP \citep{2006amos.confE..50M} that are already debiased and flatfield corrected. Our pipeline (see also \citet{2013ExA....35..329K}, \citet{shermanPAndromeda} and K13) uses the AstroWISE data management enviroment \citep{2007ASPC..376..491V}. In the first step where we ingest the frames provided by IPP into the AstroWISE database we apply multiple masks. We use most of the masking flags already provided by IPP. The quality of these original masks was not sufficient enough so we improved those masks and provided them to the IPP in order to reprocess all of our raw data using these new mask files. This step was necessary since the masks are based on the defects on the OTA cell level and applying the masks before the skycell regridding is computationally more efficient. We also use the ghost masks provided by IPP, but significantly increase the masking since the position of the ghost model is sometimes off. We also mask satellite tracks using the linear Hough transform algorithm that is implemented in the AstroWISE and we use manually created masks that we have for a few hundred frames. Saturated pixels are also masked. Since there are offsets between the OTAs we only use the dominant OTA in each skycell and mask the other OTAs. This implies that we mask up to 75\% for the rare case where four OTAs are in one frame with the same area. This OTA masking is quite extreme but it ensures that there are no systematic jumps inside a frame and that we can photometrically align the frames to each other. After applying these masks we only use those frames that have more than 20\% unmasked pixels in the central region of the frame (pixel 1661 to 4985 in x and y). The mean masking over all frames in all skycells is 64\% in the \rps and \ips bands and 65\% in the \gps band. The mean masking per skycell ranges from 48\% to 83\% in the \rps band, from 48\% to 82\% in the \ips band and from 43\% to 85\% in the \gps band. 

As can be seen in Fig. \ref{footprint_K17a_r}, Fig. \ref{footprint_K17a_i} and Fig. \ref{footprint_K17a_g}, the central six skycells are each divided into four smaller skycells. The point spread function (PSF) varies slightly over the FOV of the GPC. The PSF is broadest on the edges of the GPC but also blurred in the center of the GPC. Over the PAndromeda observing campaign the center of the GPC was pointed on different skycells. Due to this the central six skycells showed the largest PSF variation. By splitting them into smaller frames the variation inside these new skycells is significantly reduced. This ensures that our PSF photometry, which uses a constant PSF model for a skycell, performes well.    

In order to perform difference imaging we need to construct so called reference frames. A reference frame is a stack of the images with the best PSF.
The frames are photometrically aligned so that the sky background and the zero point are the same. Masked areas in one frame (if the masked region is not too large) are replaced from another frame with almost similar PSF (\citealt{ediss5984}, p. 134). The frames are assigned a weights according to the inverse of the squared product of the seeing and the mean error. The reference frame is the weighted stack of these frames. In the \rps and \ips bands we use one hundred frames for the reference frame. For the \gps band we use forty frames. Additionally we require the original position angle to be zero or a multiple of ninety degrees in the \rps and \ips band. The reason for this is that we want to limit the increased noise caused by e.g. chip gaps or bleeding to the four cardinal directions. The observing strategy required the second visit in one night to have a position angle of ninety degrees. Unfortunately there were also a significant number of frames taken with a random position angle. We disregard those for the reference frame in the \rps and \ips band, but have to use them in the \gps band because in this band we do not have enough data. In K13 we used the seventy best-seeing frames, but here we also selected according to masking. Since we increased the masking drastically, we have to make sure that the reference frames cover as much area as possible. Everything that is masked in the reference frame will be lost even if there are frames that have data in these masked regions. Due to the fact that the masking fraction and the seeing are not correlated, we select first seven times more frames for the reference frame where the masking is the lowest and then select from those frames the best-seeing frames. The frame that is used to photometrically align the other frames is one of the ten best-seeing frames that overlaps the most with the other frames. The reason for this is that the alignment works better the more area the frames have in common. Since we wanted to limit the area lost due to the new strict masking, we also require that a frame does not increase the masking in the reference frame too much. Pixels that are masked in all but one frame will still be dominant in the reference frame. During the creation of the reference frame each frame contributing to it is checked if including it to the stack would decrease the unmasked area in the reference frame by more than one percent. If that is the case the frame will be dropped from the reference frame. Therefore the final reference frame can include less than one hundred (\rps and \ips band) or forty frames (\gps band) respectively. Each reference frame is calculated twice. The first time the sources are detected using SExtractor \citep{1996A&AS..117..393B}.
The reference frame is inspected and if there are e.g. satellite tracks missed by the algorithm or ghosts etc. we create manual masks for the input frames. This first reference frame is then used to photometrically align the input frames for the creation of the final second reference frame. In the second run we use DAOPHOT \citep{1987PASP...99..191S} to detect the sources. Due to the crowding in M31 SExtractor sometimes determines a wrong position of a star, while DAOPHOT is optimized for high crowding. We perform PSF photometry on the positions identified with DAOPHOT on a background subtracted reference frame. The neighboring stars are iteratively subtracted in order to account for the crowding. Same as in K13, we fit a bicubic spline model to determine the background (sky and M31). For each skycell we determine a PSF model, but do not vary it over the reference frame. The flux calibration is done as in K13, but we do not use the Sloan Digital Sky Survey (SDSS; \citet{2009ApJS..182..543A}) for the calibration but the PS1 PV2 catalog \citep{2016arXiv161205560C} instead. The reason for this is that the SDSS catalog is very sparsely populated in the central region of M31. With the PV2 catalog we have a lot of comparison stars so that we are able to also correct for PSF variation over the field by modeling the zero point with a spline. 

For the difference imaging we also need to create so called visit stacks, which are stacks of all frames from one visit. Ideally there should have been two visits of M31 per night, one in the first half of the night and one in the last half. The data taken has significant deviations from that. Sometimes the second visit was done right after the first or there were three visits or visits with less frames. Also the total integration time per visit changed several times in order to balance the total exposure time between the \rps and \ips bands. So instead of automatically splitting the data of one night into two visit stacks we grouped all the frames by hand into visit stacks with a stack depth as homogeneous as possible. The frames of a visit stack are photometrically aligned such that the sky background and the zero point are the same. The weighted stack of these frames is photometrically aligned to the reference frame, such that the visit stack and the reference frame have the same zero point and sky background. Note that the visit stack has a smaller signal to noise ratio since the reference frame is much deeper and the PSF is different, i.e. usually worse. As described in K13, the PSF of the reference frame is aligned to the visit stack by determining a convolution kernel when calculating the difference frame. Compared to K13 we decreased the size of the area one convolution kernel uses in order to account better for differences not only in PSF size but also PSF shape. We also masked out all sources brighter than 18 magnitudes in the \rps band, since those bright sources are not relevant for our science cases and are producing residuals in the difference frame. The PSF photometry on the difference frame uses the PSF constructed from the convolved reference frame. 

In order to obtain the flux in one epoch of the light curve of a resolved source, the flux in the difference frame is added to the flux in the reference frame. Since the skycells overlap, the light curve can be constructed as a combination of measurements from different skycells. For the case where a resolved source has measurements of one epoch in multiple skycells, we use the measurement with the smallest error. For our Cepheid detection we only use epochs in the light curve that have a signal to noise ratio of larger than two and where the visit stack has more than two frames (the published light curves do include those epochs for consistency). In total we have detected 7226125 unique sources, where 3348137 unique sources have data in at least the \rps band. For all sources that have \rps band data we determine the periodicity of the light curve with SigSpec \citep{2007A&A...467.1353R}. In K13 we explained in detail how we determine the periods and the period errors. We have not changed any parameter of the period determination, since the procedure described in K13 works very well. For the period error determination we increased the number of bootstrapping samples from 1000 in K13 to now 10000. The error rescaling factor described in K13 that was necessary due to the change of the pixel scale is as expected 1.0 and therefore not necessary anymore. We use the AB magnitude system throughout this paper and in the published data.

\section{Cepheid detection and classification\label{detection}}

The goal is to find as many Cepheids as possible in an unbiased way. We could look through all periodic light curves by eye and select those that have a typical shape for a Cepheid. The problem with this approach is that different pulsators are sometimes hard to distinguish especially in a noisy light curve. Also the selection would be unreproducible and not objective. Therefore we use essentially the same procedure we developed in K13 in order to be as unbiased as possible. We apply some selection criteria and determine manually a Cepheid sample from high signal to noise light curves and use the Fourier parameters of this sample to construct a 3 dimensional parameter space. The parameters of Cepheid light curves have to be within this parameter space and are further constrained by a color cut.

In order to find the Cepheids in our data we apply several selection criteria. From the 3348137 resolved sources in the \rps band we start by selecting those that are also resolved in \ips band. We also require the source to be periodic in both bands and that the periods are similar to one percent ($| \frac{\mathrm{P}_{\rps} - \mathrm{P}_{\ips}}{\mathrm{P}_{\rps}} | < 0.01$). Additionally, we only select sources where the period in the \rps band is $1.5~\mathrm{d} < \mathrm{P}_{\rps} < 150~\mathrm{d}$. The light curves of the resulting 21313 sources are fitted in order to obtain the Fourier coefficients $A_{21}$ and $\varphi_{21}$. $A_{21}$ is the amplitude ratio of the first two Fourier components and $\varphi_{21}$ is the phase difference between those two coefficients. The Fourier decomposition is done in the same way as described in K13, however, we use less degrees of freedom for the \gps band, i.e. 3 instead of 5. This is done because we have less epochs in the \gps band. From the fit we calculate the mean magnitude by determining the mean flux from the fitted curve and convert the mean flux back to a magnitude. The determination of the error of the mean magnitude is more difficult since the error is not just the fit error of the mean magnitude. In K13 we used the photometric error determined in the reference frame as the mean magnitude error since this was the dominating error. While it is still true that the error in the reference frame is the dominating error, we can not just use this error because the light curve might consist of epochs from multiple reference frames. We determine the mean magnitude error by constructing a sample of ten thousand light curve realizations. For each realization we determine the mean magnitude by fitting the corresponding light curve. We construct a new light curve for each realization where each epoch's magnitude is the sum of the flux from the difference frame and a new flux from the corresponding reference frame. This new reference frame flux is drawn from a Gaussian distribution around the measured reference frame magnitude with a width of the measured photometric error. From the distribution of these ten thousand mean magnitudes we calculate the mean magnitude error. For example, a light curve that consists of data from four different skycells is fitted ten thousand times. Each time the four reference frame magnitudes are modified by drawing four new magnitudes from four Gaussian distributions with the width of the corresponding reference frame error. In this way we obtain a new light curve where each epoch's magnitude is modified by changing the corresponding reference frame flux. If the light curve consists almost exclusively from epochs from one skycell the error of the corresponding reference frame will dominate the mean magnitude error. This is not always the case however, hence we have to use this sophisticated procedure to determine the mean magnitude error. We also calculate the errors of $A_{21}$ and $\varphi_{21}$ simultaneously to calculating the mean magnitude error using the same technique. The errors of $A_{21}$ and $\varphi_{21}$ determined with this approach are much smaller than the fit errors of the Fourier decomposition. Therefore we use the errors obtained from the Fourier decomposition, but the other errors are included in the published data (see in the appendix). 

The Wesenheit, a reddening free magnitude (\citet{1976MNRAS.177..215M}, \citet{1982ApJ...253..575M}, \citet{1983IBVS.2425....1O}) is defined as in K13:
\begin{equation}
W_{ri} = \rps-R (\rps- \ips) \label{eqn_W}
\end{equation}
with $R=3.86$.

The selection criteria described in the next sub sections could be performed in any order. The published data also includes those light curves that do not make the selection criteria. We performed the cuts in the order described in the next sub sections. Therefore, any fit parameter that were calculated after applying a selection criterion or cut are not available for such light curves excluded before performing the corresponding fit. 

\subsection{Manual classification}

\begin{figure}
	\centering
	\includegraphics[width=\linewidth]{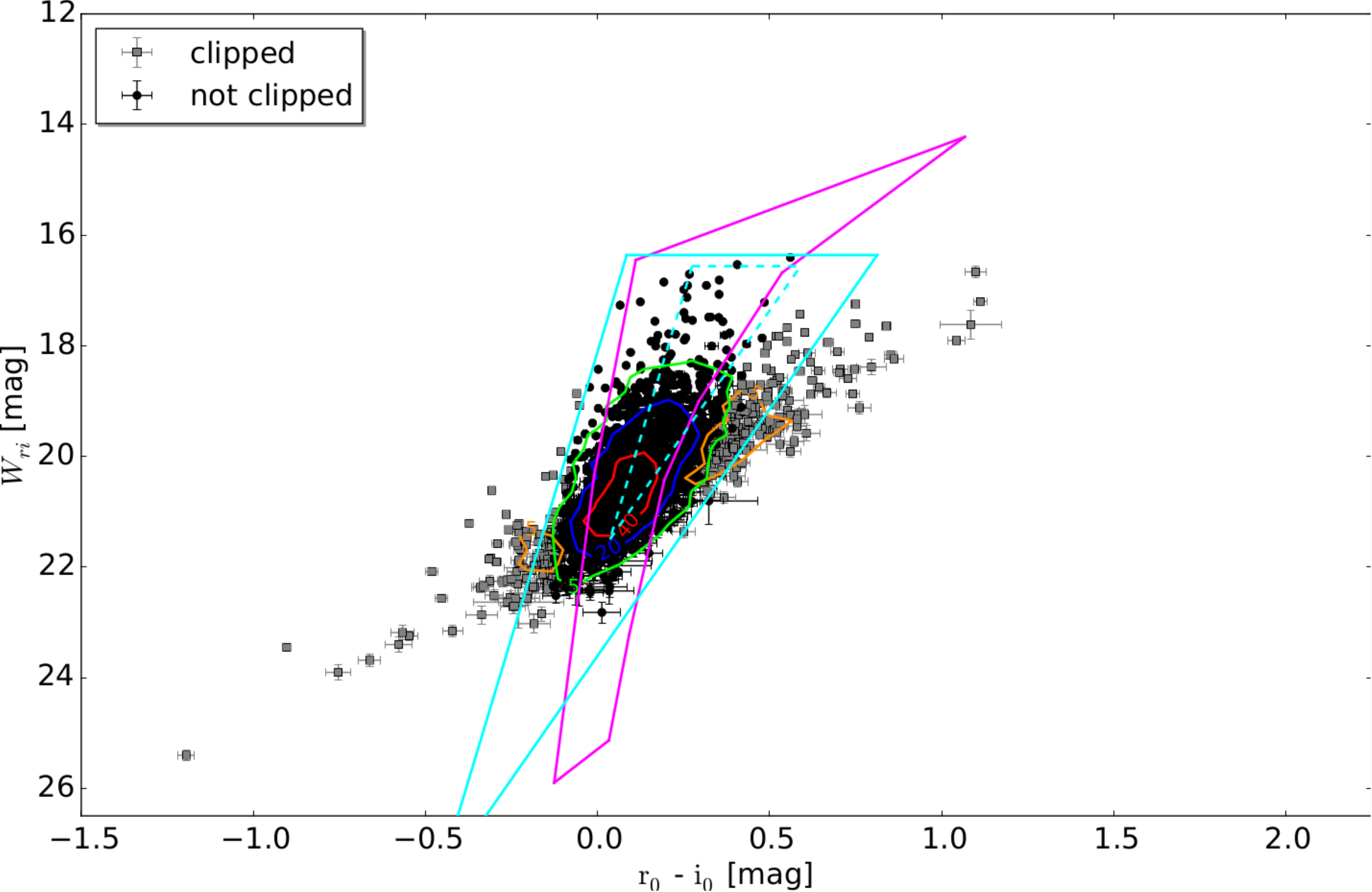}
	\caption{Color-Wesenheit cut for the manually selected Cepheid sample. 1966 sources are independently identified by four authors to be Cepheids by only considering the shape of the light curve. The color cut shown in Fig. \ref{fig_colorcutregion} is applied to this manually selected sample and 295 sources are clipped, leaving 1671 Cepheids. The extinction correction is done by using the \citet{2009A&A...507..283M} color excess map. The edges of the instability strip are used to constrain the color excess estimate. The lines are the same as in Fig. \ref{fig_colorcutregion}. The error bars shown are the photometric errors. The possible color excess range is also considered for the cut, therefore some points appear to be outside of the region defined by the magenta lines but are still classified as Cepheids. Density contour lines to the selected and clipped sample are also shown.
		\label{fig_colorWesenheitmanual}}
\end{figure}

As described previously we need to identify a certain number of Cepheids manually so that we are able to construct a three dimensional parameter space which enables us to find Cepheids in an as unbiased way as possible. Of course the manual selection of Cepheids by the light curve shape is biased to some degree. It is especially difficult to determine the variable type for noisy light curves.  As described previously, we start with 21313 sources but only select those where the period significance determined by SigSpec is larger than 25. This leaves us with 10562 light curves. From these we preselected 2807 by using only those that are not noisy and the light curves even remotely resemble Cepheids, i.e. Mira-like light curves were also selected. The selection is only based on the shape of the light curves in all available bands. So we do not look at the position in the PLR. In order to be as unbiased as possible, four of the authors independently assigned flags to those 2807 light curves. We then matched the four classifications and reviewed cases where we disagreed. We disregarded sources where we could not reach a consensus of opinion. We end up with 1966 light curves that are manually classified as Cepheids only based on their light curve shape. We apply the color cut described in the next subsection and clip 295 sources, leaving us with 1671 Cepheids. The color cut can be seen in Fig. \ref{fig_colorWesenheitmanual}. The number of clipped sources is quite high considering that those were manually identified to have a Cepheid like light curve. The color cut used in K13 shown with solid cyan lines would cut less objects, but the new color cut (magenta lines) works better in selecting Cepheids in the complete sample.   

\subsection{Instability strip cuts \label{insta}}

\begin{figure}
\centering
\includegraphics[width=\linewidth]{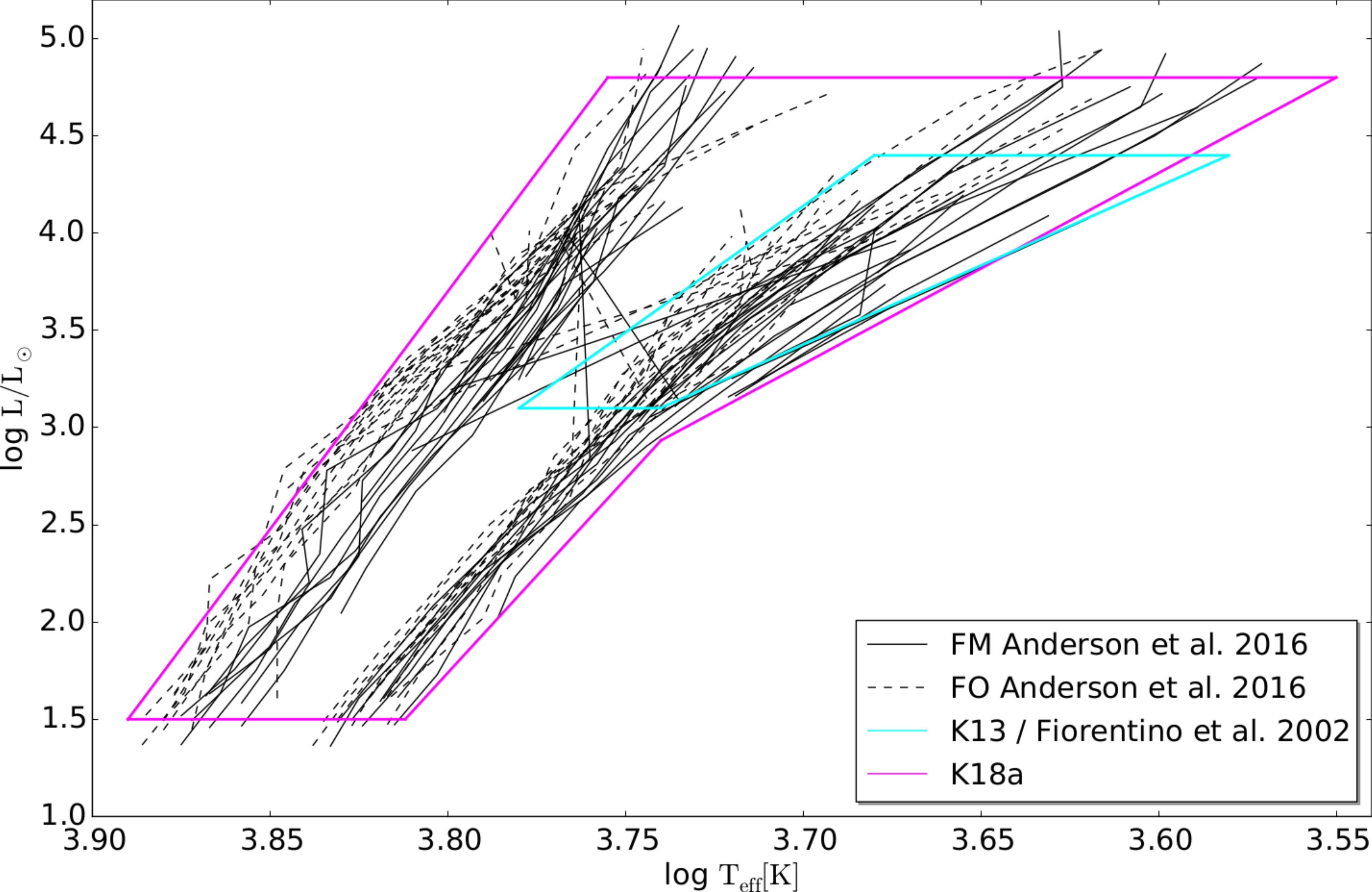}
\caption{Instability strip edges used to perform the color-Wesenheit cut. The outline of the region used in K13 (solid cyan line) is based on the \citet{Fiorentino2002} instability strip. All \citet{Anderson} instability strip edge models contained in their tables A.1 to A.6 are shown in black. This implies that we show all three different metallicities \mbox{(Z = 0.014, 0.006, 0.002)} and all three rotations ($\omega_{\mathrm{ini}}$ = 0.0, 0.5, 0.9). The FM models are shown as solid lines and the FO models as dashed lines. The magenta outlined region that we use for the color-Wesenheit cut, was chosen such that it encompasses the \citet{Anderson} models, while keeping the shape of the region as simple as possible. The region is defined by straight lines connecting the points (3.755,4.8), (3.890,1.5), (3.550,4.8), (3.740,2.934) and (3.812,1.5). \label{fig_instability}}
\end{figure}

\begin{figure}
\centering
\includegraphics[width=\linewidth]{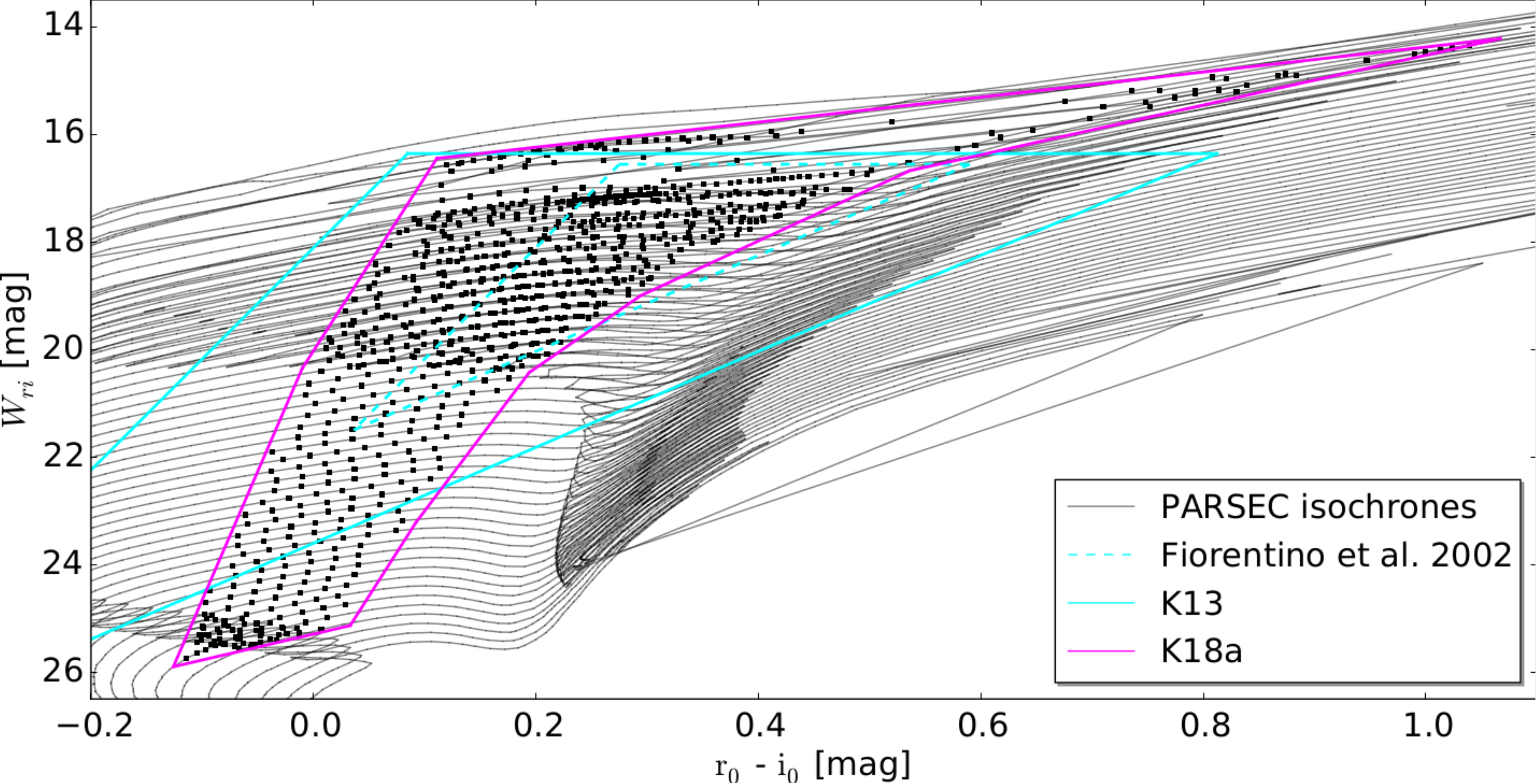}
\caption{Instability strip region used for the color-Wesenheit cut. The area enclosed by the dashed cyan triangle is defined by the \citet{Fiorentino2002} instability strip. In K13 we had to enlarge this region by 0.2 mag in all directions. The resulting region defined by the solid cyan lines also covers fainter Cepheids, that would be cut if the dashed cyan triangle would be used. The PARSEC isochrones (\citet{2012MNRAS.427..127B}, \citet{2014MNRAS.444.2525C}, \citet{2015MNRAS.452.1068C}, \citet{2014MNRAS.445.4287T}) shown in black have solar metalicity (Z = 0.0152) and ages between \mbox{7.05 $\leq$ log(age/yr) $\leq$ 9.25} and have been shifted by a distance modulus of 24.36 mag \citep{Vilardell}. The black squares show where the luminosity and temperature of the isocrones are within the magenta area defined in Fig. \ref{fig_instability}. These points define the magenta outlined region that is used to perform the color cut. For visibility reasons this plot only shows the isochrones with solar metalicity, but the magenta outlined region is chosen such that it also encompasses the other two metalicities from the \citet{Anderson} models. The color-Wesenheit cut region is defined by straight lines connecting the points (0.111,16.450), (-0.010,20.350), (-0.126,25.900), (0.033, 25.134), (0.092,23.222), (0.194,20.430), (0.294,19.003), (0.536,16.680) and (1.067,14.223).
\label{fig_colorcutregion}}
\end{figure}

As already mentioned we use a color cut for a better identification of Cepheids. We use the reddening free Wesenheit and the \rps - \ips color for the selection since we do not have \gps data for all light curves. In order to define the color cut we use the same approach as in K13. Theoretical pulsation models provide us with the information how the instability strip edges depend on the effective temperature ($T_{\mathrm{eff}}$) and the luminosity ($L$). We then use isochrones to translate the instability strip defined by $T_{\mathrm{eff}}$ and $L$ to the observable color and Wesenheit.
In K13 we used the \citet{Fiorentino2002} instability strip, but had to extend the edges by 0.2 mag in order to include also fainter, but obvious Cepheids that would otherwise not have been covered, which make the reliability of the instability strip parameters somewhat doubtful. Now we use the \citet{Anderson} instability strip edge models. As can be seen in Fig. \ref{fig_instability} those also include fainter and cooler Cepheids. We include all their models from tables A.1 to A.6, i.e. all three different metallicities \mbox{(Z = 0.014, 0.006, 0.002)} and all three rotations ($\omega_{\mathrm{ini}}$ = 0.0, 0.5, 0.9). The \citet{Anderson} instability strip edges are used to define an area in which we require our Cepheids to reside in. The area is defined by lines connecting the points (3.755,4.8), (3.890,1.5), (3.550,4.8), (3.740,2.934) and (3.812,1.5) as shown in Fig. \ref{fig_instability}. In the next step we use the PARSEC isochrones (\citet{2012MNRAS.427..127B}, \citet{2014MNRAS.444.2525C}, \citet{2015MNRAS.452.1068C}, \citet{2014MNRAS.445.4287T}) and require the $T_{\mathrm{eff}}$ and $L$ of the isochrones to be within the area defined by the instability strip edges. This defines an area in the Wesenheit-color diagram, as seen in Fig. \ref{fig_colorcutregion}. This region, which is defined by lines connecting the points (0.111,16.450), (-0.010,20.350), (-0.126,25.900), (0.033, 25.134), (0.092,23.222), (0.194,20.430), (0.294,19.003), (0.536,16.680) and (1.067,14.223), is used to perform the color cut. The area shown in Fig. \ref{fig_colorcutregion} is also valid for all metallicities used in the \citet{Anderson} models. In order not to clutter up the plot we only show the solar metallicity isochrones. The isochrones can also be used to determine the instability strip in other filter bands. We use this in K18b where we identify our Cepheid sample in HST data. Since the color cut uses the intrinsic color we have to correct for the extinction. As in K13 we use the \citet{2009A&A...507..283M} color excess map and the foreground extinction towards M31 determined by \citet{1998ApJ...500..525S} ($E(B-V)_{\mathrm{fg}}=0.062$) combined with the correction factor 0.86 from \citet{2011ApJ...737..103S}. The \citet{2009A&A...507..283M} map gives the color excess for a line of sight going completely trough M31. Therefore we correct only by half the color excess assuming the Cepheid is only affected by half the dust. This corresponds to a probability argument, where sources are in front of an infinitively thin dust layer or behind, e.g. the extinction correction in the \rps band is:
\begin{equation}
\overline{A_{\rps}} =  (0.5 \cdot E(B-V)  + 0.86 \cdot 0.062) \cdot R_{r,\rm P1} 
\end{equation}
where $R_{r,\rm P1}=2.5535$ ($R_{i,\rm P1}=1.8928$ and $R_{g,\rm P1}=3.5258$) for a reddening law of $R_V=3.1$.
The assumption that the Cepheid is obscured by only half the dust can be refined in some cases. An object satisfies the color selection if it is within the area defined above, also considering its errors. So each object is not a point in the Wesenheit-color space but rather a box, where the height of the box is the error of the Wesenheit. The width of this box is the error of the color combined with the full color excess of that object. So the right side of the box is where only the foreground extinction has been corrected and the left side where the foreground plus the full color excess is used. In an edge on view this corresponds to a location on the top of the disk where only the foreground extinction has to be considered or to the bottom where additionally the full color excess has to be taken into account. In case this box intersects the color cut area defined by the instability strip edges, half the width of the box that is inside the color cut area and is due to the color excess is used to correct the extinction. So we use the color cut area to narrow down the range of color excess the object can have.   

In the following we use the extinction corrected magnitudes, i.e. e.g. \rps denotes the extinction corrected magnitude. The uncertainty caused from the color excess is not propagated into the magnitude errors.

\subsection{Type classification \label{section_typeclass}}

\begin{figure}
\centering
\includegraphics[width=\linewidth]{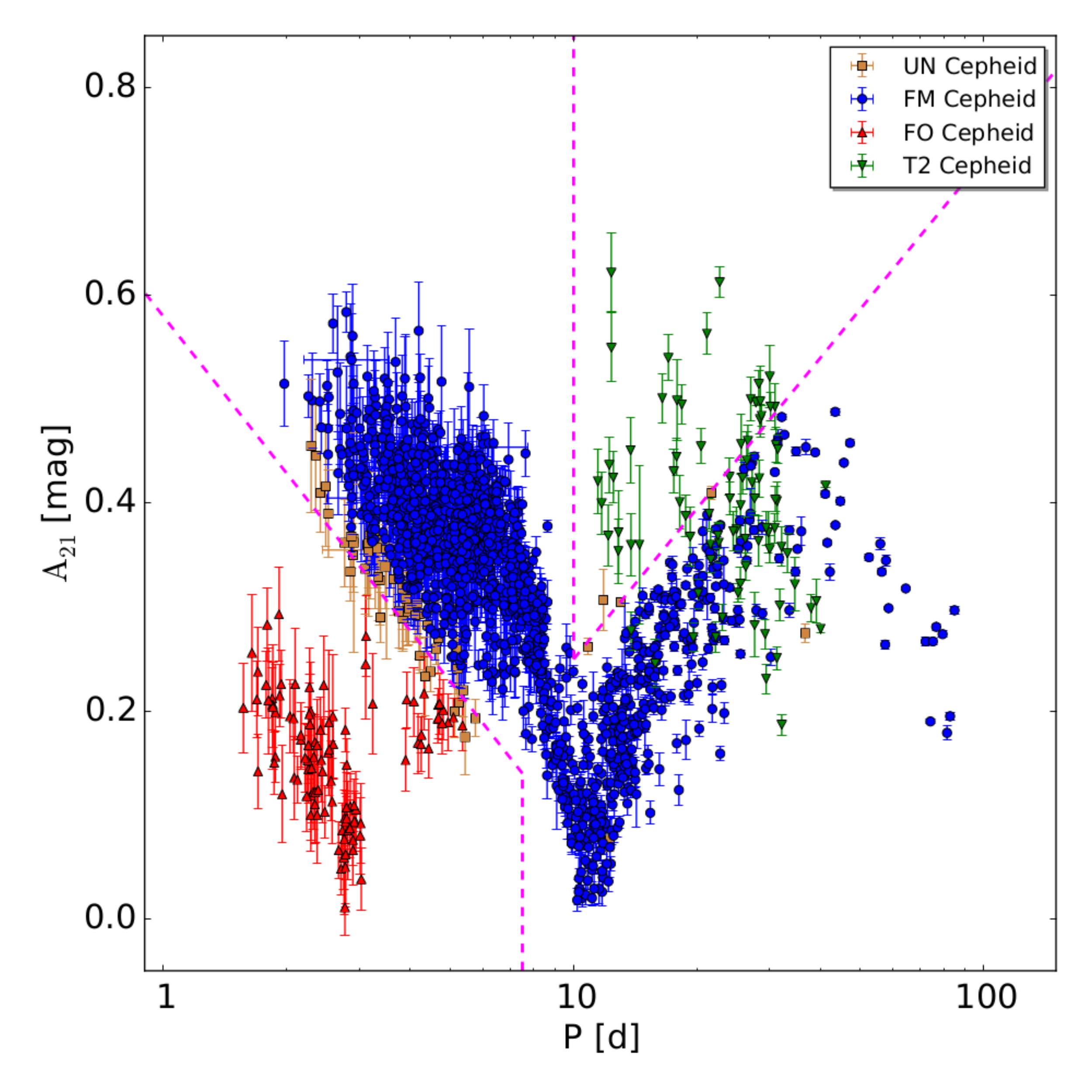}
\caption{Amplitude ratio ($A_{21}$) diagram for the manually classified Cepheid sample. This projection of the three dimensional space is mainly used to identify the FO Cepheids, but there is also a region that is used to classify a part of the T2 Cepheids. The region used to identify the FO Cepheids (equations \ref{eqn_fo1} and \ref{eqn_fo2}) is marked with a dashed magenta lines. Also the T2 region (equations \ref{eqn_t24} and \ref{eqn_t25}) is marked with dashed magenta lines. In total there are 54 UN Cepheids, 1401 FM Cepheids, 103 FO Cepheids and 82 T2 Cepheids.
\label{fig_A21manualpretype}}
\end{figure}

\begin{figure}
\centering
\includegraphics[width=\linewidth]{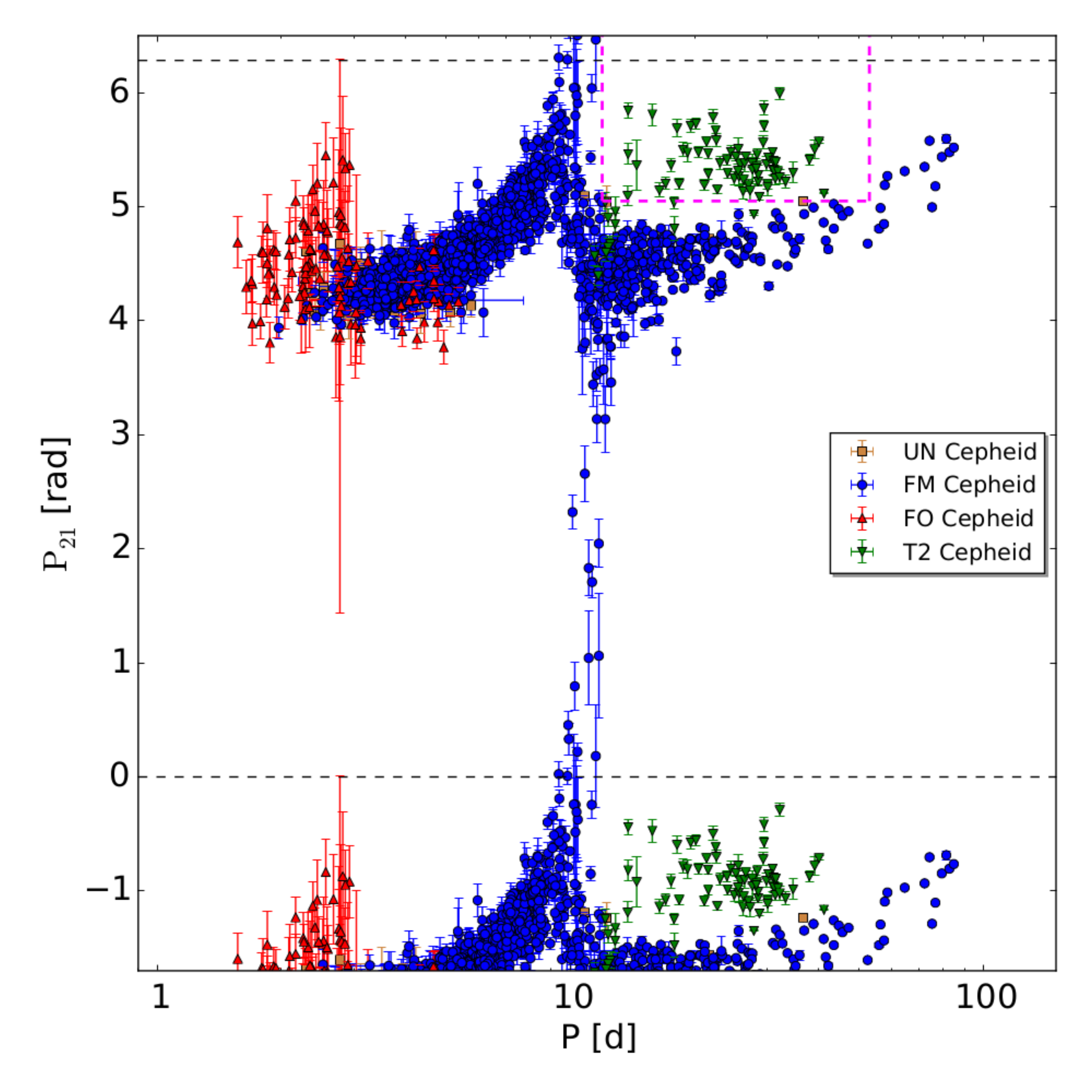}
\caption{Phase difference ($\varphi_{21}$) diagram for the manually classified Cepheid sample. This projection of the three dimensional space is used to identify the bulk of the T2 Cepheids. The region used for this classification (equations \ref{eqn_t21}, \ref{eqn_t22} and \ref{eqn_t23}) is marked with dashed magenta lines. $\varphi_{21}$ is 2$\pi$ periodic therefore the data between the dashed black lines is also plotted periodic for better visibility. The number of Cepheids in each subtype is the same as in Fig. \ref{fig_A21manualpretype}. 
\label{fig_P21manualpretype}}
\end{figure}

\begin{figure}
\centering
\includegraphics[width=\linewidth]{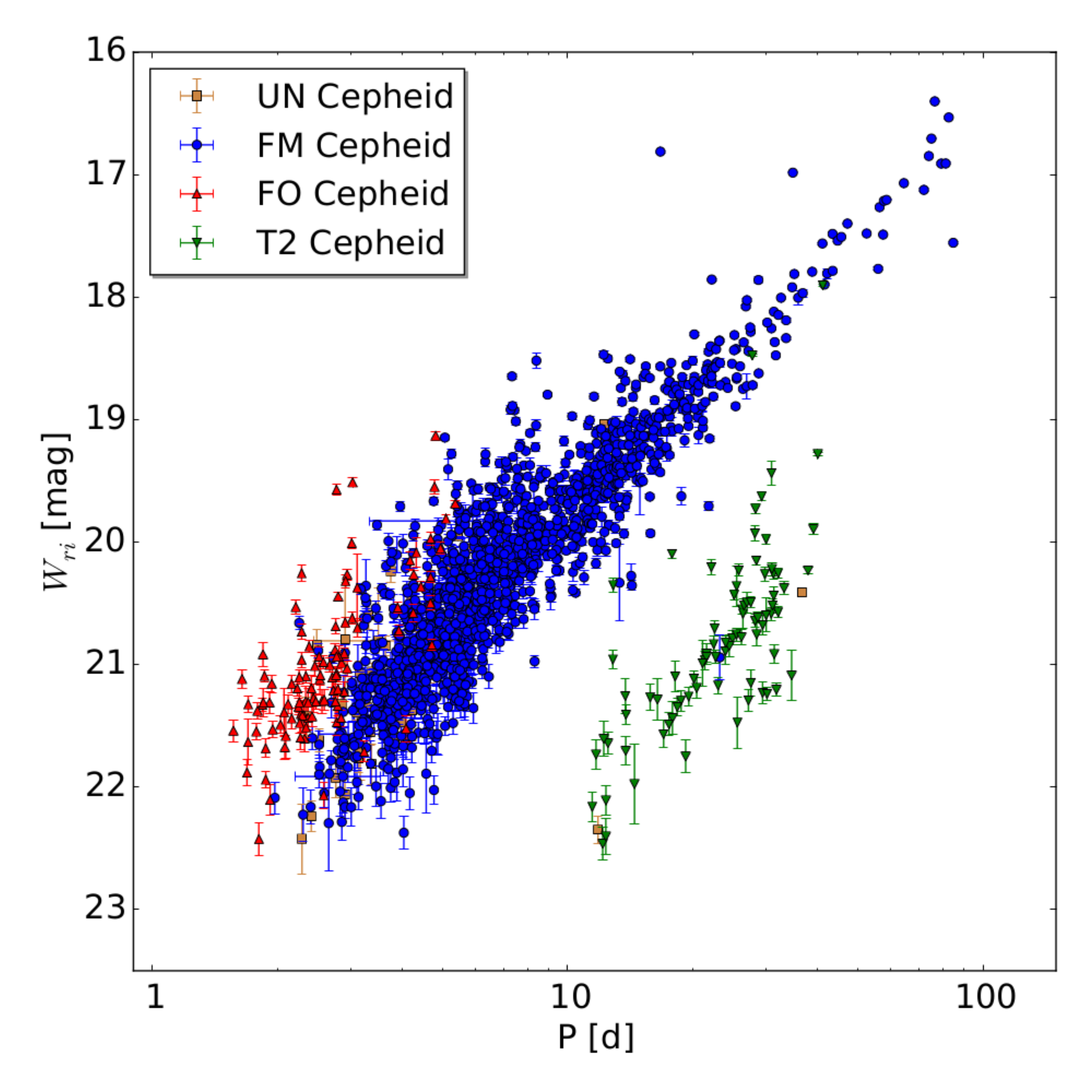}
\caption{Period Wesenheit diagram for the manually classified Cepheid sample. At this point no outlier clipping has been performed. Therefore all UN Cepheids are due to uncertainties in the type classification. The number of Cepheids in each subtype is the same as in Fig. \ref{fig_A21manualpretype}. Considering the dispersion in each relation the type classification performs well. The PLRs are well sampled and no obvious selection effects from the manual classification that is based on the shape of the light curve can be seen.
\label{fig_Wmanualpretype}}
\end{figure}

Fundamental mode (FM) Cepheids and first overtone (FO) Cepheids as well as type II (T2) Cepheids occupy different locations in the $A_{21}$-$\varphi_{21}$-$P$ space, which can be used to identify the Cepheid type (see e.g \cite{1999AcA....49..223U}, \cite{2007A&A...473..847V} and K13). As in K13 we use two projections of this space, i.e. $A_{21}$-$P$ (see Fig \ref{fig_A21manualpretype}) and $\varphi_{21}$-$P$ (see Fig. \ref{fig_P21manualpretype}) for the type classification. In K13 we used the $A_{21}$-$P$ projection to distinguish between FM and FO Cepheids and the $\varphi_{21}$-$P$ space to separate T2 and FM Cepheids. Here we additionally use an area in the $\varphi_{21}$-$P$ projection to find T2 Cepheids. The different Cepheid types could also be identified in the PLR, but by using this three dimensional space, i.e. using the Fourier parameters, we only use the light curve shape to determine the Cepheid type. FO and FM PLRs have a small separation compared to their dispersions. Therefore, defining a line in the PLR that would distinguish both Cepheid types would bias the resulting PLRs. In the three dimensional Fourier space we also do see overlaps between the different Cepheid type sequences. To address this issue we require the Cepheids to be completely inside the defined areas including their errors. Consequently some Cepheids that are in a transition region, and therefore, cannot be assigned a definite Cepheid type. We label those as the unclassified (UN) Cepheid type. In K13 we called the Cepheids that were clipped in the Wesenheit relation UN Cepheids. As we will discuss later we still perform outlier clipping in the Wesenheit PLR and call the clipped Cepheids UN Cepheids. Therefore, in this paper UN Cepheids can arise from two reasons. We do not distinguish between the two and call both UN Cepheids. We separate the areas by defining the following linear parametrization: 
\begin{eqnarray}
  m_{\mathrm{FO}} = \frac{0.14-0.34}{\log(7.5) - \log(3.0)} \\
  t_{\mathrm{FO}} = 0.34 - \log(3.0) \cdot m_{\mathrm{FO}} \\
  m_{\mathrm{T2,1}} = 0.00 \\
  t_{\mathrm{T2,1}} = 5.05 \\
  m_{\mathrm{T2,2}} = \frac{0.48 - 0.25}{\log(30.0) - \log(10.0)} \\
  t_{\mathrm{T2,2}} = 0.25 - \log(10.0) \cdot m_{\mathrm{T2,2}} 
\end{eqnarray}
In order for a Cepheid to be assigned the FO type the following two equations have to be true (see Fig. \ref{fig_A21manualpretype}):
\begin{eqnarray}  
  A_{21} + A_{21,e} < \log(P) \cdot m_{\mathrm{FO}} + t_{\mathrm{FO}} \label{eqn_fo1}\\
  P < 7.5~\mathrm{d} \label{eqn_fo2}
\end{eqnarray}
For it to be a T2 Cepheid these equations (see Fig. \ref{fig_P21manualpretype}):
\begin{eqnarray} 
P > 11.95~\mathrm{d} \label{eqn_t21}\\
\varphi_{21} - \varphi_{21,e} > \log(P) \cdot m_{\mathrm{T2,1}} + t_{\mathrm{T2,1}} \label{eqn_t22}\\
P < 53.0~\mathrm{d} \label{eqn_t23}
\end{eqnarray}
or these equations (see Fig. \ref{fig_A21manualpretype}):
\begin{eqnarray}  
A_{21} - A_{21,e} > \log(P) \cdot m_{\mathrm{T2,2}} + t_{\mathrm{T2,2}} \label{eqn_t24}\\
P > 10.0~\mathrm{d} \label{eqn_t25}
\end{eqnarray}
have to be true (i.e. (Eqn. \ref{eqn_t21} AND Eqn. \ref{eqn_t22} AND Eqn. \ref{eqn_t23}) OR (Eqn. \ref{eqn_t24} AND Eqn. \ref{eqn_t25}) ). For it to be considered a FM Cepheid at least one of these equations has to be true (see Fig. \ref{fig_A21manualpretype}):
\begin{eqnarray}  
  A_{21} - A_{21,e} > \log(P) \cdot m_{\mathrm{FO}} + t_{\mathrm{FO}} \label{eqn_fm1}\\
  P > 7.5~\mathrm{d} \label{eqn_fm2}
\end{eqnarray}
and additionally at least one of these equations has to be true (see Fig. \ref{fig_P21manualpretype}):
\begin{eqnarray} 
 P < 11.95~\mathrm{d} \label{eqn_fm3}\\
 \varphi_{21} + \varphi_{21,e} < \log(P) \cdot m_{\mathrm{T2,1}} + t_{\mathrm{T2,1}} \label{eqn_fm4}\\
 P > 53.0~\mathrm{d} \label{eqn_fm5}
\end{eqnarray}
and also at least one of those equations has also to be true (see Fig. \ref{fig_P21manualpretype}):
\begin{eqnarray} 
 A_{21} + A_{21,e} < \log(P) \cdot m_{\mathrm{T2,2}} + t_{\mathrm{T2,2}} \label{eqn_fm6}\\
 P < 10.0~\mathrm{d} \label{eqn_fm7}
\end{eqnarray}
I.e. for a FM Cepheid the logical connection is: (Eqn. \ref{eqn_fm1} OR Eqn. \ref{eqn_fm2}) AND (Eqn. \ref{eqn_fm3} OR Eqn. \ref{eqn_fm4} OR Eqn. \ref{eqn_fm5}) AND (Eqn. \ref{eqn_fm6} OR Eqn. \ref{eqn_fm7}).
In case the Cepheid can not be classified as FM, FO or T2 it is classified as UN.
Fig. \ref{fig_Wmanualpretype} shows that the type classification of the manually selected Cepheid sample performs reasonably well considering the dispersion in the PLRs.

\subsection{Selection criteria \label{selection}}

In order to facilitate the automatic classification of Cepheids additional selection criteria are needed. The combination of color cut and three dimensional parameter space selection alone does not suffice to obtain a clean Cepheid sample. Table \ref{tabcuts-ri} summarizes the selection criteria and provides information on how many objects are rejected in each step. The cuts could be applied in any order without changing the resulting sample. The published Cepheid catalog also includes the tables and light curves of the clipped objects as well as information on the clipping reason in form of a bit flag. Some fits or, e.g., the final type determination are performed after certain cuts have been applied in order to save computation time. Therefore the catalog of the clipped objects might not have all information. The selection criteria are applied to the manually classified Cepheid sample as well as the sample that is selected using the three dimensional parameter space (3d sample). Table \ref{tabcuts-ri} provides the number of clipped objects for each sample. The selection criteria are applied first to the manual sample so that the three dimensional parameter space can be defined (as explained in the next section). The selection criteria are then applied again to the resulting 3d sample. The selection criteria in Table \ref{tabcuts-ri} are applied to the \rps and \ips band while the criteria in Table \ref{tabcuts-g} are applied to the \gps band. Objects clipped due to the criteria in Table 
\ref{tabcuts-ri} are removed from the final Cepheid sample while objects clipped only in the \gps band (Table \ref{tabcuts-g}) are flagged and therefore not used for the \gps PLR or the \gps sample in general.

Criterion \Rmnum{1}, the color cut discussed previously, is an important selection criterion for the 3d sample. Selection criterion \Rmnum{2} is a magnitude cut that ensures that the fitted light curve does not overshoot. As can be seen in Table \ref{tabcuts-ri} this cut is not used for the \rps and \ips band, but the cut is relevant for the \gps band (Table \ref{tabcuts-g}). This is not surprising since the \gps band has relatively few epochs and the light curves therefore have gaps which facilitate overshoots in the light curve fitting. Criterion \Rmnum{3} is an amplitude cut, where we reject small amplitudes because those are difficult to distinguish from noise. Amplitudes are smaller for longer wavelengths \citep{1991PASP..103..933M} and therefore we have a different upper amplitude cut in the \gps band. In criterion \Rmnum{4} we make a noise cut by comparing the median absolute deviation of the observed magnitudes with the fitted magnitudes in relation to the amplitude of the light curve. Large gaps in the light curve cause problems in fitting the light curve. As mentioned previously, the gaps might cause overshoots, but even without overshoots the shape of the light curve might not be determined well enough for the type classification or the determination of the Fourier parameters $A_{21}$ and $\varphi_{21}$. Therefore we introduce criterion \Rmnum{5} where we require that the largest gap is smaller than a quarter phase of the light curve. As we mentioned previously we perform ten thousand bootstrapping realizations for determining the magnitude error and the errors of the Fourier parameters $A_{21}$ and $\varphi_{21}$. For each of those realizations we determine the Cepheid type and require in criterion \Rmnum{6} that the Cepheid type is the same in more than nine thousand realizations. There are also two additional flags (64 and 128) not mentioned in Tables \ref{tabcuts-ri} and \ref{tabcuts-g}. Flag 64 is assigned to seven manually classified Cepheids and they are removed from the Cepheid sample because of their location in the three dimensional parameter space. The reason for this is explained in the next subsection. Flag 128 is assigned 19 times in the final Cepheid catalog (so those Cepheids are still included in the catalog) in order to mark Cepheids where the light curve from visual inspection does not look like a typical Cepheid light curve. We keep those Cepheids in the sample so that the selection of Cepheids is as objective as possible. The results do not change if we disregard those Cepheids.

\begin{table}[h!]
  \centering
  \caption{Selection criteria used in the \rps and \ips bands. The numbers of remaining Cepheids for the manual and 3d sample do not have to add up to the final number since some cuts are performed at the same time and some objects fulfill multiple criteria. The columns manual and 3d provide the number of clipped objects and the total number of objects still present at this step. The flag column provides a bit flag so that the reason for the clipping can be found in the published data. The magnitude of the fitted light curve is denoted as m(t) while the observed magnitudes are called $m_i$. $\Theta_i$ is the phase of the i-th observed epoch and $\Theta_{i+1}$ the next largest phase (which is not necessarily the next epoch).}
  \begin{tabular}{c|c|c|c|c|c}
  \quad & band & Selection criterion & manual & 3d & flag\\[2ex]
  \tableline
  \Rmnum{1} & all          & color cut                                                            & 295/1966 & 1922/3390 & 1\\[2ex]
  \tableline
  \Rmnum{2} & \rps \& \ips & $15~\mathrm{mag} \leq m(t) \leq 25~\mathrm{mag}$                     &   0/1671 &    0/1468 & 2\\[2ex]
  \tableline
  \Rmnum{3} & \rps \& \ips & $0.1~\mathrm{mag} \leq \max(m(t))-\min(m(t)) \leq 1.75~\mathrm{mag}$ &   1/1671 &   72/1468 & 4\\[2ex]
  \tableline
  \Rmnum{4} & \rps \& \ips & $\frac{\mathrm{median}(|m(t)-m_i|)}{\max(m(t))-\min(m(t))} \leq 0.3$           &  20/1671 &  343/1468 & 8\\[2ex]
  \tableline
  \Rmnum{5} & \rps \& \ips & $\max(\Theta_i-\Theta_{i+1}) \leq 0.25$                              &   0/1671 &    1/1468 & 16\\[2ex]
  \tableline
  \Rmnum{6} & all          & same type from lightcurve bootstrap $\geq 90\%$                      &   3/1671 &    73/1468 & 32\\[2ex]
  \tableline
  \quad     & \rps \& \ips & final sample                                                         &     1640 &       1046 & \quad \\[2ex]
  \end{tabular}
  \label{tabcuts-ri}
\end{table}

\begin{table}[h!]
	\centering
	\caption{Selection criteria used in the \gps band, otherwise see table \ref{tabcuts-ri}.}
	\begin{tabular}{c|c|c|c|c|c}
		\quad & band & Selection criterion & manual & 3d & flag\\[2ex]
		\tableline
		\Rmnum{1} & all          & color cut                                                            & 295/1966 & 1922/3390 & 1\\[2ex]
		\tableline
		\Rmnum{2} & \gps         & $15~\mathrm{mag} \leq m(t) \leq 25~\mathrm{mag}$                     & 162/1671 &  391/1468 & 2\\[2ex]
		\tableline
		\Rmnum{3} & \gps         & $0.1~\mathrm{mag} \leq \max(m(t))-\min(m(t)) \leq 2.00~\mathrm{mag}$ & 198/1671 &  504/1468 & 4\\[2ex]
		\tableline
		\Rmnum{4} & \gps         & $\frac{\mathrm{median}(|m(t)-m_i|)}{\max(m(t))-\min(m(t))} \leq 0.3$           &  24/1671 &  362/1468 & 8\\[2ex]
		\tableline
		\Rmnum{5} & \gps         & $\max(\Theta_i-\Theta_{i+1}) \leq 0.25$                              & 175/1671 &  246/1468 & 16\\[2ex]
		\tableline
		\Rmnum{6} & all          & same type from lightcurve bootstrap $\geq 90\%$                      &   3/1671 &    73/1468 & 32\\[2ex]
		\tableline
		\quad & \gps             & final sample                                                         &     1302 &        595 & \quad \\[2ex]
	\end{tabular}
	\label{tabcuts-g}
\end{table}

\subsection{3d classification\label{3dcut}}

\begin{figure}
\centering
\includegraphics[width=\linewidth]{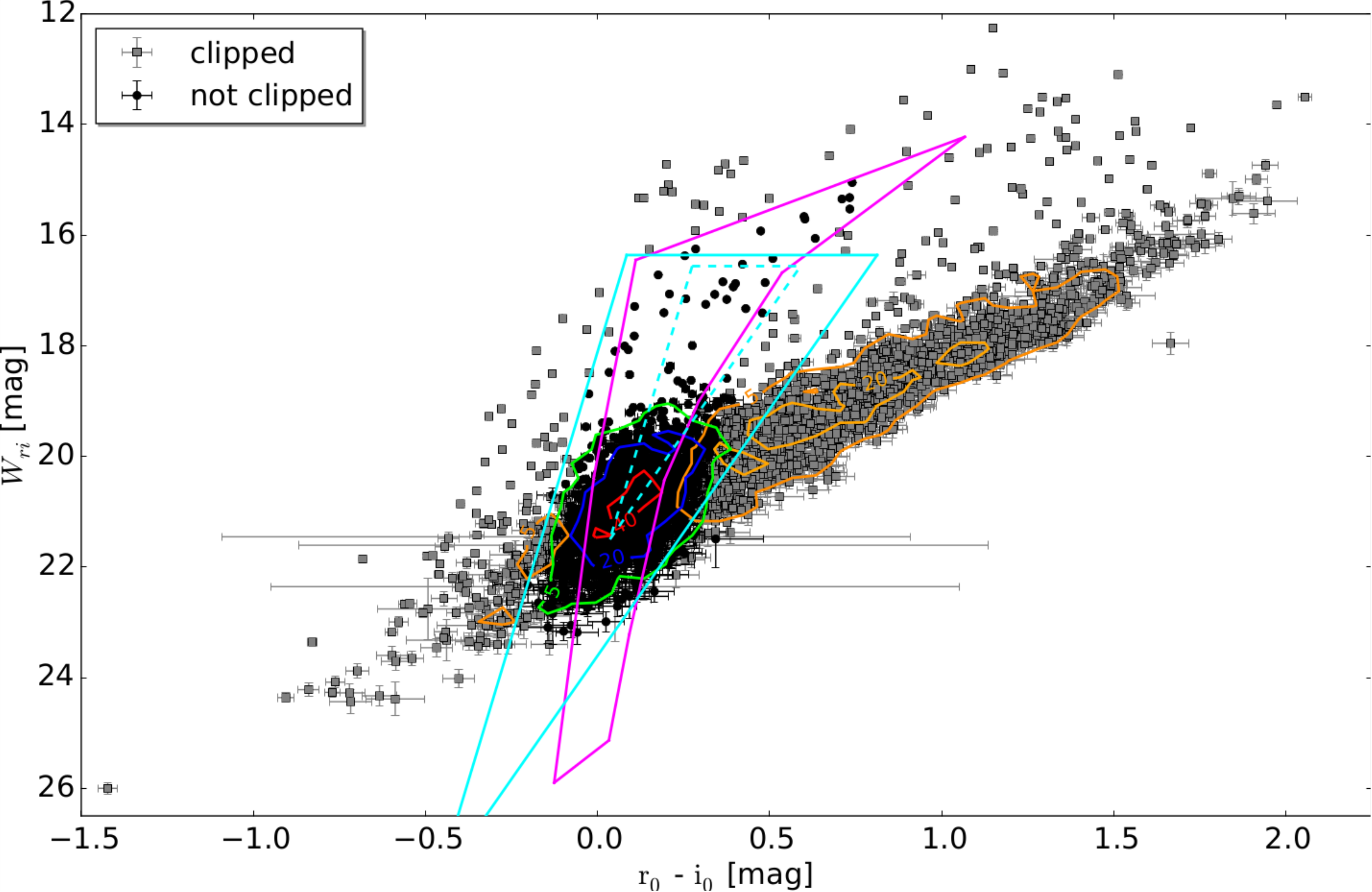}
\caption{Color-Wesenheit cut for the Cepheid sample selected by the three dimensional parameter space. 3390 light curves are selected by the three dimensional parameter space classification. As for Figure \ref{fig_colorWesenheitmanual} the color cut is applied and 1922 sources are clipped, with 1468 sources remaining after the color cut. The extinction correction is done by using the \citet{2009A&A...507..283M} color excess map. The edges of the instability strip are used to constrain the color excess estimate. The lines are the same as in Fig. \ref{fig_colorcutregion}. The error bars shown are the photometric errors. The possible color excess range is also considered for the cut, therefore some points appear to be outside of the region defined by the magenta lines but are still classified as Cepheids.
\label{fig_colorWesenheit3d}}
\end{figure}

Similar to K13 we use the three dimensional parameter space $A_{21}$-$\varphi_{21}$-$P$ to select light curves that have a shape typical for a Cepheid. The advantage of this approach is that it is less biased than visual selection and that the selection is reproducible. Nevertheless the approach requires the manual sample in order to define this 3d space. In K13 we span a grid in this 3d space and selected those grid points where one manually classified Cepheid resides in. By doing this we had to choose six parameters, i.e. the grid size and the grid starting point for each dimension. Here we improve the method by using only three parameters. We define a sphere around each point in the manual sample. In order for an object to be selected it has to be inside at least one of the spheres that are made up by the manual sample in the $A_{21}$-$\varphi_{21}$-$P$ space. One of the free parameters is the radius of the sphere and the other two are the normalizations of the other two dimensions, i.e. how each dimension is scaled compared to the other dimensions. We choose $A_{21}$-$(2\pi)^{-1}\varphi_{21}$-$\log(P)$ as the new three dimensional space. We choose it such that the values of the manual sample have than a similar range. The reason for this is that we can use an sphere and do not have to use an ellipsoid. E.g. if we use a radius of 0.2 in $A_{21}$-$\varphi_{21}$-$P$ space, that would be a large difference in $A_{21}$ while it is only a small change in $\varphi_{21}$ and almost no change in $P$. In order to determine the radius of the sphere we use the manual sample. We calculate for each Cepheid in the manual sample the distance in $A_{21}$-$(2\pi)^{-1}\varphi_{21}$-$\log(P)$ space to the next closest Cepheid. The cumulative distribution function of this distance is than used to choose the radius of the sphere. We select a radius of 0.088 for the sphere (99.5\% of Cepheids are closer than this distance to their next neighbor). Seven Cepheids have a larger distance to the next neighbor. We choose to cut those seven Cepheids because they are in a region of the 3d space that is sparsely sampled and if they were to be included the radius of the sphere would have to be increased significantly which makes the selection worse (a larger radius means a more fuzzy selection). If the seven Cepheids were left in the sample but the 0.088 radius would be used then the seven Cepheids would be rejected by the 3d selection process if the manual sample would be subjected to the 3d selection (which it is not because it is used to define it). So for consistency reasons the seven Cepheids are excluded. As described in the previous subsection the seven Cepheids are flagged with the 64 bit flag. In the Appendix the location of those seven Cepheids in the 3d space can bee seen in Figures \ref{fig_manualdistcutA21}, \ref{fig_manualdistcutP21} and \ref{fig_manualdistcutW}. The 3d classification finds an additional 3390 Cepheid candidates. As can be seen in Figure \ref{fig_colorWesenheit3d} 1922 of those are clipped due to the color cut.

\subsection{Outlier clipping \label{clipping}}

\begin{figure}
\centering
\includegraphics[width=\linewidth]{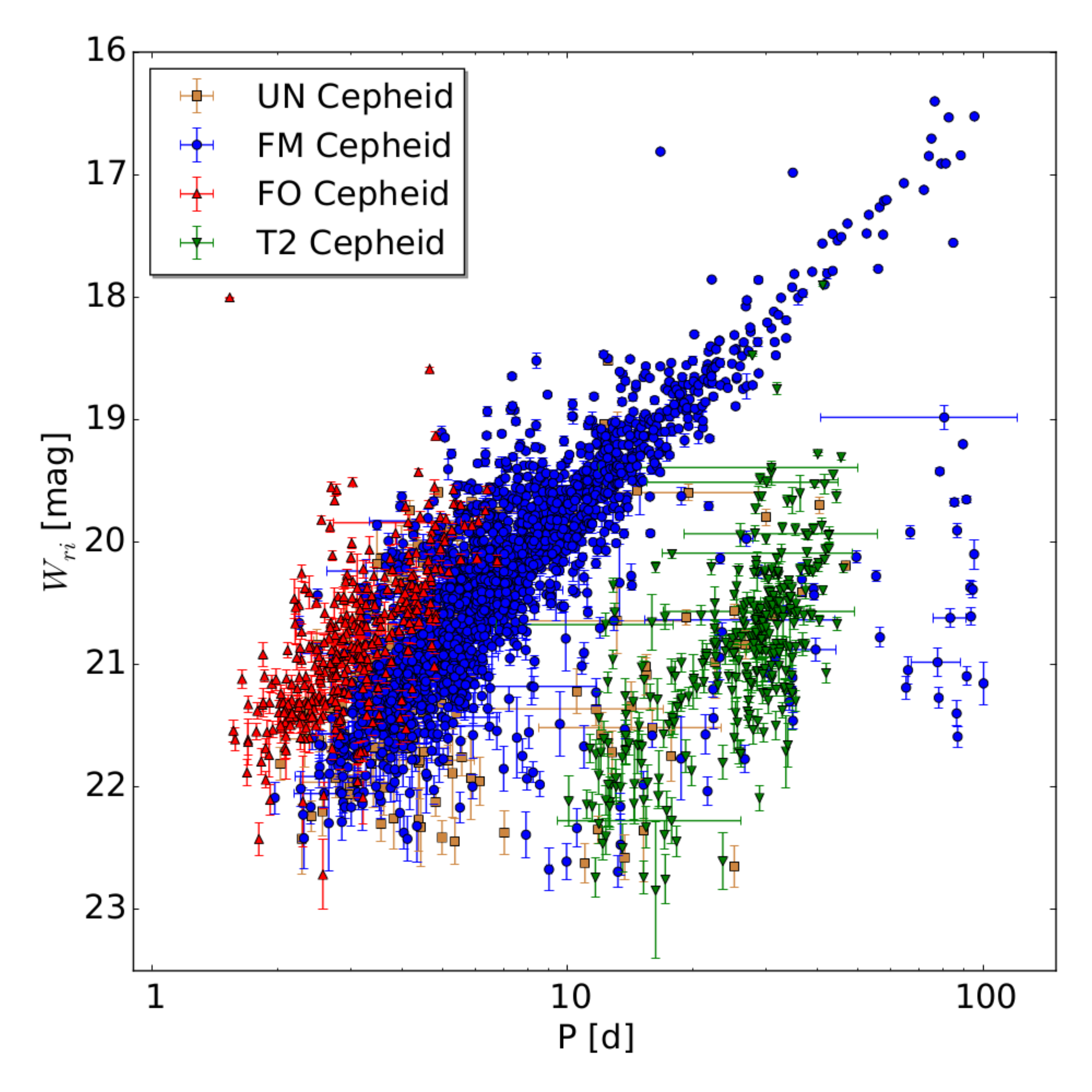}
\caption{Period Wesenheit diagram for the complete sample (i.e. the manual sample combined with the 3d sample) without outlier clipping. All UN Cepheids are due to uncertainties in the type classification. Most outliers are due to misclassification and crowding. In total there are 203 UN Cepheids, 1851 FM Cepheids, 331 FO Cepheids and 301 T2 Cepheids.
\label{fig_Wallpretype}}
\end{figure}

\begin{figure}
\centering
\includegraphics[width=\linewidth]{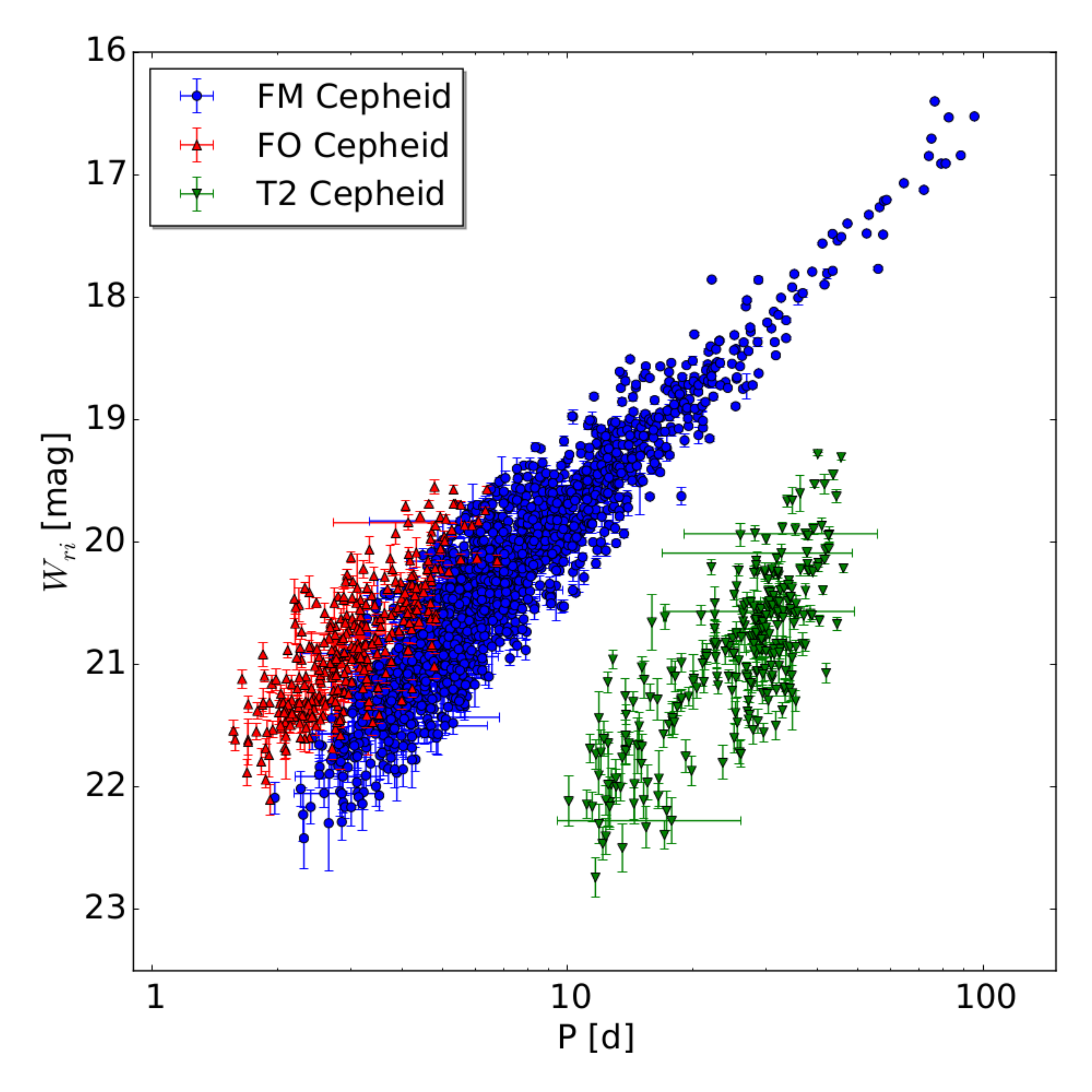}
\caption{Period Wesenheit diagram for the complete sample after the outlier clipping. The outlier clipping was performed with the method we developed in K15. As can be seen the outliers still present in Figure \ref{fig_Wallpretype} are clipped. In total there are 1662 FM Cepheids, 307 FO Cepheids and 278 T2 Cepheids.
\label{fig_Wallfinaltype}}
\end{figure}

\begin{figure}
\centering
\includegraphics[width=\linewidth]{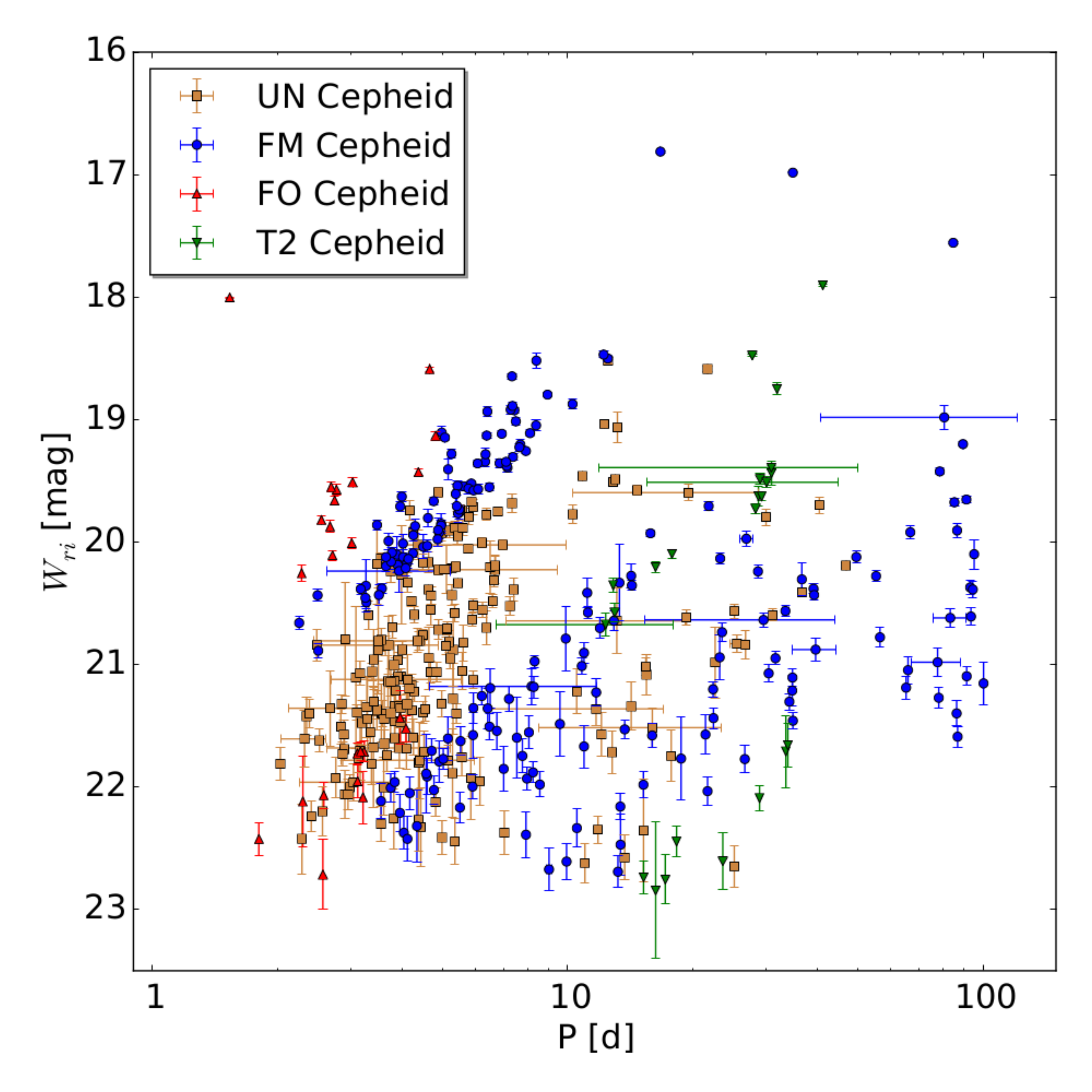}
\caption{Period Wesenheit diagram of the Cepheids clipped from the complete sample. The UN Cepheids shown here are due to uncertainties in the type classification. All Cepheids shown in this plot are assigned the UN type after the clipping is performed and we do not distinguish between the two reasons for an UN Cepheid (type classification and clipping) anymore. Shown are 203 UN Cepheids, 189 FM Cepheids, 24 FO Cepheids and 23 T2 Cepheids.
\label{fig_allclippedfinaltype}}
\end{figure}

\begin{figure}
\centering
\includegraphics[width=\linewidth]{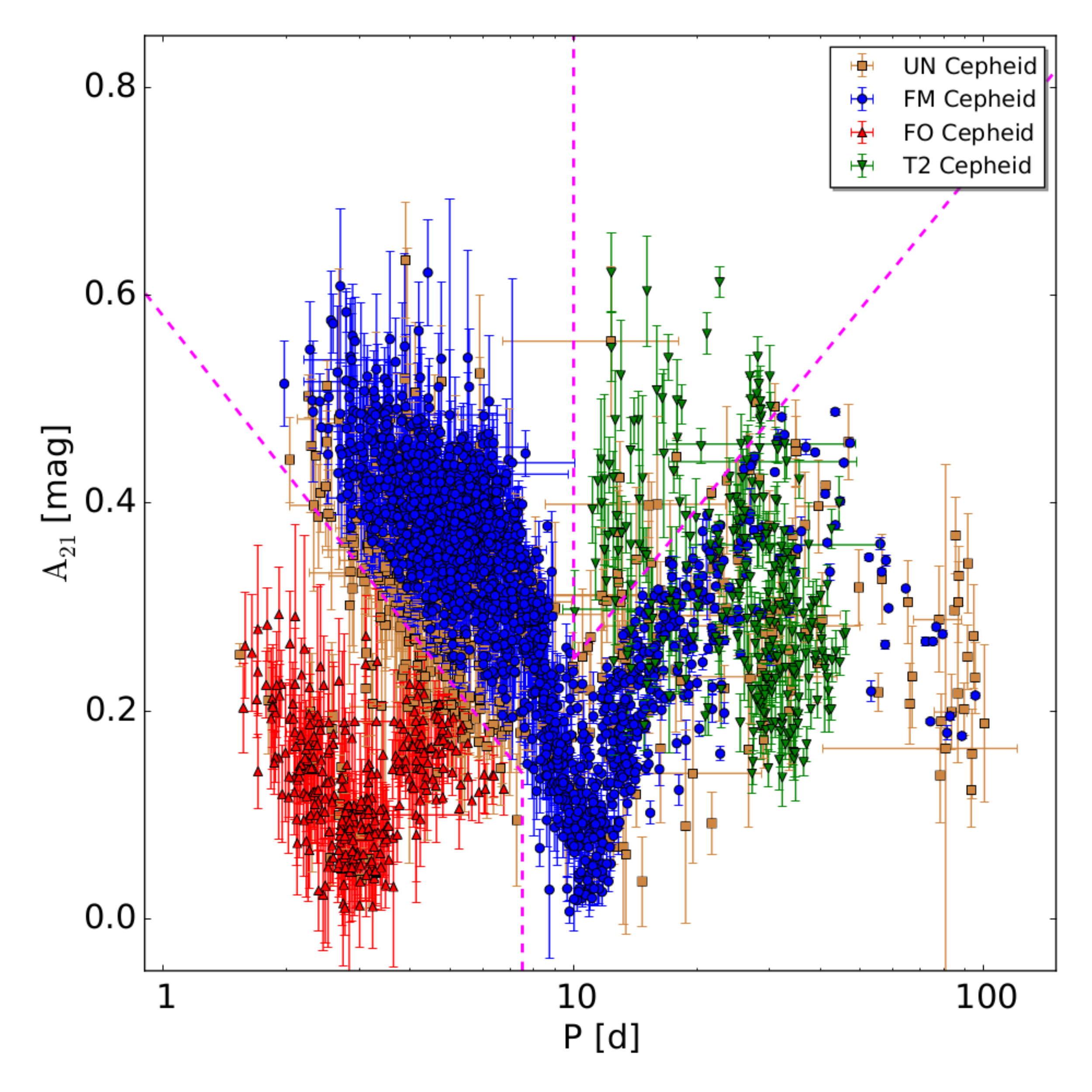}
\caption{Amplitude ratio ($A_{21}$) diagram for the final sample. The parameter space is better sampled than for the manual sample in Figure \ref{fig_A21manualpretype}. The shape of the three dimensional parameter space projection is similar to that of the manual sample since the additional Cepheids have been selected using this parameter space. The dashed magenta lines are the same as in Figure \ref{fig_A21manualpretype} and are used for the type classification. In total there are 439 UN Cepheids, 1662 FM Cepheids, 307 FO Cepheids and 278 T2 Cepheids.
\label{fig_A21}}
\end{figure}

\begin{figure}
\centering
\includegraphics[width=\linewidth]{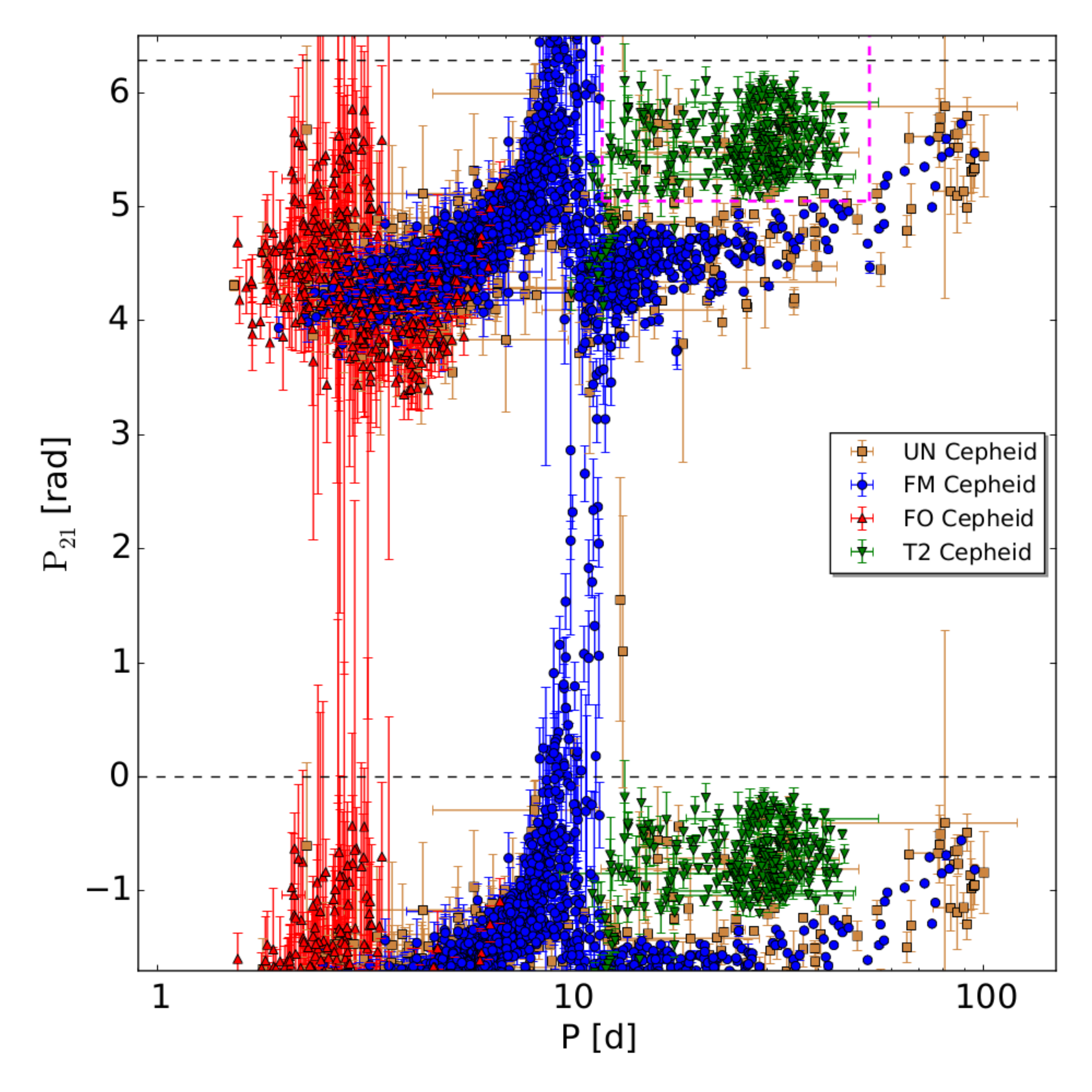}
\caption{Phase difference ($\varphi_{21}$) diagram for the final sample. Similar to Figure \ref{fig_A21} this projection of the three dimensional parameter space is also more densely populated than in the manual sample shown in Figure \ref{fig_P21manualpretype}. The dashed magenta lines are the same as in Figure \ref{fig_P21manualpretype} and also used to determine the Cepheid type. The data between the black dashed lines is periodic and therefore also plotted periodic. The number of Cepheids is the same as in Figure \ref{fig_A21}.
\label{fig_P21}}
\end{figure}

Figure \ref{fig_Wallpretype} shows the Wesenheit PLR of the combined manual and 3d sample after all selection criteria have been applied. As can be seen there are outliers. There are different reasons for these outliers: crowding, misclassification, misidentification, blending and extinction. In K15 we discuss those reasons in more detail, but we want to point out that the crowding problem is significantly more severe in this ground based optical data than in the HST data. In order to remove the outliers from the sample, we perform an iterative $\kappa$-$\sigma$ clipping with the median absolute deviation of the residuals as the magnitude error. This method that we developed in K15 is very robust and is also used in K18b as well as in \citet{2016ApJ...826...56R} and \citet{2016ApJ...830...10H}. As in K15 and K18b we use a $\kappa = 4$. The clipped Wesenheit PLR shown in Figure \ref{fig_Wallfinaltype} has no outliers anymore. In Figure \ref{fig_allclippedfinaltype} we show all Cepheids that were clipped as well as the UN Cepheids. The UN Cepheids in this figure were assigned the UN type because the Cepheids are occupying a transitional region where the type classification is unsecure (see subsection \ref{section_typeclass}). In K13 the UN type was assigned only to the clipped Cepheids. Here the UN type is assigned to all Cepheids shown in Figure \ref{fig_allclippedfinaltype}. After this point we do not distinguish between the two reasons a Cepheid could have been assigned the UN type. The three dimensional parameter space of the final clipped sample is shown in Figures \ref{fig_A21} and \ref{fig_P21}. In these figures we already do not distinguish between the two UN type origins. The three dimensional parameter space is more densely sampled and of course it is similar to the manual sample shown in Figures \ref{fig_A21manualpretype} and \ref{fig_P21manualpretype} since the additional Cepheids have been selected by this 3d space. It can also be seen that in the final sample the transitional region between Cepheid types is more prominently populated and therefore there are more UN Cepheids in these regions than in the manual sample.

\section{PAndromeda Cepheid catalog\label{cat}}

\begin{figure}
\centering
\includegraphics[width=\linewidth]{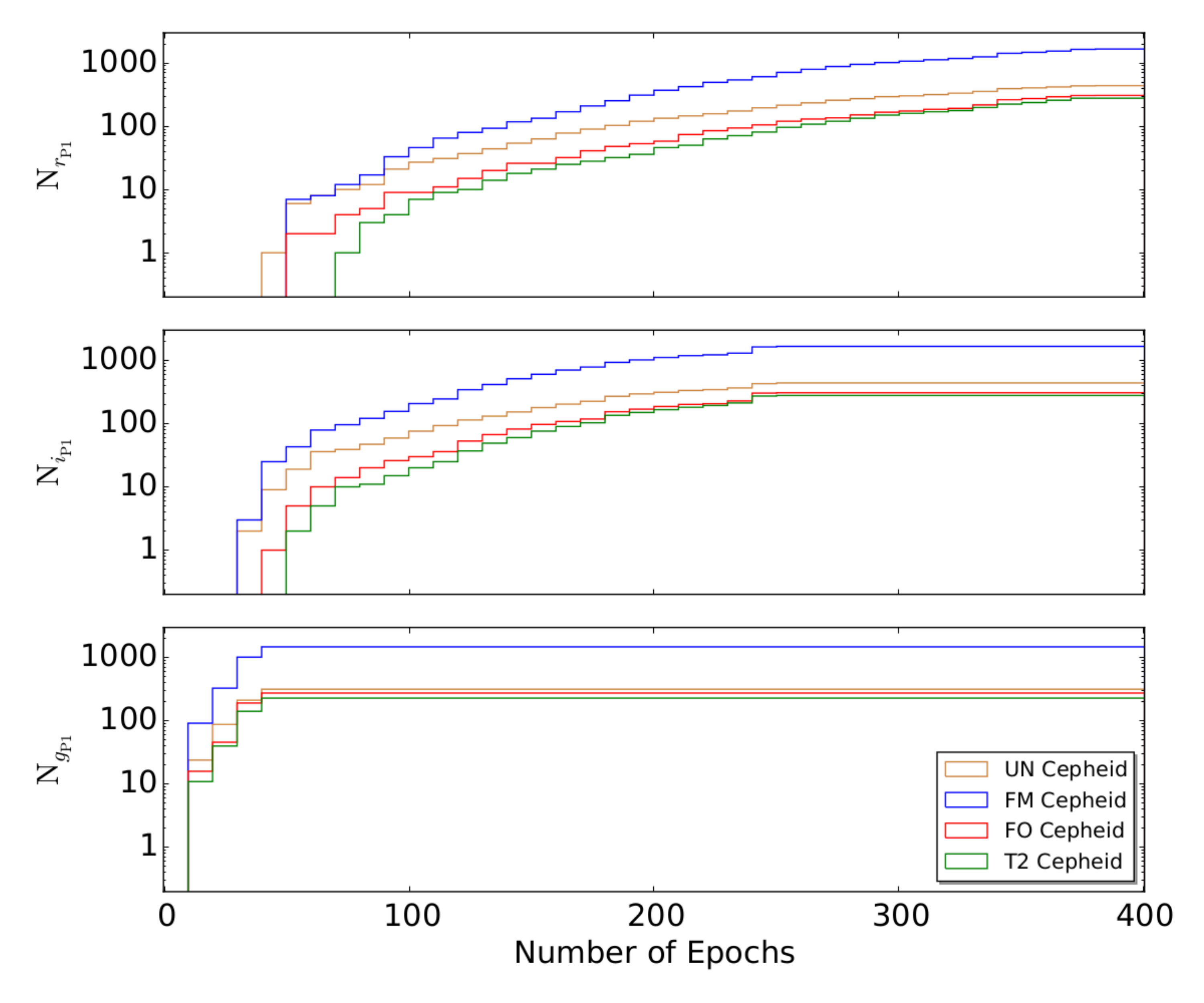}
\caption{Cumulative histogram of the epoch distribution for the \rps, \ips and \gps band of the PAndromeda Cepheid catalog. The bin size is 10 epochs. In total the PAndromeda data consists of up to 420 epochs in the $\rps$-band, up to 262 epochs in the $\ips$-band and up to 56 epochs in the $\gps$-band. As can be seen these numbers are not reached because we require the Cepheid epochs in the light curves to have a signal to noise ratio of larger than two and a visit stack depth of more than two frames.		
\label{fig_epochs}}
\end{figure}

\begin{figure}
\centering
\includegraphics[width=\linewidth]{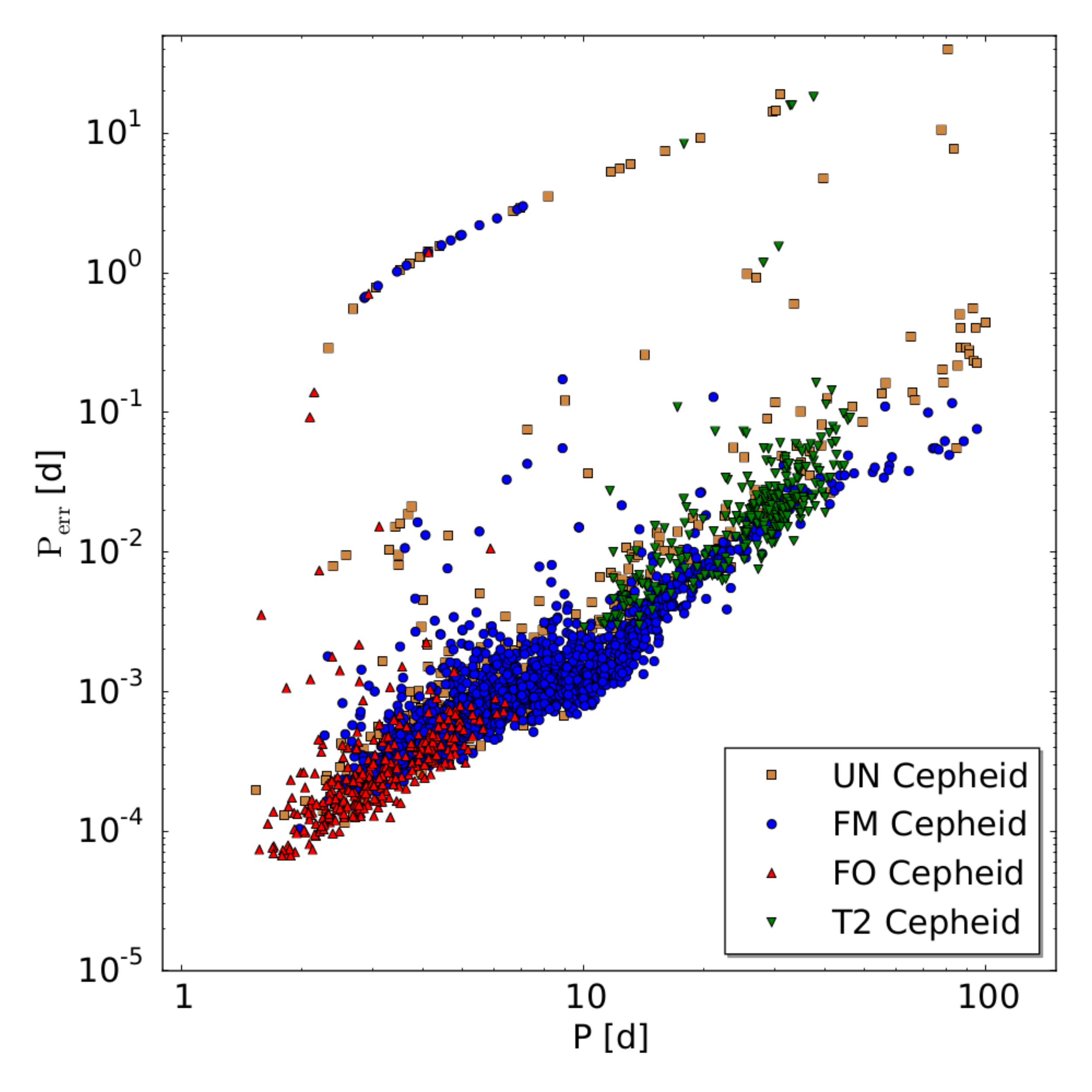}
\caption{Period error digram. The period errors are determined trough the bootstrapping method. The sequence with the larger period errors is caused by aliasing. The period errors are very small since the PAndromenda Cepheid sample has a lot of epochs. 
\label{fig_pererr}}
\end{figure}

\begin{figure}
\centering
\includegraphics[width=0.9\linewidth]{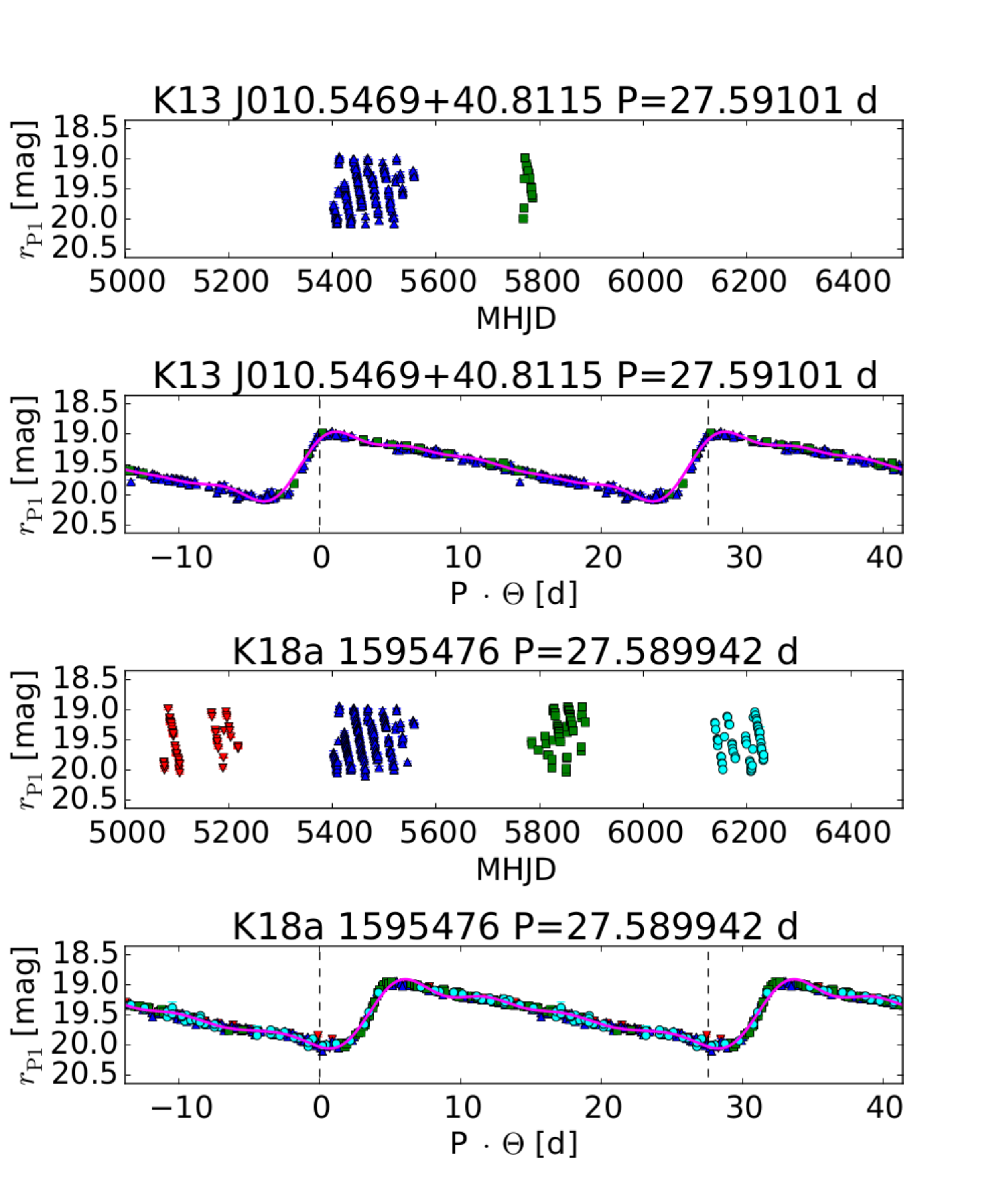}
\caption{Example light curve of a FM Cepheid in the \rps band. The two upper panels show the K13 data and the two bottom panels the new data. Each upper panel of the two samples shows the unfolded light curve with the modified heliocentric julian date on the x axis. Each bottom panel shows the folded light curve and the fit to the light curve as a magenta line. Here the x axis is the product of the period with the phase $\Theta$. The zero point of the phase is arbitrary. Each seasons observations are marked in a different color.
\label{fig_bsplc}}
\end{figure}

\begin{figure}
\centering
\includegraphics[width=\linewidth]{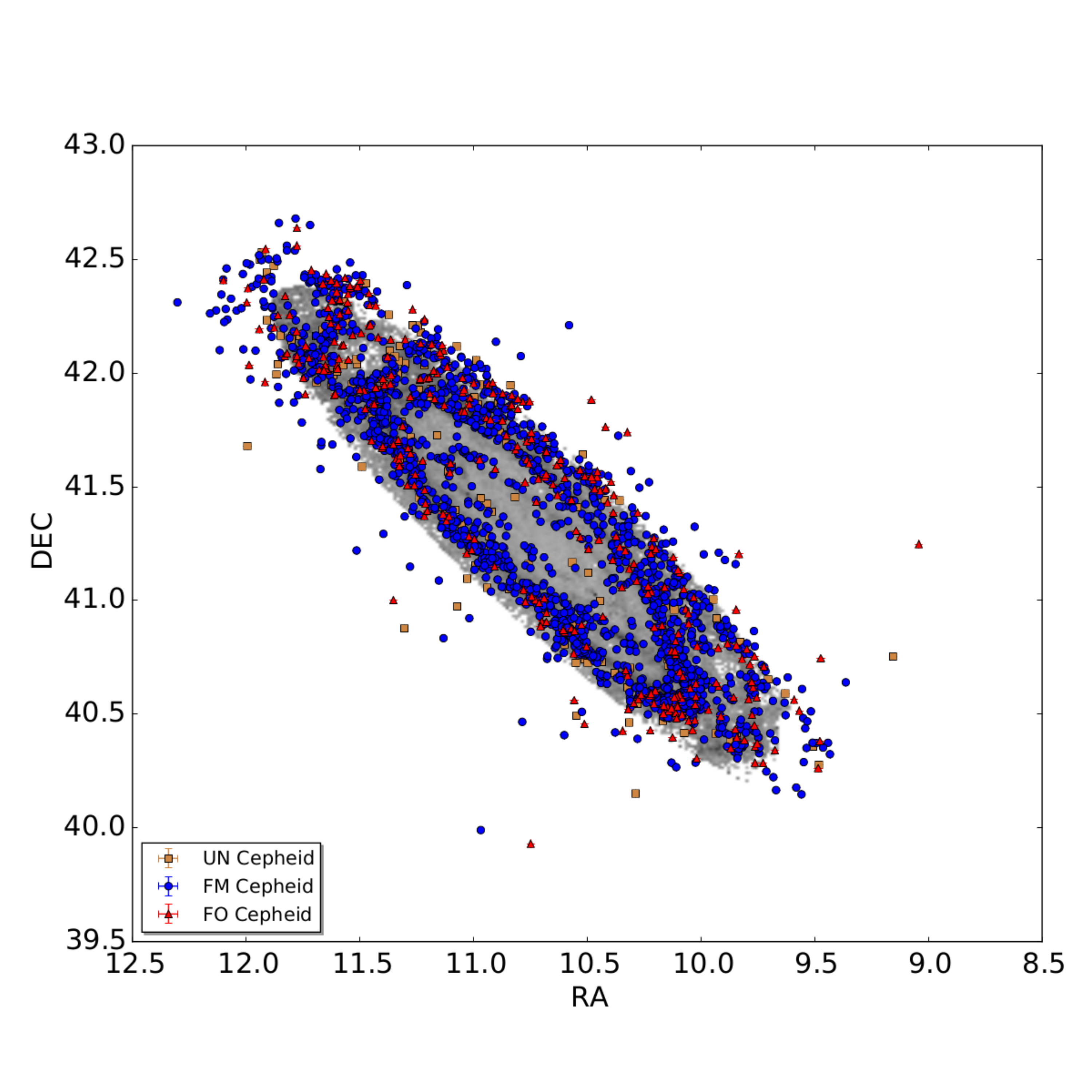}
\caption{Spatial distribution of FM and FO Cepheids of the PAndromeda Cepheid catalog. The Cepheids are plotted over the E(B-V) map of \citet{2009A&A...507..283M} and they follow the ring structure of M31.
\label{fig_radec}}
\end{figure}

\begin{figure}
\centering
\includegraphics[width=\linewidth]{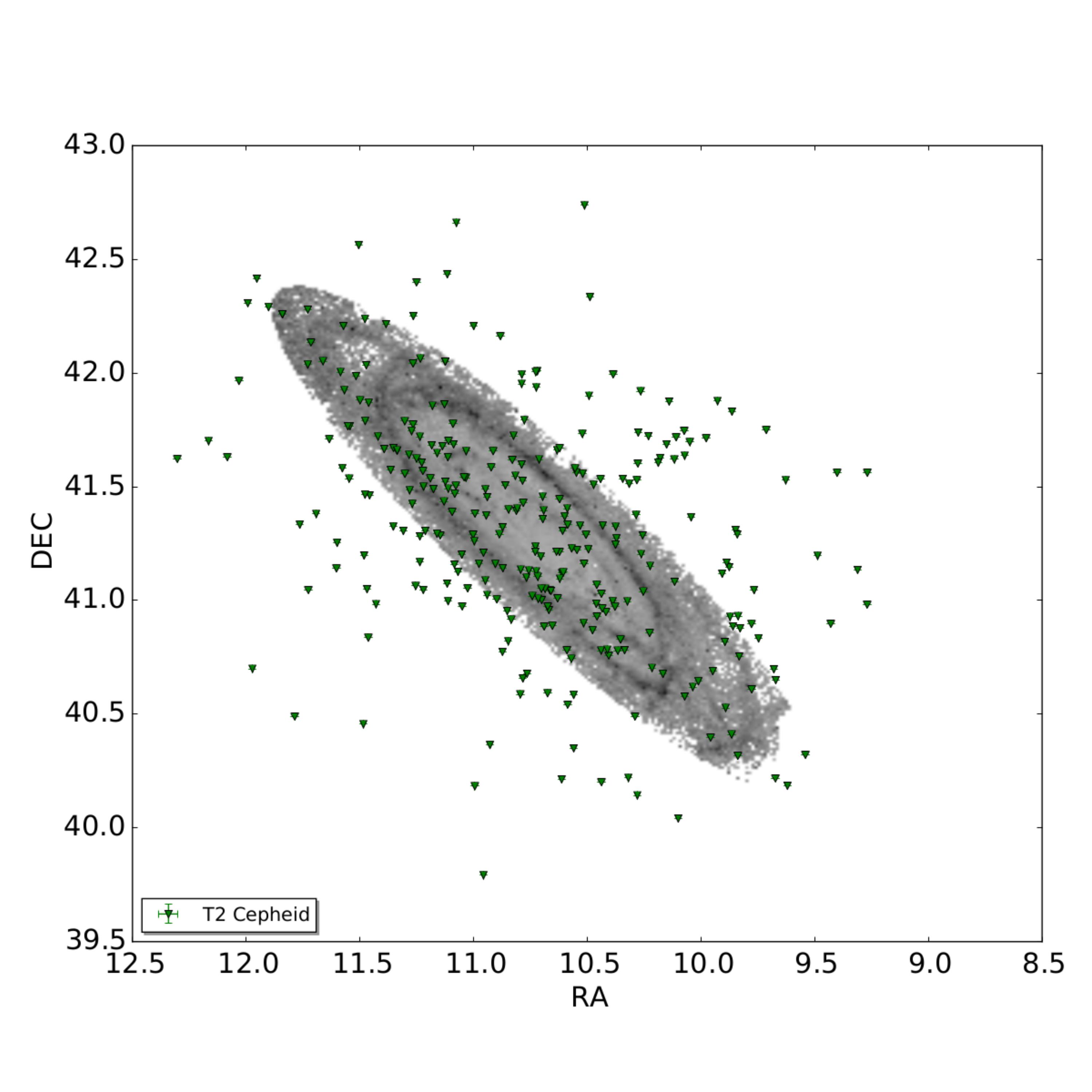}
\caption{Spatial distribution of T2 Cepheids of the PAndromeda Cepheid catalog. Same as in Figure \ref{fig_radec} the Cepheids are plotted over the \citet{2009A&A...507..283M} E(B-V) map. In contrast to the FM and FO Cepheids, the T2 Cepheids follow the M31 halo.
\label{fig_radect2}}
\end{figure}

The PAndromeda Cepheid catalog consists of 1662 FM Cepheids, 307 FO Cepheids, 278 T2 Cepheids and 439 UN Cepheids and is the largest and most homogeneous catalog of Cepheids in M31. For all Cepheids we provide the light curves as well as the fitted parameters in electronic form. In the appendix we show an excerpt of the published tables and briefly discuss the fit parameters. We also provide the light curves and fit parameters of 2670 sources that have been cut for the various reasons discussed in the previous section. The appendix also includes a comparison to the K13 data. As can be seen in Figure \ref{fig_epochs} many Cepheids in the \rps band have a lot of epochs and therefore the periods can be determined very accurately as seen in Figure \ref{fig_pererr}. An example light curve is shown in Figure \ref{fig_bsplc}. Typically the K13 data has less epochs as it is also the case in the example light curve and the phase zero point is arbitrary and therefore it will be different. Note that because of the strict masking some epochs present in K13 might be missing here. In some extreme cases there are no epochs from a certain year of the observing campaign due to the masking. This is in part the reason that some Cepheids identified in K13 are missing in this sample. Some areas are completely masked and those Cepheids can no longer be recovered. 140 Cepheids from K13 have no data due to masking. 148 Cepheids do not make the initial cut where we require the periods in the \rps and \ips band to be similar to one percent and in a period range between 1.5 and 150 days. Those Cepheids are not included in the published light curves of the selected sources, because we do not publish the sources that do not make the first cut due to their large quantity. 148 Cepheids are a rather large number of missing Cepheids, since all K13 should make this simple cut. The strict masking can remove a significant number of epochs in one band such that the period can not be determined precisely anymore. This is then the reason that the Cepheids do not make the first cut. Of the remaining 1721 Cepheids from K13, 1453 are part of the PAndromeda Cepheid catalog. The other 268 Cepheids are cut for various reasons. As with the other sources that have been cut, the light curves are also published and the reason for the cut is encoded in the bit flag. A cross-identification table is provided with the published data. In the appendix we compare the Cepheid type of the two samples. The spatial distribution of FM and FO Cepheids shown in Figure \ref{fig_radec} follows the ring structure in M31, while the T2 Cepheids shown in Figure \ref{fig_radect2} trace the M31 halo.

\section{Results\label{results}}

\begin{figure}
\centering
\includegraphics[width=0.9\linewidth]{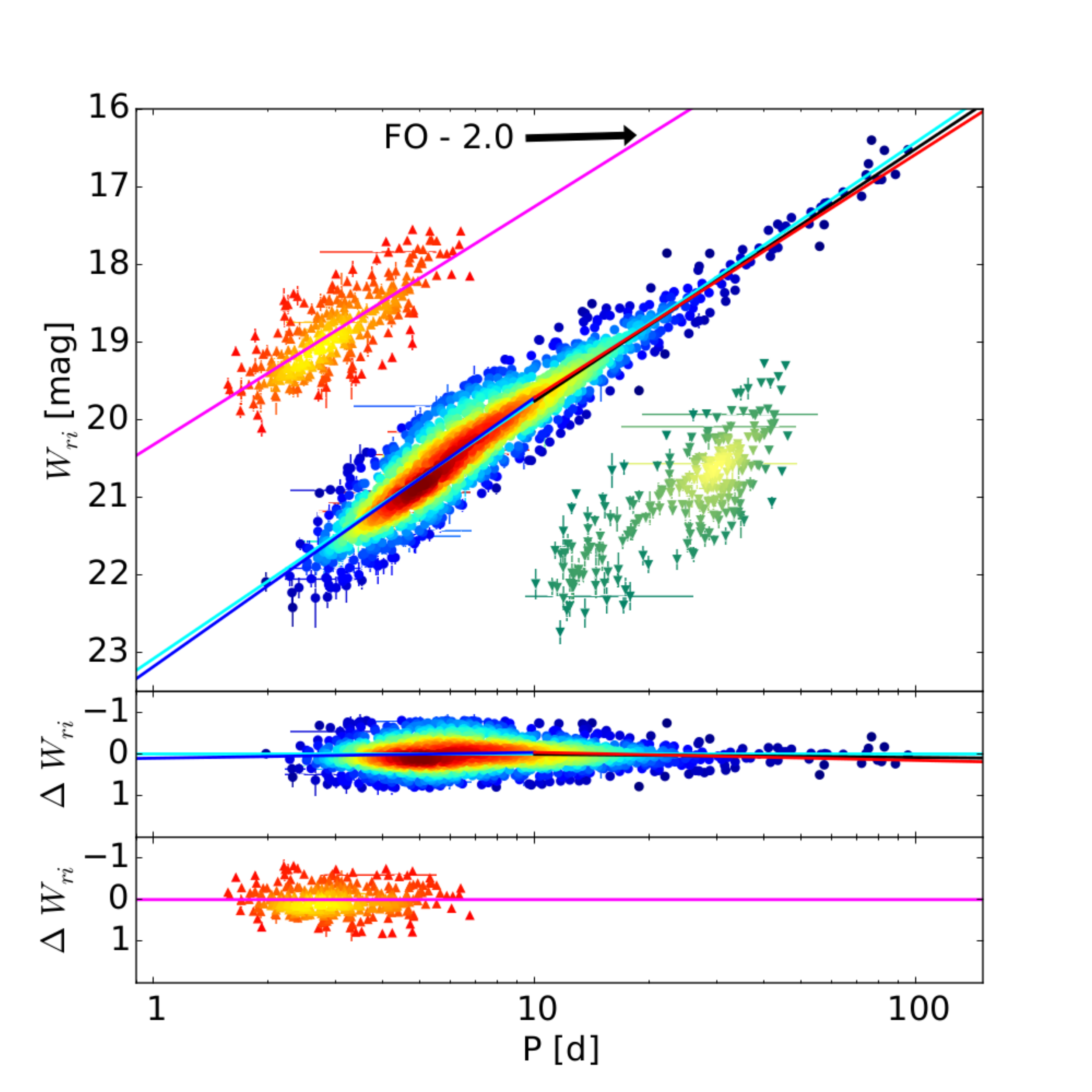}
\caption{Period Wesenheit diagram. In total there are 1662 FM Cepheids, 307 FO Cepheids and 278 T2 Cepheids. In the top panel the FO Cepheids have been shifted by two magnitudes for better visibility. The FM Cepheids are shown as circles, the FO Cepheids are shown as upwards pointing triangles, while the T2 Cepheids are shown as downwards pointing triangles. The color is assigned according to the density of Cepheids in this part of the PLR by using the kernel density estimate. The solid cyan line corresponds to fit \#1 in Table \ref{table_PLRs}, while the black solid line is fit \#2 (only long period Cepheids). The magenta line is fit \#3 in the same table. The blue and red solid lines are from the broken slope fit \#1 in Table \ref{table_PLRs_broken}. The two bottom panels show the residuals of the fit to the cyan and magenta solid line respectively.	
\label{fig_W}}
\end{figure}

\begin{figure}
\centering
\includegraphics[width=0.9\linewidth]{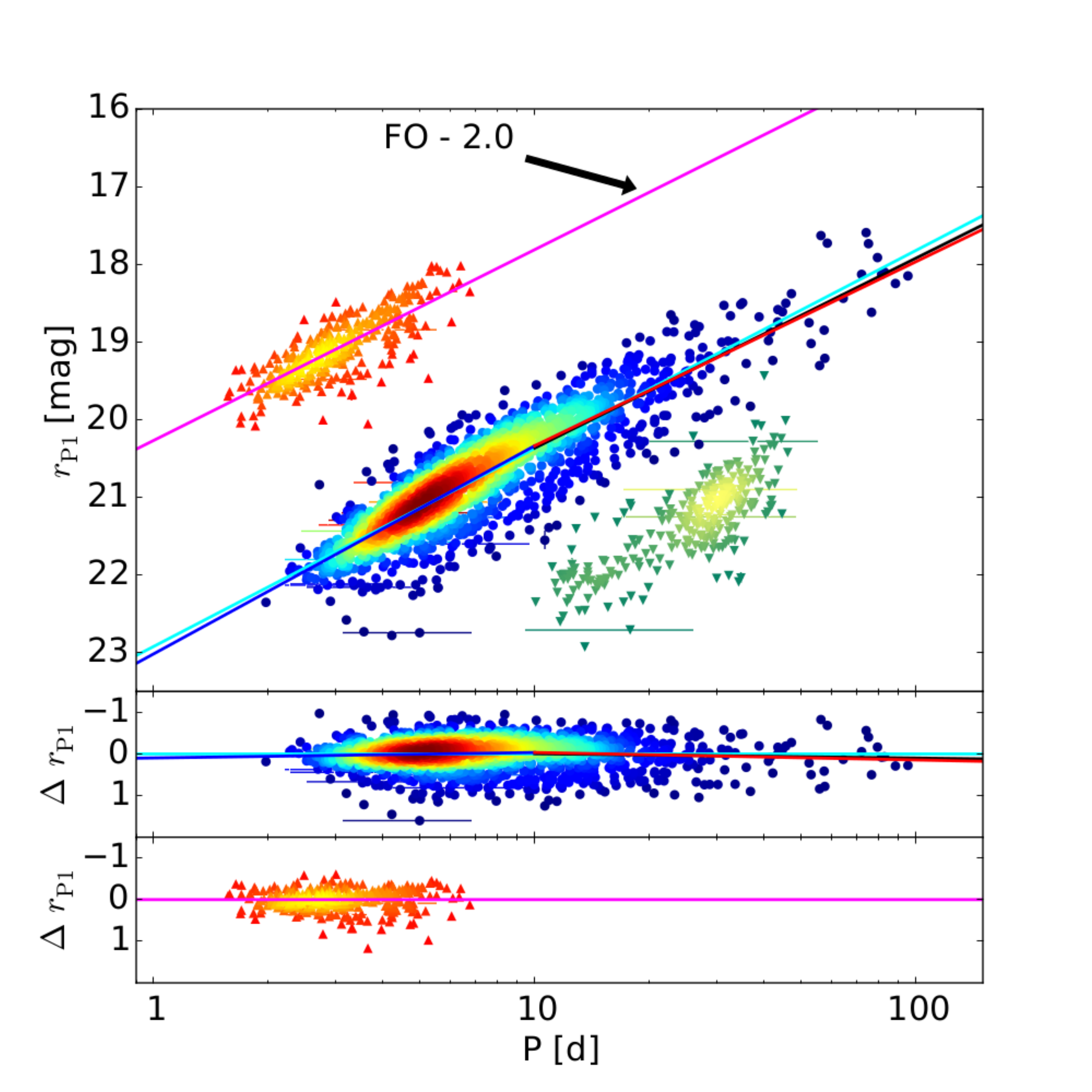}
\caption{\rps band PLR. The number of Cepheids is the same as in Figure \ref{fig_W}. The FO Cepheids have been offset by two magnitudes for better visibility. The FM Cepheids are shown as circles, the FO Cepheids are shown as upwards pointing triangles, while the T2 Cepheids are shown as downwards pointing triangles. The FM Cepheid PLR is shown as solid cyan line and corresponds to the fit \#4 in Table \ref{table_PLRs}. The fit to the long period FM Cepheids is shown as black solid line (fit \#5 in Table \ref{table_PLRs}), while the FO Cepheid PLR is shown as magenta solid line (fit \#6 in Table \ref{table_PLRs}). The blue and red solid lines are the fit of a broken slope with a common suspension point shown as fit \#2 in Table \ref{table_PLRs_broken}. Same as in \ref{fig_W} the two bottom panels show the residuals to the fits \#4 and \#6 in Table \ref{table_PLRs} respectively. 
\label{fig_r}}
\end{figure}

\begin{figure}
\centering
\includegraphics[width=0.9\linewidth]{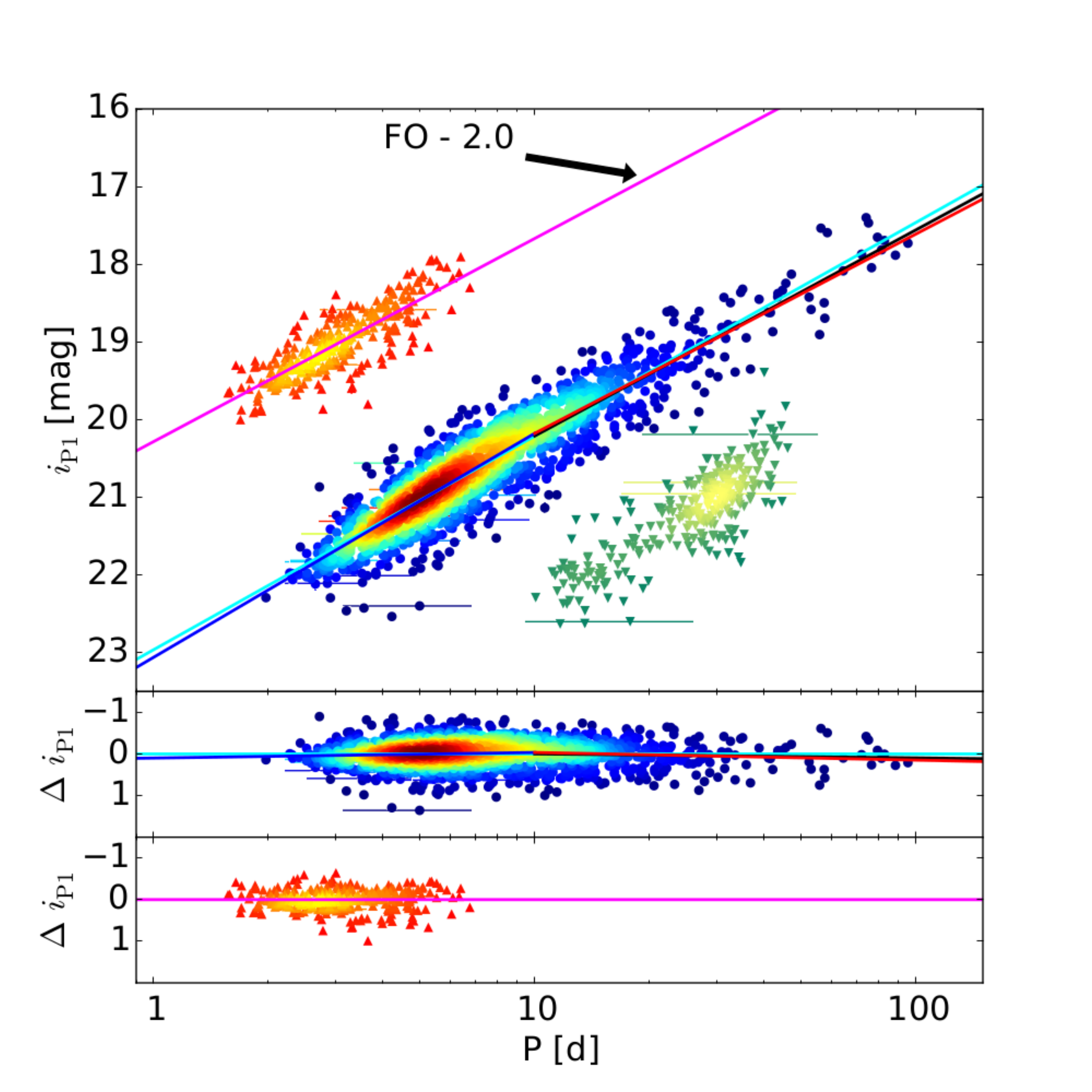}
\caption{\ips band PLR. Shown are the fits \#7, \#8 and \#9 from Table \ref{table_PLRs} and fit \#3 from Table \ref{table_PLRs_broken}. Otherwise analog to Figures \ref{fig_W} and \ref{fig_r}. 
\label{fig_i}}
\end{figure}

\begin{figure}
\centering
\includegraphics[width=0.9\linewidth]{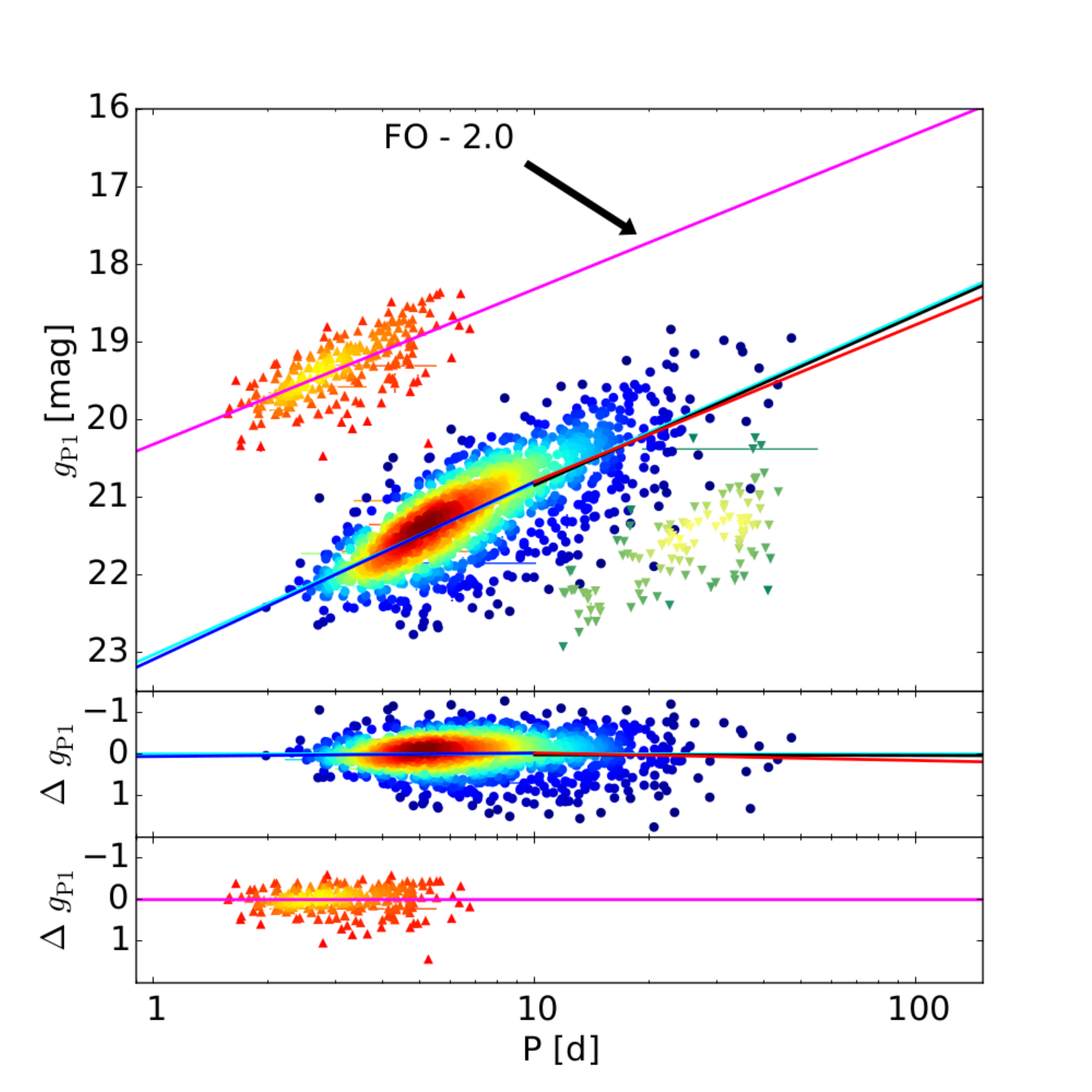}
\caption{\gps band PLR. Analog to Figures \ref{fig_W} and \ref{fig_r}, but shown are the fits \#10, \#11 and \#12 from Table \ref{table_PLRs} and fit \#4 from Table \ref{table_PLRs_broken}.
\label{fig_g}}
\end{figure}

\begin{figure}
\centering
\includegraphics[width=0.9\linewidth]{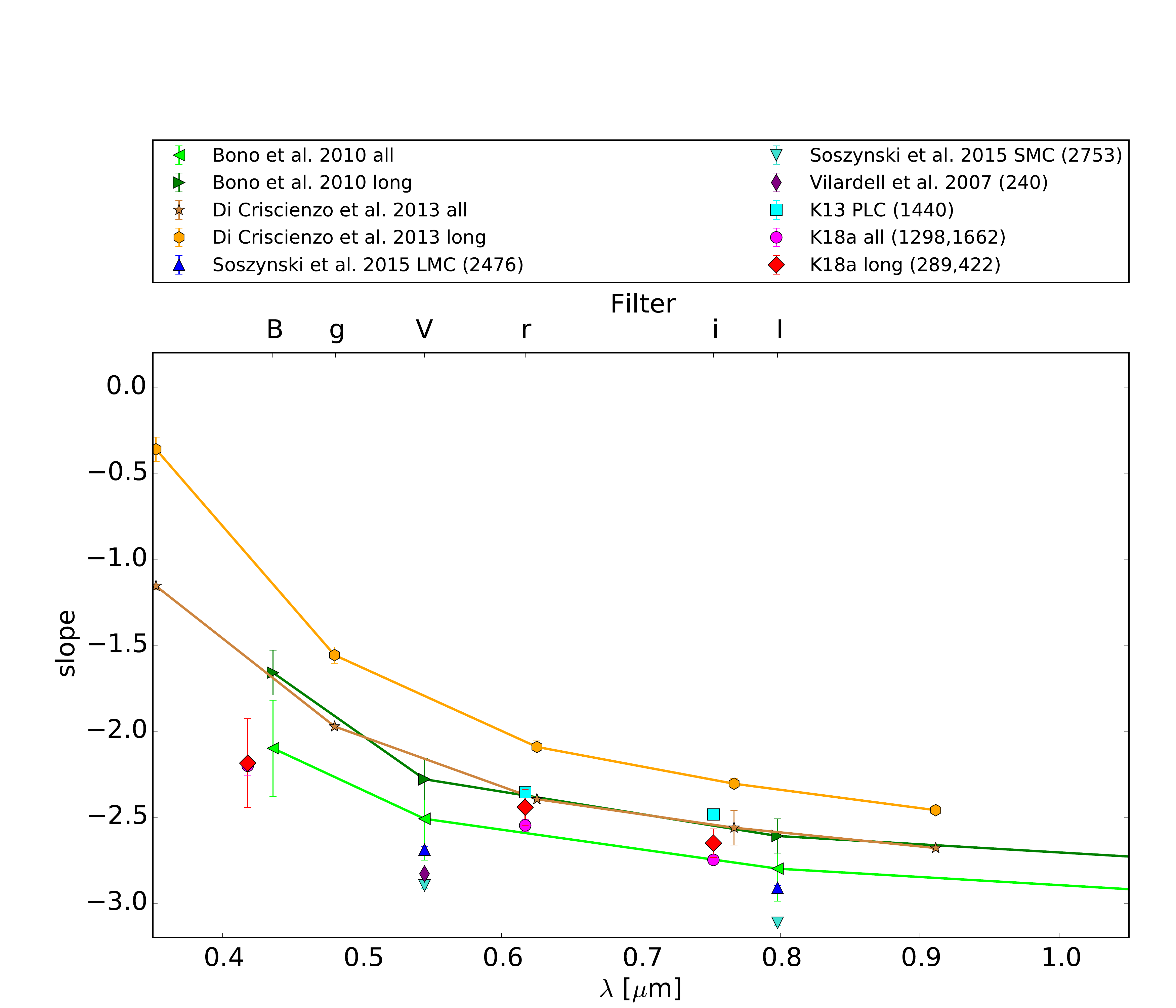}
\caption{PLR slope comparison with the literature for different wavelengths. Shown are the theoretical \citet{2010ApJ...715..277B} predictions from their table 2 for $\mathrm{log(Z/X)} = -1.55$ and $0.4 \leq \mathrm{log(P)} \leq 2.0$ (all) and $\mathrm{log(P)} > 1.0$ (long). Also shown are the theoretical predictions from \citet{2013MNRAS.428..212D} table 2 for $\mathrm{Z} = 0.02$ and $\mathrm{log(P)} \leq 2.0$ (all) and $\mathrm{log(P)} \geq 1.0$ (long). Also shown are the LMC and SMC slopes from \citet{2015AcA....65..297S} table 2 for the V and I band as well as the slope obtained by \citet{2007A&A...473..847V}. The FM PLC slopes from K13 table 3 are also included (the slope errors in K13 have not been determined with the bootstrapping method). Also shown are the slopes of the complete sample (\#4, \#7 and \#10) and the slopes of the sample of long period Cepheids (\#5, \#8 and \#11) from table \ref{table_PLRs}. Our slopes are steeper than in K13, but shallower than the LMC and SMC slopes. As the theoretical predictions our long period Cepheid sample slopes are shallower than for the complete sample.
\label{fig_slopelambda}}
\end{figure}

The PLR fit parameter are summarized in Table \ref{table_PLRs}, while the corresponding fits are shown in Figures \ref{fig_W}, \ref{fig_r}, \ref{fig_i} and \ref{fig_g}. The T2 Cepheids are not subject of this work, so they are shown without a fit. Fit parameter for long period Cepheids ($\log(P) > 1$) are also provided. Due to the fact that the intrinsic dispersion of the PLR is much larger than the typical photometric error we do not use the photometric errors shown in the Figures \ref{fig_W}, \ref{fig_r}, \ref{fig_i} and \ref{fig_g} as a weight in the fit. We rather assign the same weight to each Cepheid. An alternative to this approach is the method used in \citet{2014MNRAS.440.1138E}, namely to quadratically add an internal dispersion factor $\sigma_{\mathrm{int}}$ to all photometric errors such that the $\chi^2$ of the fit is 1. The resulting PLR fit parameter of this method are within the errors shown in Table \ref{table_PLRs}. This is not surprising since the resulting $\sigma_{\mathrm{int}}$ is much larger than the typical photometric error and thus the weight is almost the same for each Cepheid. The error of each fit parameter is the $\pm\sigma$ range around the fit parameter determined from a distribution of 10000 bootstrapping samples. As mentioned previously not all Cepheids have color excess information from the \citet{2009A&A...507..283M} map and therefore some Cepheid magnitudes are only corrected for foreground extinction. If we remove those from the PLR fit the result stays the same within the given errors. Of course some Cepheids might not be included in our sample because they are rejected in the color cut due to the missing color excess information. The PLR dispersion is significantly reduced (by $\approx$ 25\%) in the Wesenheit compared to K13. Also the dispersion for the long period sample is greatly reduced. The FM and FO PLR dispersion is also reduced or stays the same as in the case of the \rps band PLR. Figure \ref{fig_slopelambda} shows a comparison of the slope values to other work. Compared to K13 our new slopes are shallower. The theoretical predictions of \citet{2010ApJ...715..277B} are quite close to our slopes. Same as \citet{2010ApJ...715..277B} and \citet{2013MNRAS.428..212D} we also observe that the long period sample slope is steeper than the slope of the complete sample. Due to the large scatter in the \gps PLR the slopes of the long period sample slope and the complete sample are almost identical. Our slopes are steeper than the slopes for LMC and SMC in \citet{2015AcA....65..297S}.

\begin{figure}
\centering
\includegraphics[width=\linewidth]{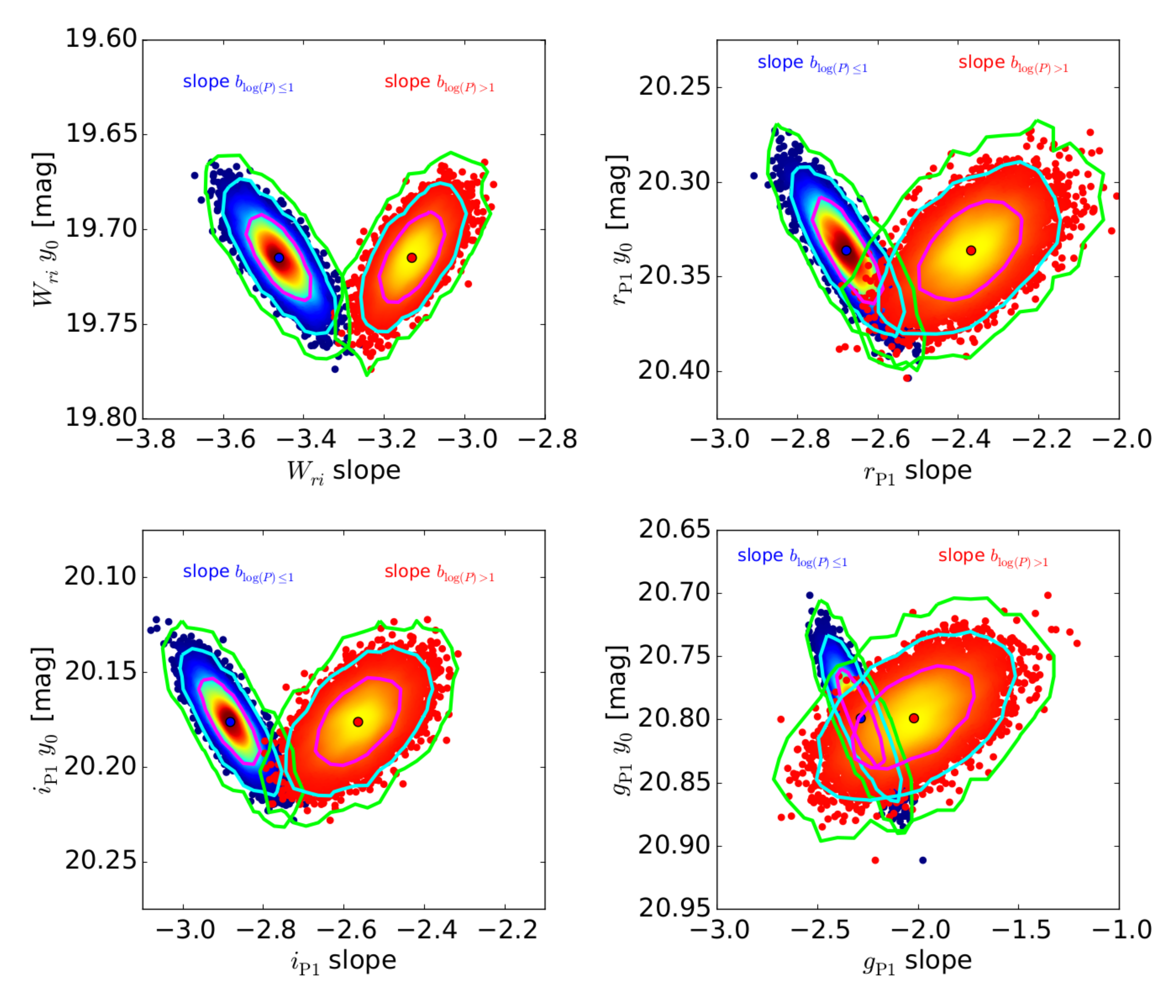}
\caption{Bootstrapping of the broken slope fit for the Wesenheit, \rps, \ips and \gps PLRs. Shown is the common suspension point $y_0$ vs. the slope. The points are colored according to the kernel density estimate. The $1\sigma$, $2\sigma$ and $3\sigma$ contour lines are shown as solid lines. The two distributions are clearly disjunct except in the \gps band, which means that we find a broken slope, although the broken slope is not as significant as in K15 since the contour lines overlap.
\label{fig_all_susp_boot}}
\end{figure}

In K15 we found a broken slope in HST F110W and F160W data. The large dispersion in the ground based data could make the identification of the broken slope more difficult. Table \ref{table_PLRs_broken} shows the broken slope fit with a common suspension point at $\log(P)=1$. These fits are already shown in Figures \ref{fig_W}, \ref{fig_r}, \ref{fig_i} and \ref{fig_g}. As in K15 we test how significant the broken slope is by using the bootstrapping method. Figure \ref{fig_all_susp_boot} shows that the broken slope is not as significant as in K15 since the contour lines overlap. But it also shows that there are clearly two distinct distributions respectively except in the \gps band. This is consistent to the significantly disjunct slopes shown in Table \ref{table_PLRs_broken}. In the \gps band the error of the slope for the long period Cepheids is large, which is consistent with the fact that the long period Cepheids in the \gps band have a large scatter and there are relatively few of them. So the fact that we do not observe the broken slope in the \gps band might be due to the few epochs in the \gps band, since a large error of the long period slope can mask a possible broken slope in the \gps band.

While the dispersion of the FO PLR decreased compared to K13 it is still not as nicely separated from the FM Cepheid PLR as in the HST data in K15 and K18b. This can best be seen in Figure \ref{fig_allclippedfinaltype} where some of the FO Cepheids overlay the FM Cepheids and vice versa (in Figure \ref{fig_W} we offset the FO Cepheids so that the two relations can be seen without the overlap). The FO to FM ratio is 0.18 and therefore twice the ratio in K13. This is closer to the ratio of 0.27 reported by \citet{2007A&A...473..847V} for M31 and the ratio of 0.13 from the General Catalog of Variable Stars \citep{GCVS2012} for the Milky Way. As already described in K13 lower metallicity objects like the SMC and LMC have higher FO to FM ratios in the order of 0.4 -- 0.7 \citep{1999AcA....49..223U}. However there are significant uncertainties associated with our FO to FM ratio, mostly stemming from the type classification and the fact that we do not perform a completeness analysis. Especially the lower period cut is influencing our result.
 
\begin{figure}
\centering
\includegraphics[width=\linewidth]{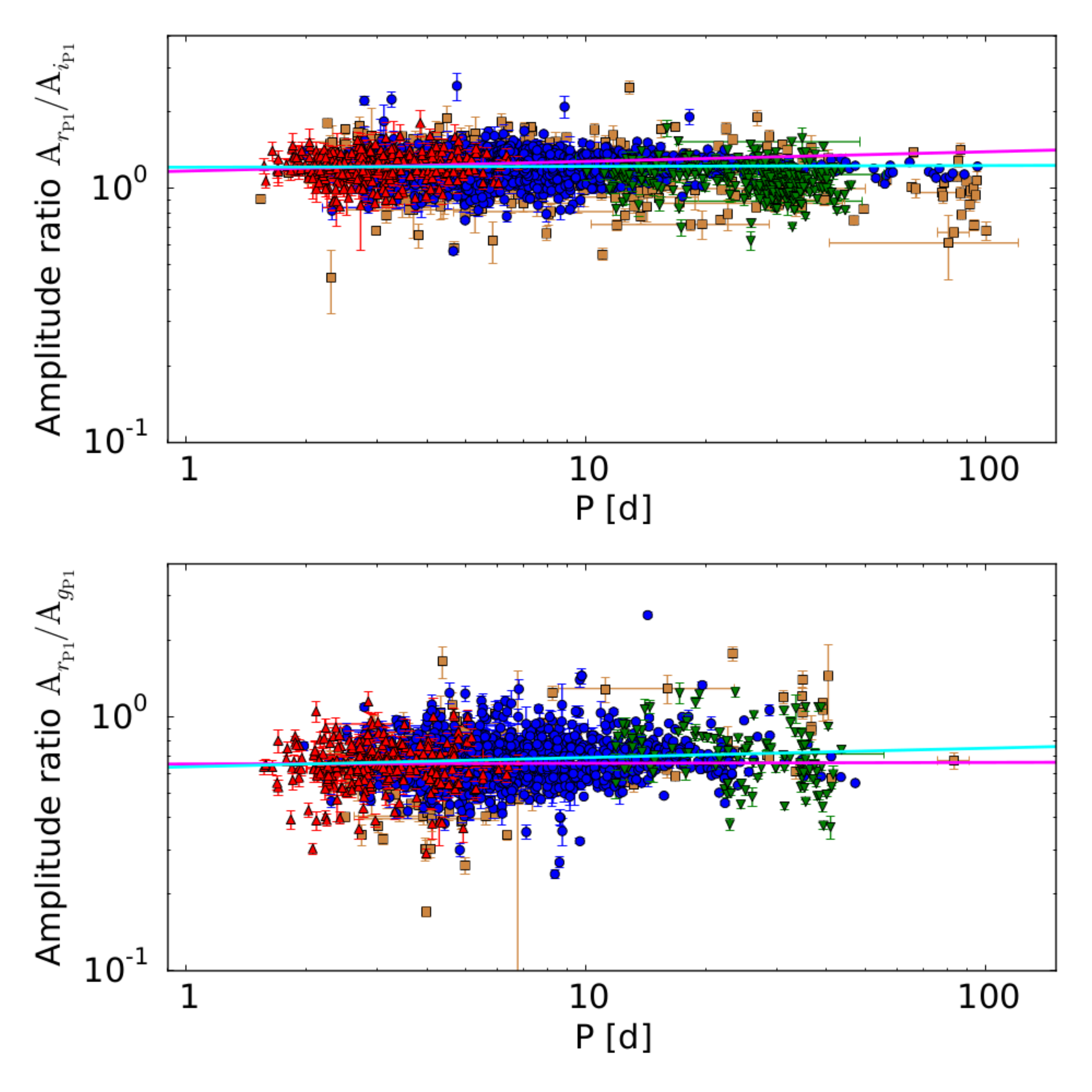}
\caption{Amplitude ratio diagram. Shown are the two amplitude ratios $\mathrm{A}_{\rps} / \mathrm{A}_{\ips}$ (top panel) and $\mathrm{A}_{\rps} / \mathrm{A}_{\gps}$ (bottom panel). The fits from Table \ref{table_ampratio} are also shown (cyan line for FM Cepheids and magenta line for FO Cepheids). The amplitude ratios show no or only a slight dependence on period. \label{fig_ampratio}}
\end{figure}

\begin{figure}
\centering
\includegraphics[width=\linewidth]{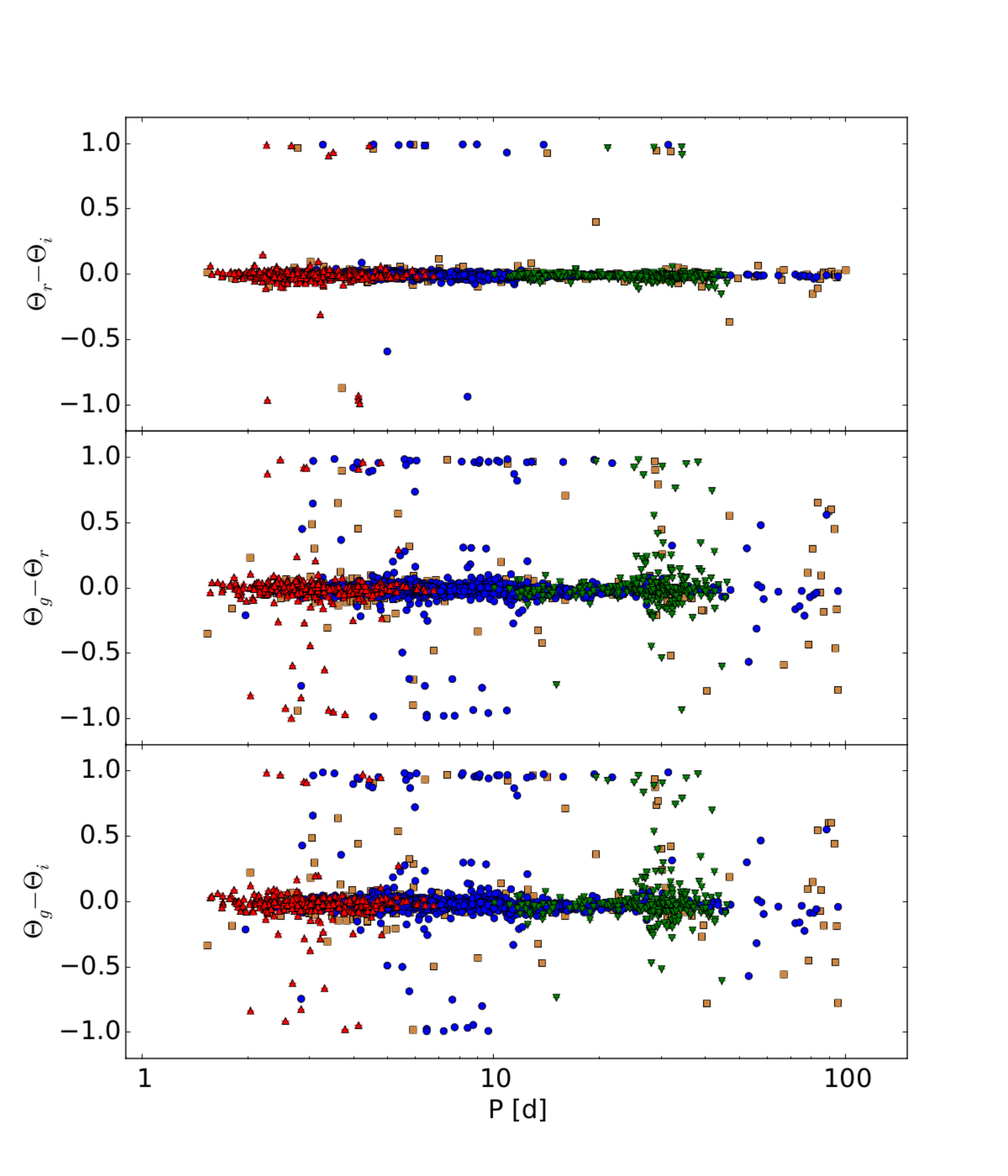}
\caption{Phase lag between the \rps and \ips band (top panel), \gps and \rps band (middle panel) and \gps and \ips band (bottom panel). The phase lag shows no dependence on period. The median phase lag is -0.007, -0.011 and -0.019 respectively.
\label{fig_phaselag-all}}
\end{figure}

We also determine amplitude ratios for our sample. Table \ref{table_ampratio} summarizes the fits shown in Figure \ref{fig_ampratio}. The amplitude ratio
between the \rps and \ips band decreases slightly compared to K13. As expected the amplitude ratio between \rps and \gps band is smaller than one since the Cepheid amplitudes decrease with increasing wavelength \citep{1991PASP..103..933M}. The obtained fits show no or only slight dependence on period. We also determine the phase lag between the three different bands. As can be seen in Figure \ref{fig_phaselag-all} the phase lag is zero and independent of the period. This fact enables us in K18b to correct the random phased HST data.

\begin{deluxetable}{cccccccc}
\tabletypesize{\scriptsize}
\rotate
\tablecaption{PLR fit parameters\label{table_PLRs}}
\tablewidth{0pt}
\tablehead{
\colhead{$\#$} & \colhead{band} & \colhead{type} & \colhead{range} & \colhead{$N_{fit}$} & \colhead{a (log P = 1)} & \colhead{slope b} & \colhead{$\sigma$}
}
\startdata
1 & Wesenheit & FM & all & 1662 & 19.752 (0.008) & -3.323 (0.026) & 0.327 \\
2 & Wesenheit & FM & log(P) $>$ 1 &  422 & 19.760 (0.020) & -3.247 (0.058) & 0.265 \\
3 & Wesenheit & FO & all &  307 & 19.256 (0.071) & -3.065 (0.131) & 0.317 \\
4 & \rps & FM & all & 1662 & 20.371 (0.011) & -2.548 (0.035) & 0.354 \\
5 & \rps & FM & log(P) $>$ 1 &  422 & 20.365 (0.028) & -2.443 (0.105) & 0.398 \\
6 & \rps & FO & all &  307 & 19.811 (0.065) & -2.456 (0.115) & 0.259 \\
7 & \ips & FM & all & 1662 & 20.212 (0.009) & -2.749 (0.029) & 0.304 \\
8 & \ips & FM & log(P) $>$ 1 &  422 & 20.210 (0.023) & -2.651 (0.083) & 0.322 \\
9 & \ips & FO & all &  307 & 19.667 (0.060) & -2.615 (0.105) & 0.243 \\
10 & \gps & FM & all & 1298 & 20.822 (0.018) & -2.202 (0.058) & 0.433 \\
11 & \gps & FM & log(P) $>$ 1 &  289 & 20.843 (0.051) & -2.186 (0.262) & 0.533 \\
12 & \gps & FO & all &  246 & 20.319 (0.087) & -1.997 (0.152) & 0.298 \\
\enddata
\tablecomments{The magnitude errors were set to the same value. The errors of the fitted parameters were determined with the bootstrapping method.
}
\end{deluxetable}

\begin{deluxetable}{ccccccc}
\tabletypesize{\scriptsize}
\rotate
\tablecaption{broken slope PLR fit parameters\label{table_PLRs_broken}}
\tablewidth{0pt}
\tablehead{
\colhead{$\#$} & \colhead{band} & \colhead{$N_{fit}$} & \colhead{$b_{\log(P)\leq1}$} & \colhead{$b_{\log(P)>1}$} &\colhead{$a_{\log(P)=1}$} & \colhead{$\sigma$}
}
\startdata
1 & Wesenheit & 1662 & -3.461 (0.051) & -3.131 (0.050) & 19.715 (0.014) & 0.326 \\
2 & \rps & 1662 & -2.679 (0.054) & -2.368 (0.086) & 20.336 (0.017) & 0.353 \\
3 & \ips & 1662 & -2.883 (0.046) & -2.565 (0.068) & 20.176 (0.014) & 0.303 \\
4 & \gps & 1298 & -2.286 (0.077) & -2.021 (0.189) & 20.799 (0.025) & 0.433 \\
\enddata
\tablecomments{Two lines with a common suspension point at $\log(P)=1$ are fitted. The magnitude errors were set to the same value. The errors of the fitted parameters were determined with the bootstrapping method.}
\end{deluxetable}

\begin{deluxetable}{cccccccc}
	\tabletypesize{\scriptsize}
	\rotate
	\tablecaption{Amplitude ratio fit parameters\label{table_ampratio}}
	\tablewidth{0pt}
	\tablehead{
		\colhead{$\#$} & \colhead{Amplitude ratio} & \colhead{type} & \colhead{range} & \colhead{$N_{fit}$} & \colhead{a (log P = 1)} & \colhead{slope b} & \colhead{$\sigma$}
	}
	\startdata
	1 & $\mathrm{A}_{\rps} / \mathrm{A}_{\ips}$ & FM & all & 1662 &  1.224 (0.003) &  0.009 (0.012) & 0.137 \\
	2 & $\mathrm{A}_{\rps} / \mathrm{A}_{\gps}$ & FM & all & 1298 &  0.694 (0.005) &  0.059 (0.018) & 0.137 \\
	3 & $\mathrm{A}_{\rps} / \mathrm{A}_{\ips}$ & FO & all &  307 &  1.286 (0.032) &  0.113 (0.059) & 0.151 \\
	4 & $\mathrm{A}_{\rps} / \mathrm{A}_{\gps}$ & FO & all &  246 &  0.656 (0.032) &  0.005 (0.060) & 0.133 \\
	\enddata
	\tablecomments{The errors of the fitted parameters were determined with the bootstrapping method.}
\end{deluxetable}

\section{Conclusion\label{conclusion}}

We use the PAndromeda data to obtain the largest and most homogeneous Cepheid sample in M31. We find 1662 fundamental mode Cepheids, 307 first-overtone Cepheids, 278 type II Cepheids and 439 Cepheids with unclear Cepheid type. Similar to K13 we use a combination of a three dimensional Fourier parameter space, a color cut and other selection criteria to find the Cepheids in an as unbiased and reproducible way as possible. We improved this approach by implementing a new type classification and by using an improved color cut using the \citet{Anderson} instability strip edge models. The \citet{Anderson} instability strip edge models describe our data much better than the \citet{Fiorentino2002} instability strip edge models. Although, the \citet{Anderson} instability strip does not cover a substantial fraction of the Cepheids that we classified manually (see Figure \ref{fig_colorWesenheitmanual}).

The Period-Luminosity relations we obtain in the \rps, \ips and \gps bands (see Figures \ref{fig_W}, \ref{fig_r}, \ref{fig_i} and \ref{fig_g}) have smaller dispersions than in K13. This is achieved by a rigorous masking of the Pan-STARRS1 data and an improved data reduction. As in K15 with HST data we find a broken slope (see Figure \ref{fig_all_susp_boot}), except in the \gps band. The broken slope is not as significant as in K15. 

We use the Cepheid sample obtained here in \citet{K18b} where we obtain space-based HST Period-Luminosity relations in two near-infrared and two optical bands. The \citet{K18b} Cepheid sample is the largest Cepheid sample in M31 with HST data. The PLR dispersions in the HST optical bands is similar to the dispersions obtained here with ground based data. This means that the crowding effect on our ground-based but mean phase corrected photometry is comparable to the effect of using random phased data in the HST sample. In K18b we show that the color selection can cause the broken slope. Therefore, the new improved color cut could contribute to the observed broken slope.
 
The complete Cepheid catalog including the light curves and objects clipped during the selection process can be found electronically in the CDS database. 

\acknowledgments
This research was supported by the DFG cluster of excellence Origin and Structure of the Universe’ (www.universe-cluster.de).

\clearpage

\section{Appendix}

\subsection{Skycell layout}

\begin{figure}
	\centering
	\includegraphics[width=0.85\linewidth]{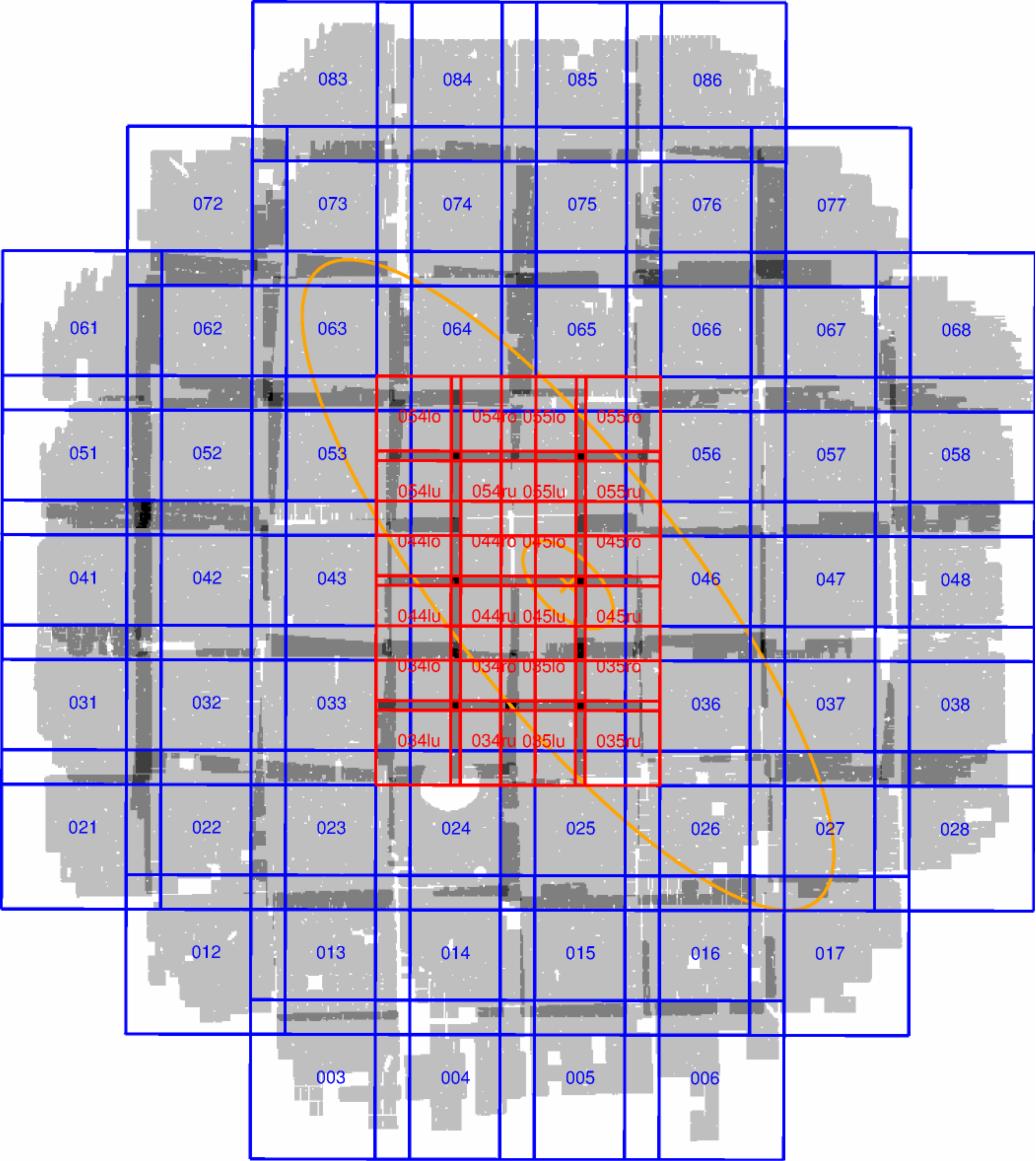}
	\caption{Skycell layout in the $\rps$-band. Each skycell covers 6647 px $\times$ 6647 px (i.e. 27.7\arcmin~$\times$ 27.7\arcmin) and overlaps with the neighboring skycells. The six central skycells (red) are each divided into four smaller skycells with 3524 px $\times$ 3524 px (i.e. 14.7\arcmin~$\times$ 14.7\arcmin) with an overlapp between the skycells. The background in grey shows the unmasked area of the reference frames in the $\rps$-band. Darker shades of grey show the overlapp between the skycells, with a maximum overlap of four skycells (black). The area covered is $\sim$~7~deg$^2$. The orange cross marks the center of M31 and the orange ellipses are the $\mu_r = 20.05$ mag and $\mu_r = 23.02$ mag surface brightness profiles from \citet{1987AJ.....94..306K}.
		\label{footprint_K17a_r}}
\end{figure}

\begin{figure}
	\centering
	\includegraphics[width=0.85\linewidth]{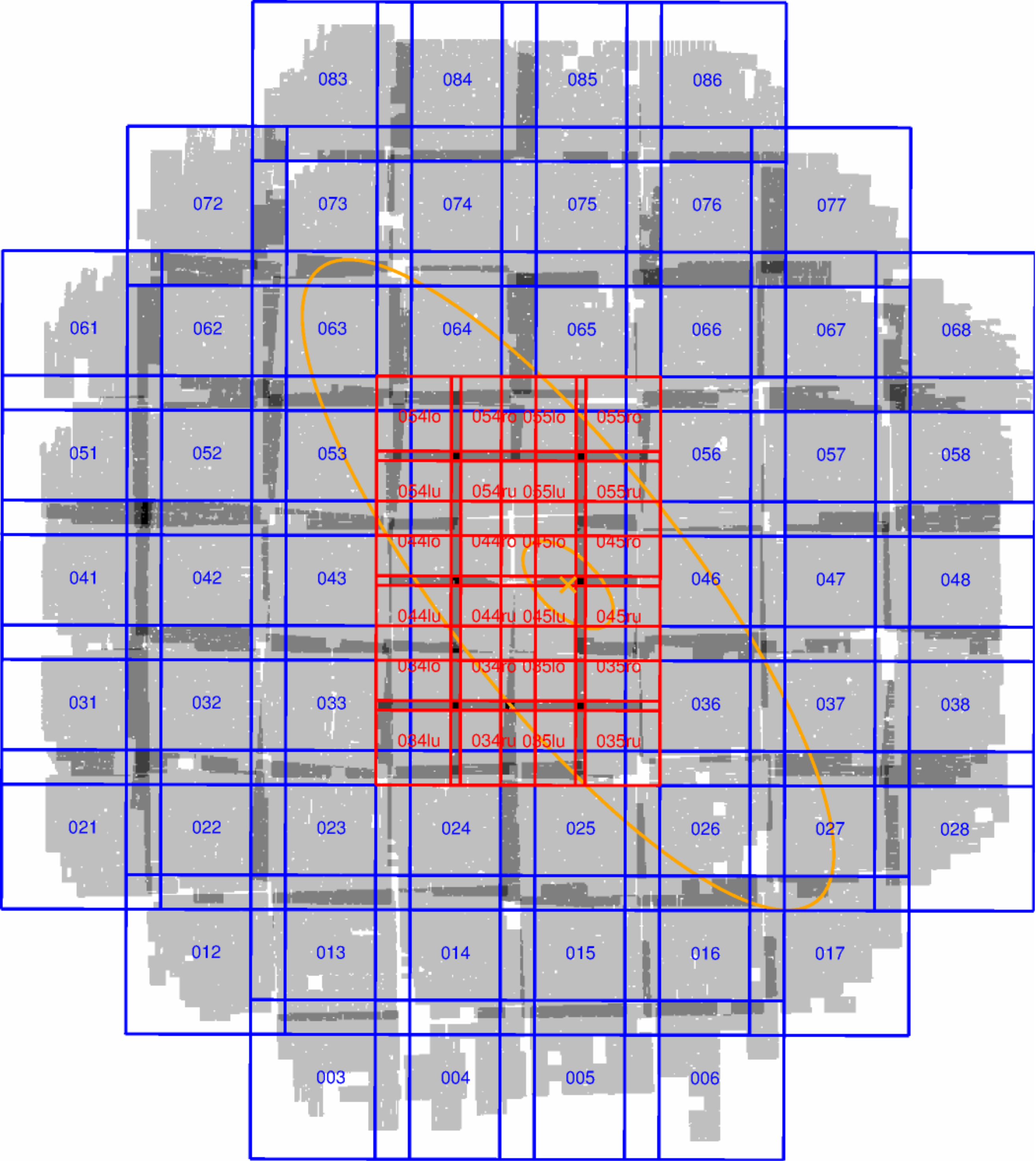}
	\caption{Skycell layout in the $\ips$-band. Otherwise same as in Fig. \ref{footprint_K17a_r}, but with the $\ips$-band reference frames in the background. The area covered is $\sim$7~deg$^2$.
		\label{footprint_K17a_i}}
\end{figure}

\begin{figure}
	\centering
	\includegraphics[width=0.85\linewidth]{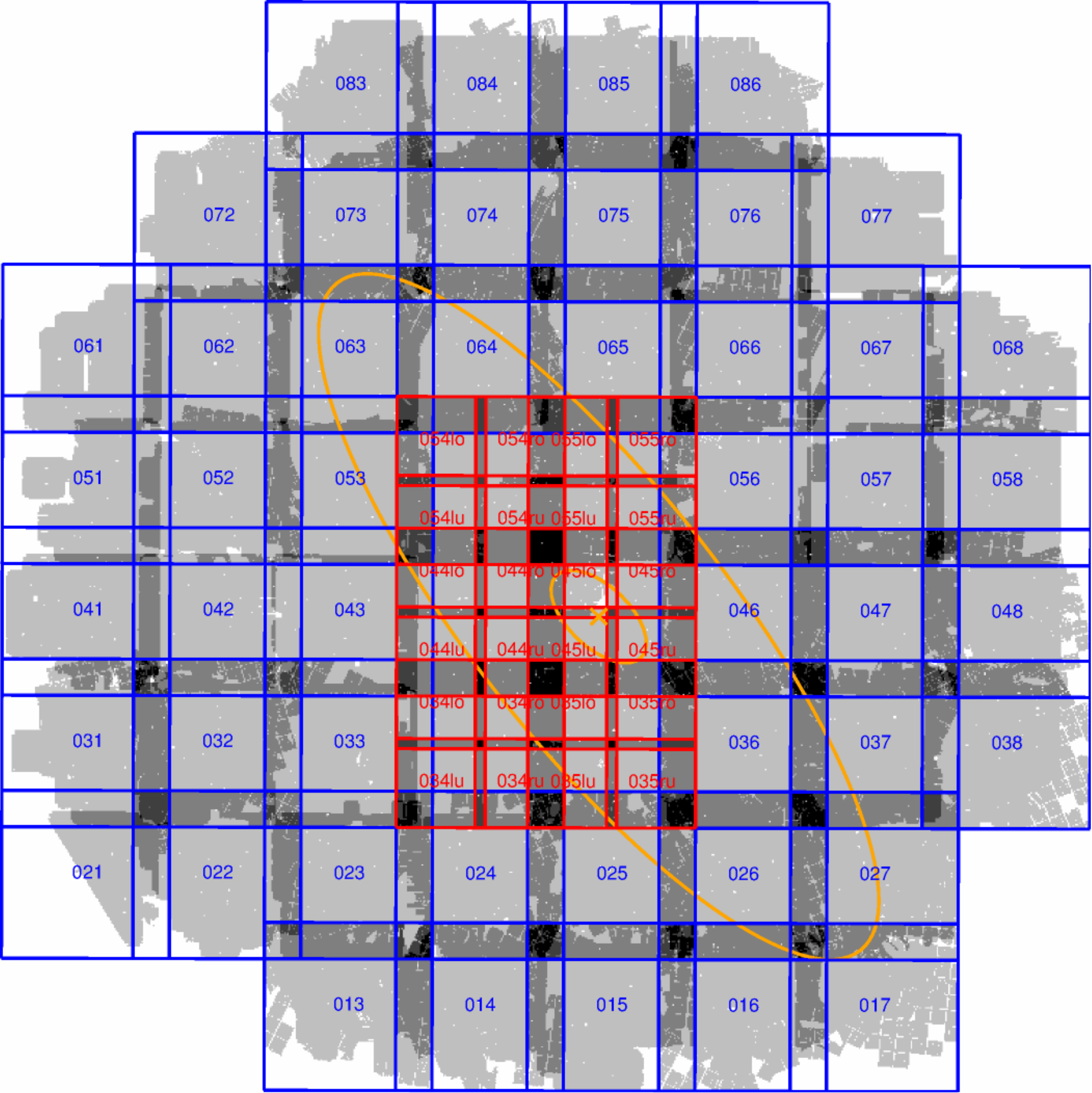}
	\caption{Skycell layout in the $\gps$-band. The \gps reference frames are shown in the background and the skycells 003, 004, 005, 006, 012 and 028 are not used in the $\gps$-band since there is almost no data. The area covered is $\sim$~6.8~deg$^2$.
		\label{footprint_K17a_g}}
\end{figure}

\begin{figure}
	\centering
	\includegraphics[width=0.85\linewidth]{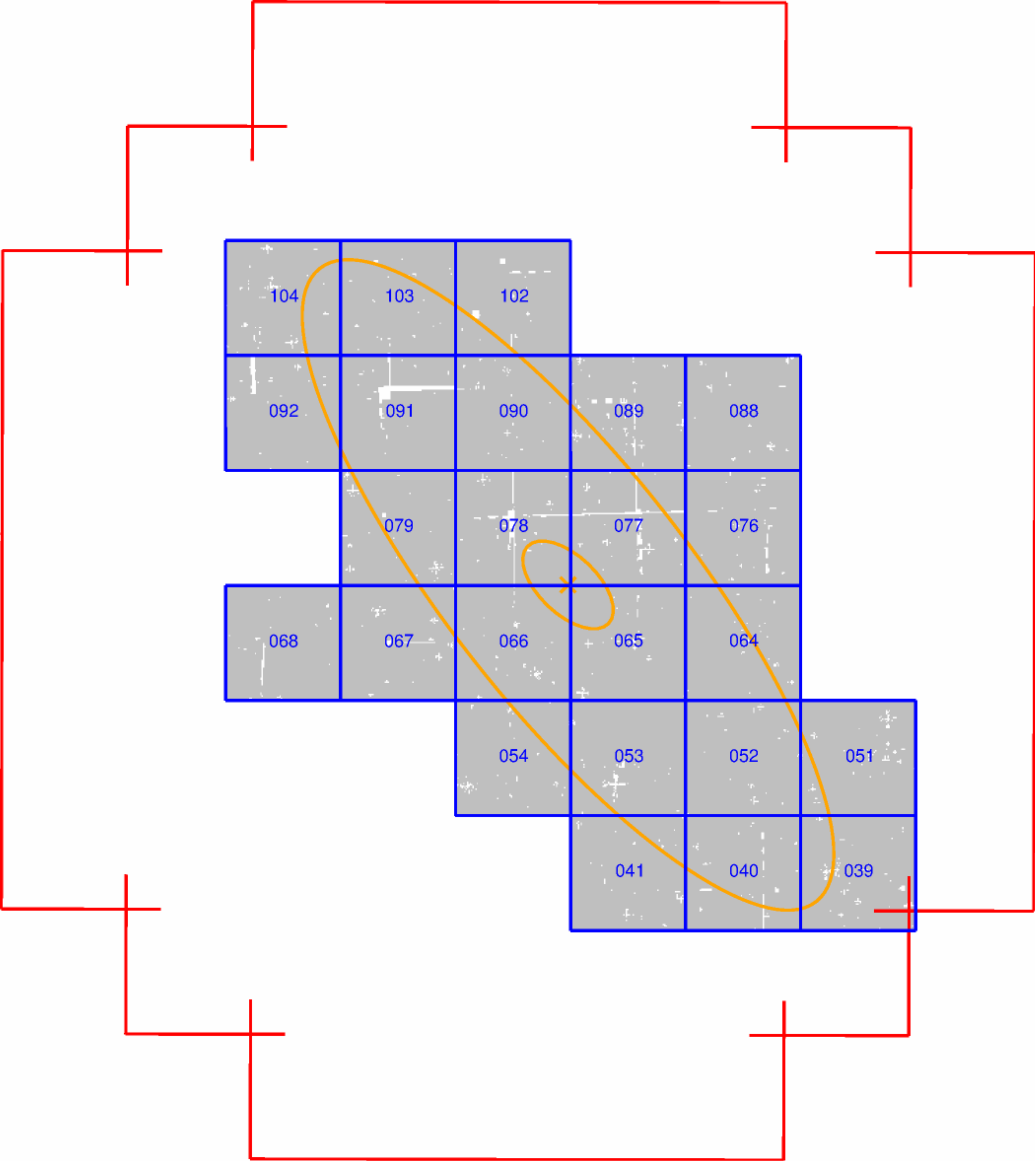}
	\caption{Skycell layout in the $\rps$-band in K13. In contrast to the data used here the skycells have no overlap and cover 6000 px $\times$ 6000 px (i.e. 20.0\arcmin~$\times$ 20.0\arcmin) in K13. The grey background shows the area covered in the $\rps$-band reference frames in K13. The area covered is $\sim$~2.6~deg$^2$. The orange cross marks the center of M31 and the orange ellipses are the $\mu_r = 20.05$ mag and $\mu_r = 23.02$ mag surface brightness profiles from \citet{1987AJ.....94..306K}. The red outline is the border of the skycells used in K18a (see Figure \ref{footprint_K17a_r}).
		\label{footprint_K13_r}}
\end{figure}

The skycell layout discussed in section \ref{survey} is shown in Figures \ref{footprint_K17a_r}, \ref{footprint_K17a_i} and \ref{footprint_K17a_g} for K18a. Figures \ref{footprint_K13_r} and \ref{footprint_K13_r} show the skycell layout in K13. 

\begin{figure}
	\centering
	\includegraphics[width=0.85\linewidth]{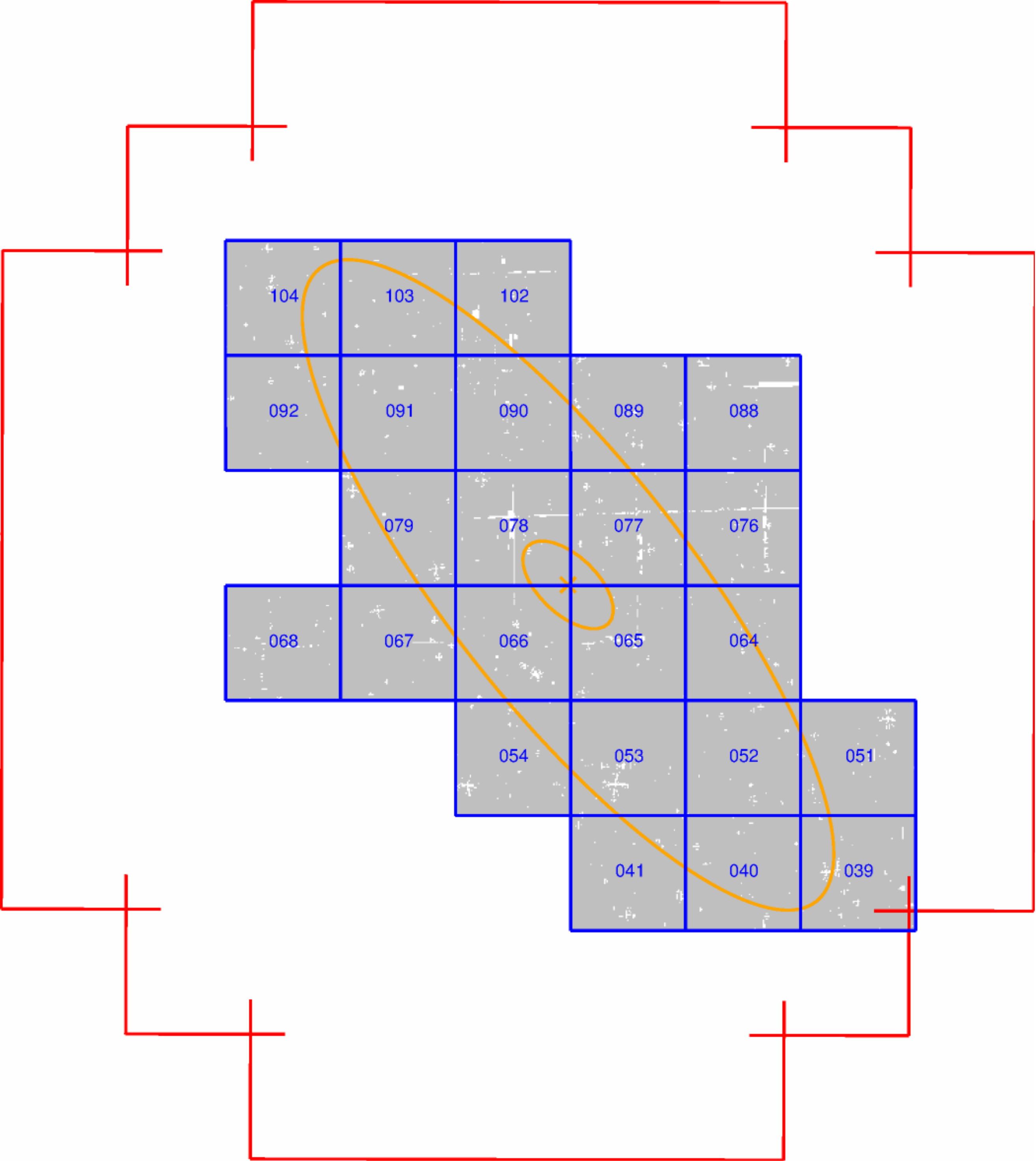}
	\caption{Skycell layout in the $\ips$-band in K13. The background shown the area covered in the $\ips$-band reference frames in K13. Otherwise same as in Fig. \ref{footprint_K13_r}. The area covered is $\sim$~2.6~deg$^2$.
		\label{footprint_K13_i}}
\end{figure}

\subsection{Flag 64}

\begin{figure}
\centering
\includegraphics[width=\linewidth]{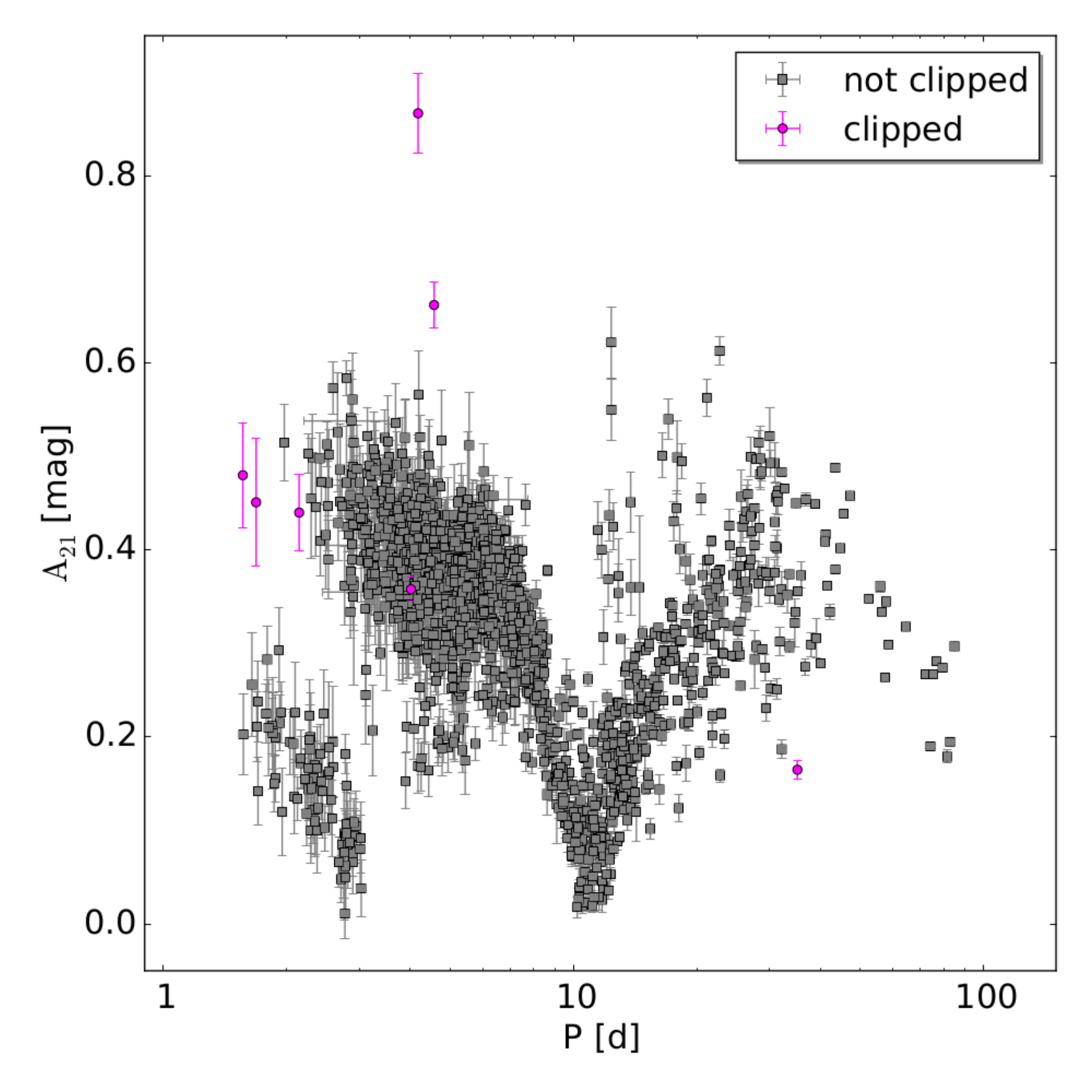}
\caption{Amplitude ratio ($A_{21}$) digram of the manually classified Cepheid sample. Shown are the 1640 Cepheids that are used for the final sample and the 7 Cepheids that are assigned the bit flag 64. The reason that this bit flag is assigned is that those seven Cepheids have a large distance to their closest neighbor in the three dimensional space. 
\label{fig_manualdistcutA21}}
\end{figure}

\begin{figure}
\centering
\includegraphics[width=\linewidth]{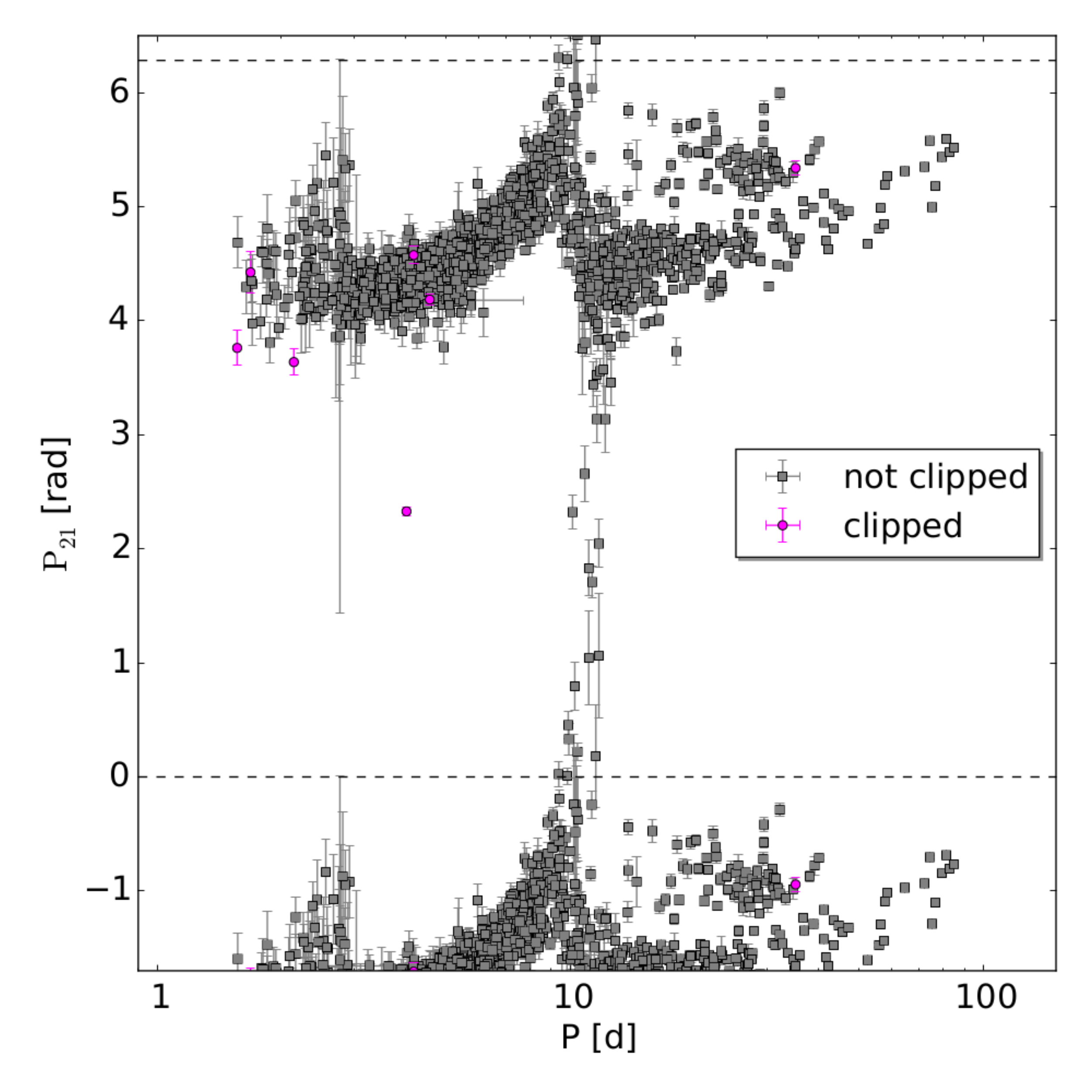}
\caption{Phase difference ($\varphi_{21}$) diagram. Same as in Figure \ref{fig_manualdistcutA21} the 7 Cepheids with the bit flag 64 are shown with the 1640 manually classified Cepheids that are used in the final sample.
\label{fig_manualdistcutP21}}
\end{figure}

\begin{figure}
\centering
\includegraphics[width=\linewidth]{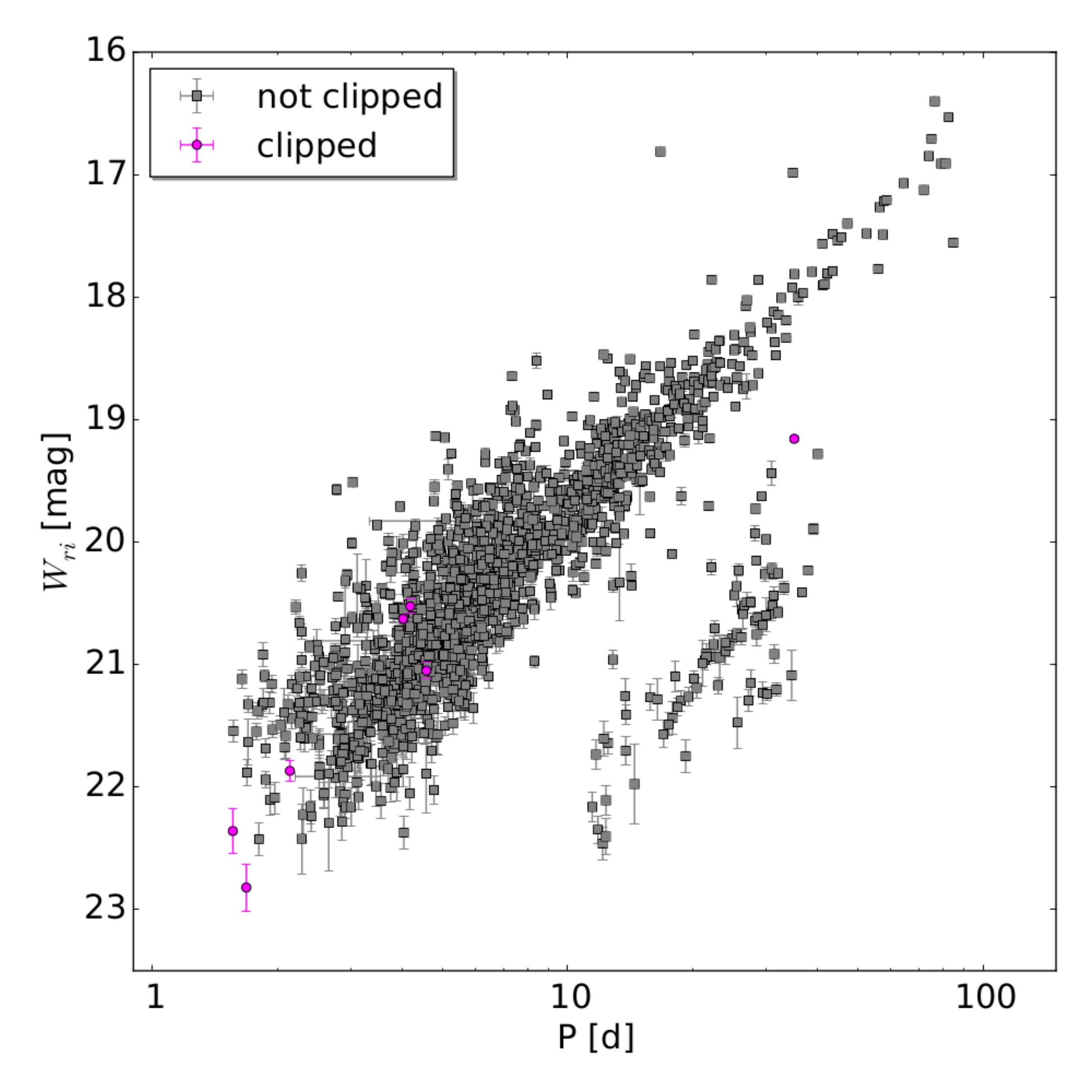}
\caption{Period-Wesenheit diagram. Shown are the 7 Cepheids with the bit flag 64 and the 1640 Cepheids that are used for the final sample.
\label{fig_manualdistcutW}}
\end{figure}

The bit flag 64 is assigned seven times as explained in section \ref{3dcut}. The seven manually classified Cepheids have a larger distance to their respective next Cepheid neighbor in the three dimensional space. Figures \ref{fig_manualdistcutA21} and \ref{fig_manualdistcutP21} shows the location of these seven Cepheids in the amplitude ratio ($A_{21}$) and phase difference ($\varphi_{21}$) diagram. Figure \ref{fig_manualdistcutW} shows where these seven Cepheids reside in the Period-Wesenheit diagram.

\subsection{Comparison with K13}

\begin{figure}
\centering
\includegraphics[width=\linewidth]{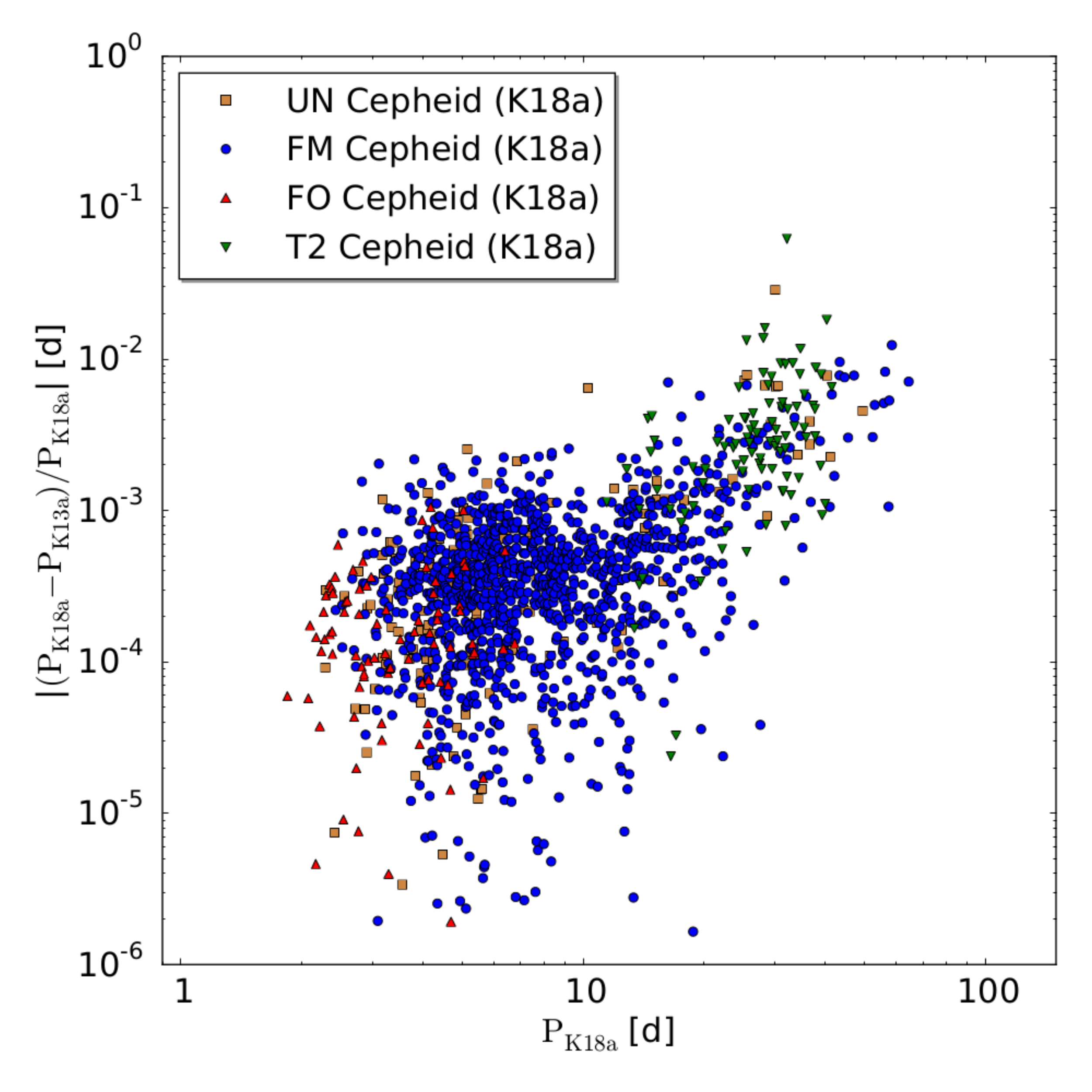}
\caption{Comparison of the relative period change between this sample and K13. The long periods change by the largest amount since we now cover multiple pulsations.\label{fig_compare_K13_per}}
\end{figure}

\begin{figure}
\centering
\includegraphics[width=\linewidth]{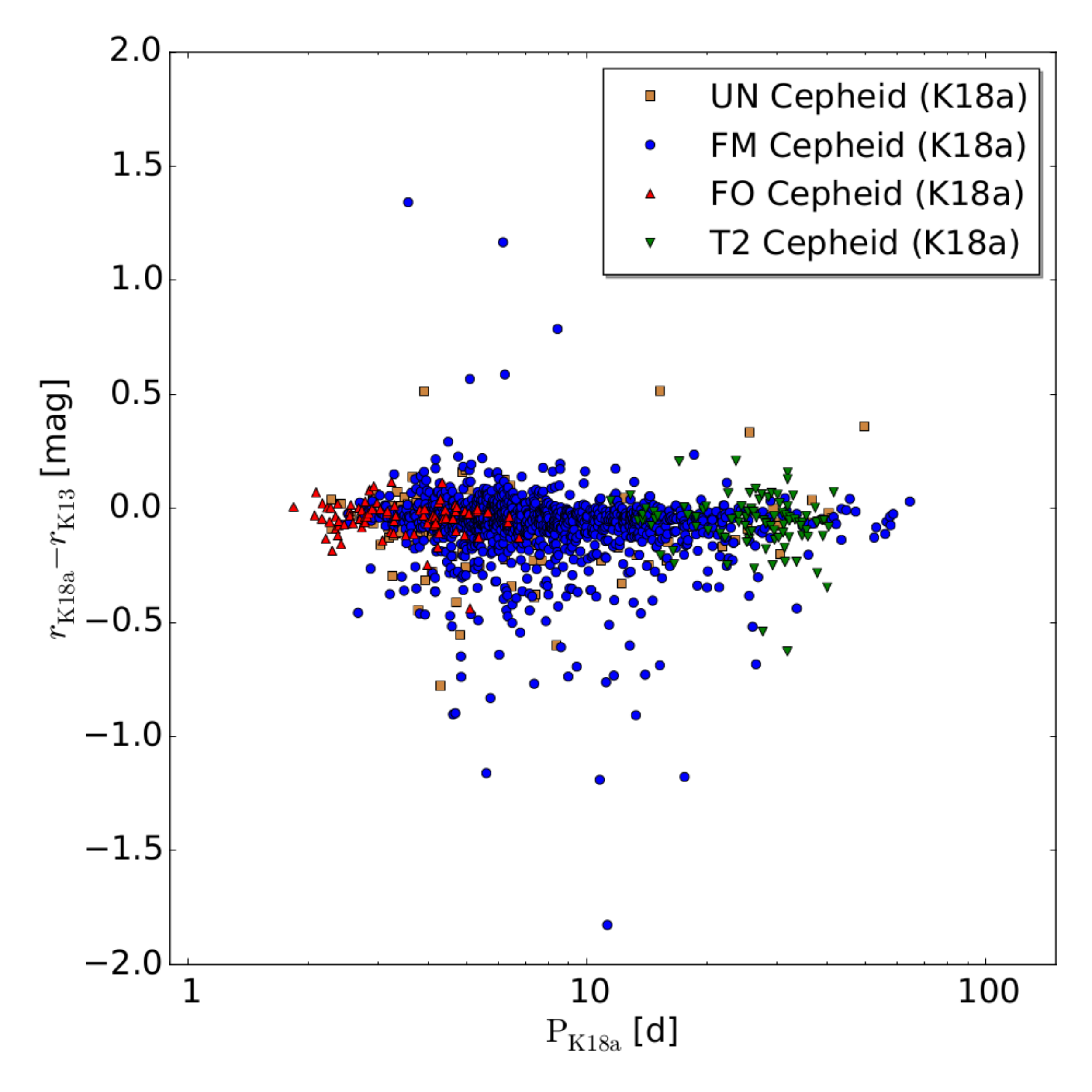}
\caption{\rps band comparison between this sample and K13 for the 1445 Cepheids that both samples have in common. The magnitudes here are not extinction corrected. The large magnitude differences are explained with a bad SExtractor position (on which the forced photometry was carried out) in K13 due to crowding. Most of those had a source position between two stars and the brighter magnitude from this sample is now correct since the position is correctly identified. But there are also a few cases where the new magnitude is fainter. This happens when there are two very close stars where the PSFs overlap and previously the pulsation was wrongly attributed to the brighter source.
\label{fig_compare_K13_mag_r}}
\end{figure}

\begin{figure}
\centering
\includegraphics[width=\linewidth]{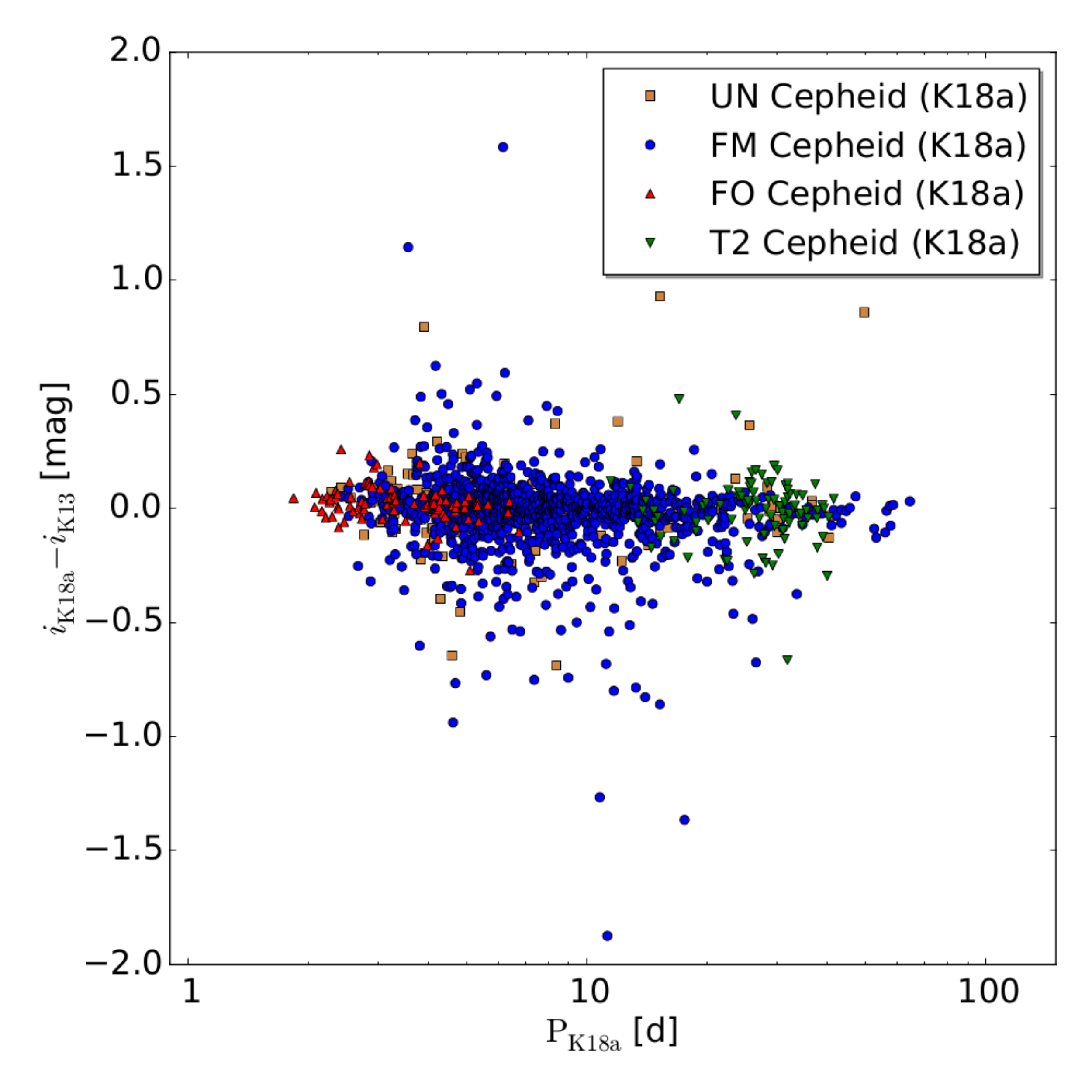}
\caption{\ips band comparison, otherwise the same as Figure \ref{fig_compare_K13_mag_r}. 
\label{fig_compare_K13_mag_i}}
\end{figure}

\begin{figure}
\centering
\includegraphics[width=\linewidth]{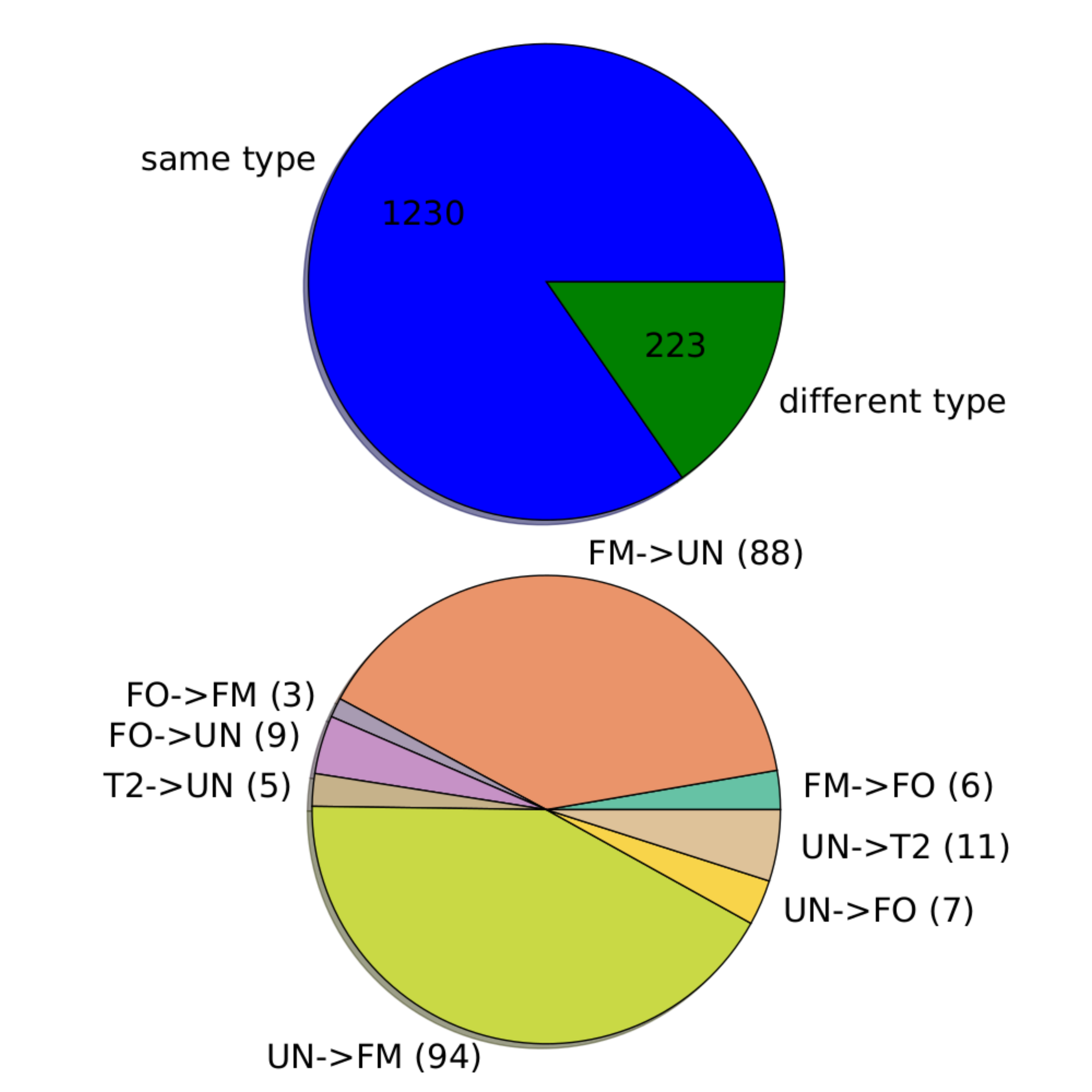}
\caption{Cepheid type comparison for the 1445 Cepheids that are in this sample and in K13. The Cepheid type changes for 221 Cepheids. Almost all of those changes except for nine cases involve the UN type.
\label{fig_compare_type}}
\end{figure}

Since our data has more epochs and covers a larger time span than K13 the periods are also more precise. The long periods show a larger relative change in Figure \ref{fig_compare_K13_per} since for these periods we now cover multiple pulsation cycles. We compare the \rps and \ips band magnitudes in Figures \ref{fig_compare_K13_mag_r} and \ref{fig_compare_K13_mag_i} for the 1445 Cepheids that are in both samples. There is no offset, but some Cepheids show a large magnitude change. The large magnitude differences are explained with a bad SExtractor position (on which the forced photometry was carried out) in K13 due to crowding. Most of those had a source position between two stars and the brighter magnitude from this sample is now correct since the position is correctly identified. But there are also a few cases where the new magnitude is fainter. This happens when there are two very close stars where the PSFs overlap and previously the pulsation was wrongly attributed to the brighter source. In Figure \ref{fig_compare_type} we show how the Cepheid type changes. Of the 1445 Cepheids which are in both samples, 1224 do not change the Cepheid type. As can be seen most change from UN to FM and from FM to UN and almost all changes except for 9 cases involve the UN type. A cross-match table is provided on the CDS.

\subsection{Excerpt of the published tables}

Tables \ref{table_maindat}, \ref{table_lcinfo} and \ref{table_other} show excerpts of the data published on the CDS. The light curves are also published.
The tables of the clipped sources are also available, but not all fits have necessarily been performed for those objects depending at which point the object was cut. In those cases the columns show -1 or -99 as default value. Some values can also be so small that they are shown as 0.0 in the tables.
The errors of some values have been determined by different methods. We use the errors shown in table \ref{table_maindat}. The method used to determine the magnitude errors is described in section \ref{detection}. A distribution of magnitudes is obtained and the error is calculated from this distribution in two slightly different approaches. The first method is to determine the error from the $\pm \sigma/2$ range around the mean magnitude from the light curve fit. The other method is to determine the error from the $1\sigma$ range of the distribution regardless of the fitted mean magnitude from the light curve. The two errors are the same if the mean of the distribution is the same as the mean magnitude from the light curve fit. This also implies that the difference of these two error estimates is a tracer of how symmetric the distribution is around the mean magnitude from the light curve. A good example for this is the period error where the distribution is obtained by using the bootstrapping method on the light curve and the error is also calculated with the two methods. The period calculated from the light curve can be so much on the edge of the distribution of periods obtained from the bootstrapping, that it is outside the $1\sigma$ range of the distribution and therefore the error from the $\pm \sigma/2$ range around the mean period is not defined anymore. The period of a light curve that is changed drastically when bootstrapping the light curve would show this behavior. Note that these Cepheids do not have poor period estimates but rather that there all epochs are important for the period determination. The error from the $1\sigma$ range can always be calculated and therefore it is used as the period error. The $\pm \sigma/2$ range period error would not be defined for all Cepheids. The different error estimates are not used in this work to perform additional cuts, but we include them in the published data for the interested reader. 

\begin{figure}
\centering
\includegraphics[width=0.85\linewidth]{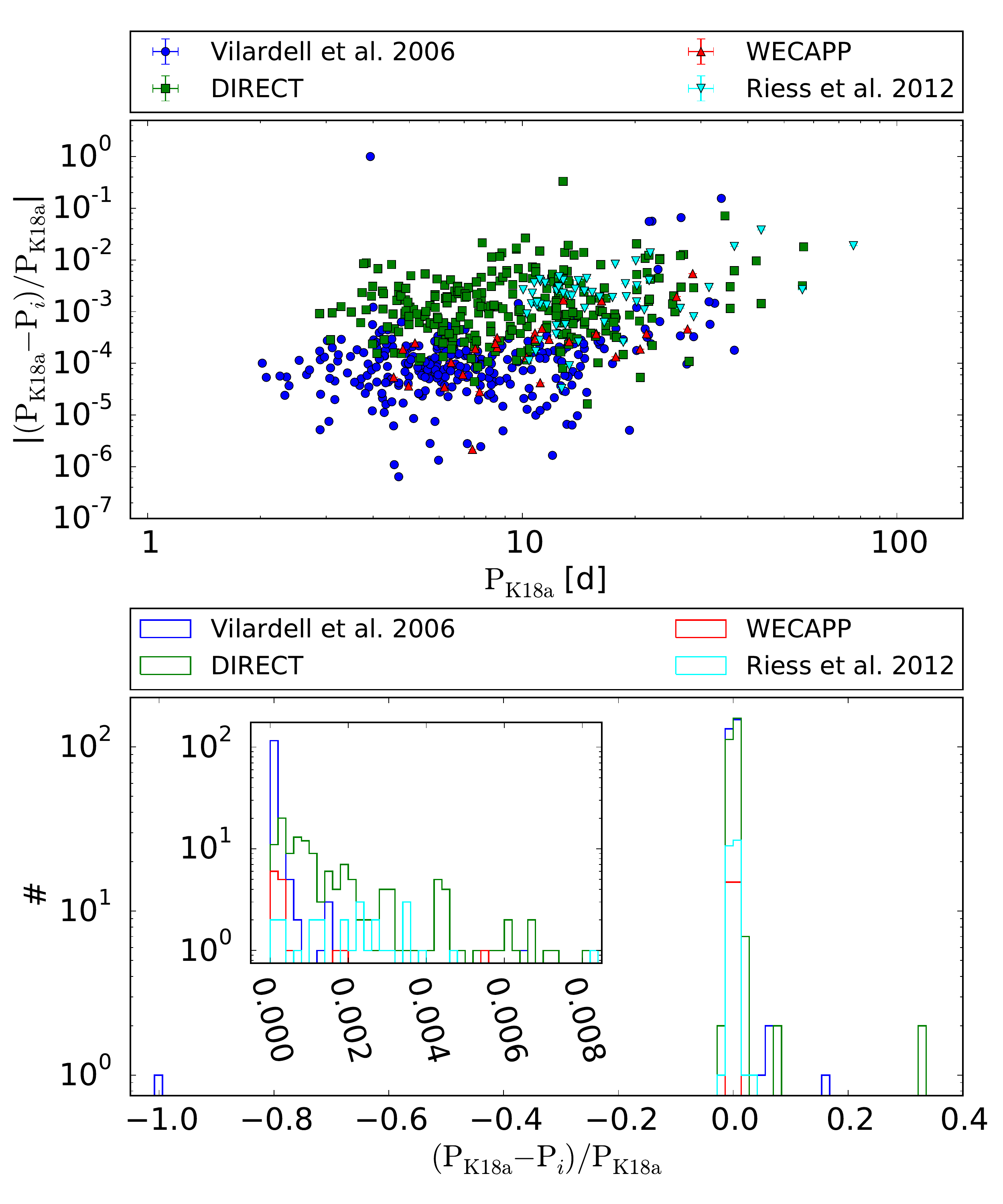}
\caption{Comparison of the periods of our K18a sample with literature samples. The top panel shows the absolute value of the relative period difference, while the bottom panel shows a histogram of the relative period difference. The \citet{2006A&A...459..321V} sample and the WECAPP sample have a lot of epochs and therefore the periods match very well to our K18a periods. The K18a baseline is longer than for the literature samples and therefore the period difference increases for larger periods.
\label{fig_matching}}
\end{figure}

Additionally to the K13 cross-match table mentioned in the previous subsection, we also provide cross-matching tables to the \citet{2006A&A...459..321V} sample with 416 Cepheids, the DIRECT sample with 332 Cepheids (\citet{1998AJ....115.1894S}, \citet{1998AJ....115.1016K}, \citet{1999AJ....117.2810S}, \citet{1999AJ....118..346K}, \citet{1999AJ....118.2211M}, \citet{2003AJ....126..175B}), the WECAPP sample \citep{2006A&A...445..423F} with 126 Cepheids and the \citet{2012ApJ...745..156R} sample with 68 Cepheids. Due to the masking discussed in section \ref{survey} we can not match all literature Cepheids. With a matching radius of 2\arcsec~we match 349 Cepheids of the \citet{2006A&A...459..321V} sample of 416 Cepheids. 8 of the \citet{2006A&A...459..321V} have two matches within the 2\arcsec~radius. 277 Cepheids are in the K18a sample and 72 are clipped. Of the 332 DIRECT Cepheids we match 316. 14 Cepheids have two matches inside the 2\arcsec~matching radius and 267 are in the K18a sample while 49 are clipped. Of the 126 WECAPP Cepheids we match 64. 30 are in the K18a sample while 34 are clipped and 2 Cepheids have two matches inside the 2\arcsec~matching radius. We match 61 of the 68 \citet{2012ApJ...745..156R} Cepheids. 3 have two matches within the 2\arcsec~radius and 55 are in K18a while 9 are clipped. In Figure \ref{fig_matching} we show the comparison of the literature periods with those that are in our K18a sample (i.e. the clipped Cepheids are not included). The \citet{2006A&A...459..321V} sample and the WECAPP sample have the most epochs of the literature samples and therefore the periods match very well to our K18a periods. Since our K18a sample covers a larger baseline the long periods are determined more precisely than in the literature samples.

\begin{deluxetable}{cccccccccccc}
\tabletypesize{\footnotesize}
\rotate
\tablecaption{main.dat\label{table_maindat}}
\tablewidth{0pt}
\tablehead{
\colhead{PSO id} & \colhead{id} & \colhead{RA} & \colhead{Dec} & \colhead{$\mathrm{P}_{\rps}$} & \colhead{$\mathrm{P}_{e,\rps}$} & \colhead{$\nrps$} & \colhead{$\rpserr$}  & \colhead{$\nips$} & \colhead{$\ipserr$} & \colhead{$\ngps$} & \colhead{$\gpserr$}
}
\startdata
PSO J009.0415+41.2478 &  704730 & 09.04155 & 41.24777 & 1.535896 & 0.000198 & 18.3081 & 0.0010 & 18.2293 & 0.0013 & 18.5515 & 0.0051 \\
PSO J009.5682+40.5154 & 2933467 & 09.56821 & 40.51542 & 1.566910 & 0.000074 & 21.6953 & 0.0168 & 21.6556 & 0.0202 & 21.9194 & 0.0336 \\
PSO J011.7772+42.5608 & 3012073 & 11.77718 & 42.56078 & 1.583140 & 0.003552 & 21.6403 & 0.0147 & 21.6324 & 0.0219 & 21.8493 & 0.0293 \\
PSO J011.6012+42.0350 & 2108691 & 11.60119 & 42.03497 & 1.643588 & 0.000113 & 21.3597 & 0.0139 & 21.2983 & 0.0160 & 21.4916 & 0.0250 \\
PSO J011.4574+42.2976 & 2493844 & 11.45745 & 42.29759 & 1.689511 & 0.000076 & 22.0400 & 0.0218 & 21.9999 & 0.0227 & 22.3310 & 0.0332 \\
\enddata
\tablecomments{PSO id: Pan-STARRS1 identifier, id: identifier used throught the K18a data, RA: right ascention (J2000.0), Dec: declination (J2000.0), $\mathrm{P}_{\rps}$: period in the \rps band (determined with SigSpec), $\mathrm{P}_{e,\rps}$: period error in the \rps band determined with half the width of the 1$\sigma$ interval of the period distribution obtained by the bootstraping method, \nrps: extinction corrected \rps band magnitude (mean flux of the light curve converted to magnitudes), \rpserr: magnitude error in the \rps band determined by the method discribed in section \ref{detection}, \nips: extinction corrected \ips band magnitude, \ipserr: magnitude error in the \ips (see section \ref{detection}), \ngps: extinction corrected \gps band magnitude, \gpserr: magnitude error in the \gps (see section \ref{detection}).
}
\end{deluxetable}

\addtocounter{table}{-1}

\begin{deluxetable}{ccccccccccc}
\tabletypesize{\footnotesize}
\rotate
\tablecaption{main.dat (continued)}
\tablewidth{0pt}
\tablehead{
\colhead{id} & \colhead{$\mathrm{W}$} & \colhead{$\mathrm{W}_e$} & \colhead{$\mathrm{A}_{21}$} & \colhead{$\mathrm{A}_{e,21}$} & \colhead{$\mathrm{\varphi}_{21}$} & \colhead{$\mathrm{\varphi}_{e,21}$} & \colhead{sample} & \colhead{$\mathrm{flag}_{\rps,\ips}$}  & \colhead{$\mathrm{flag}_{\gps}$} & \colhead{type}
}
\startdata
704730  & 18.0049 & 0.0058 & 0.254 & 0.011 & 4.312 & 0.046 &     3d & 128 & 132 & UN \\
2933467 & 21.5411 & 0.0915 & 0.203 & 0.043 & 4.687 & 0.223 & manual &   0 &   0 & FO \\ 
3012073 & 21.6105 & 0.0945 & 0.263 & 0.051 & 4.183 & 0.208 &     3d &   0 &   0 & FO \\
2108691 & 21.1181 & 0.0735 & 0.256 & 0.056 & 4.297 & 0.235 & manual &   0 &   0 & FO \\
2493844 & 21.8819 & 0.1076 & 0.211 & 0.037 & 4.350 & 0.189 & manual &   0 &   0 & FO \\
\enddata
\tablecomments{$\mathrm{W}$: Wesenheit magnitude as defined in equation \ref{eqn_W}, $\mathrm{W}_e$: Wesenheit magnitude error (determined with the same method as \rpserr, see section \ref{detection}), $\mathrm{A}_{21}$: amplitude ratio of the first two Fourier components in the \rps band (see K13 equation 5), $\mathrm{A}_{e,21}$: error of the amplitude ratio $\mathrm{A}_{21}$ obtained from the error of the Fourier decomposition, $\mathrm{\varphi}_{21}$: phase difference of the first two Fourier components in the \rps band (see K13 equation 6), $\mathrm{\varphi}_{e,21}$: error of $\mathrm{\varphi}_{21}$ (same method used as for $\mathrm{A}_{e,21}$), sample: manual in case the Cepheid was classified manually and 3d in case the Cepheid was classified with the three dimensional parameter space, $\mathrm{flag}_{\rps,\ips}$: bit flag for the \rps and/or \ips band as described in section \ref{selection}, $\mathrm{flag}_{\gps}$: bit flag in the \gps band (see section \ref{selection}), type: Cepheid type (see section \ref{section_typeclass})
}
\end{deluxetable}

\begin{deluxetable}{cccccccccccc}
\tabletypesize{\footnotesize}
\rotate
\tablecaption{lc-info.dat\label{table_lcinfo}}
\tablewidth{0pt}
\tablehead{
\colhead{id} & \colhead{$\mathrm{epo}_{\rps}$} & \colhead{$\mathrm{epo}_{\ips}$} & \colhead{$\mathrm{epo}_{\gps}$} & \colhead{$\mathrm{P}_{\ips}$} & \colhead{$\mathrm{P}_{\gps}$}  & \colhead{$\mathrm{sig}_{\rps}$} & \colhead{$\mathrm{sig}_{\ips}$} & \colhead{$\mathrm{sig}_{\gps}$} & \colhead{$\mathrm{epo}_{\mathrm{use},\rps}$} & \colhead{$\mathrm{epo}_{\mathrm{use},\ips}$} & \colhead{$\mathrm{epo}_{\mathrm{use},\gps}$}
}
\startdata
704730  & 159 & 103 & 19 & 1.535951 & -1.000000 & 17.42 & 14.58 & -1.00 & 155 & 101 & 19  \\
2933467 & 224 & 143 & 40 & 1.566921 & 1.566582  & 31.05 & 16.46 & 6.31  & 219 & 141 & 39  \\ 
3012073 & 132 & 77  & 40 & 1.583693 & 1.583230  & 18.56 & 10.25 & 7.30  & 128 & 74  & 39  \\
2108691 & 259 & 252 & 45 & 1.643664 & -1.000000 & 26.20 & 18.11 & -1.00 & 250 & 245 & 39  \\
2493844 & 345 & 223 & 41 & 1.689490 & 1.688993  & 46.92 & 24.80 & 6.51  & 337 & 220 & 39  \\
\enddata
\tablecomments{$\mathrm{epo}_{\rps}$: number of epochs in the \rps band light curve, $\mathrm{epo}_{\ips}$: number of epochs in the \ips band light curve, $\mathrm{epo}_{\gps}$: number of epochs in the \gps band light curve, $\mathrm{P}_{\ips}$: period in the \ips band, $\mathrm{P}_{\gps}$: period in the \gps band, $\mathrm{sig}_{\rps}$: SigSpec significance of the \rps band period (see section 3 in K13), $\mathrm{sig}_{\ips}$: significance of the \ips band period, $\mathrm{sig}_{\gps}$: significance of the \gps band period, $\mathrm{epo}_{\mathrm{use},\rps}$: number of epochs used for the Cepheid light curve in the \rps band (all epochs are required to have a signal to noise ratio of larger than two and the corresponding visit stack has to have more than 2 frames), $\mathrm{epo}_{\mathrm{use},\ips}$: number of epochs used for the Cepheid light curve in the \ips band, $\mathrm{epo}_{\mathrm{use},\gps}$: number of epochs used for the Cepheid light curve in the \gps band
}
\end{deluxetable}

\addtocounter{table}{-1}

\begin{deluxetable}{ccccccccccccccc}
\tabletypesize{\footnotesize}
\rotate
\tablecaption{lc-info.dat (continued)}
\tablewidth{0pt}
\tablehead{
\colhead{id} & \colhead{$\mathrm{min}_{r}$} & \colhead{$\mathrm{min}_{i}$} & \colhead{$\mathrm{min}_{g}$} & \colhead{$\mathrm{max}_{r}$} & \colhead{$\mathrm{max}_{i}$} & \colhead{$\mathrm{max}_{g}$} & \colhead{$\rps$} & \colhead{$\ips$} & \colhead{$\gps$} & \colhead{$\mathrm{A}_{21,\ips}$} & \colhead{$\mathrm{A}_{21,\gps}$} & \colhead{$\mathrm{A}_{e,21,\ips}$} & \colhead{$\mathrm{A}_{e,21,\gps}$}
}
\startdata
704730  & 18.39 & 18.28 & 18.72 & 18.49 & 18.39 & 18.76 & 18.4442 & 18.3302 & 18.7395 & 0.309 & 0.710 & 0.016 & 0.363 \\
2933467 & 21.64 & 21.60 & 21.74 & 22.07 & 21.95 & 22.42 & 21.8315 & 21.7565 & 22.1074 & 0.105 & 0.231 & 0.073 & 0.119 \\ 
3012073 & 21.57 & 21.51 & 21.75 & 22.01 & 21.92 & 22.43 & 21.7765 & 21.7333 & 22.0373 & 0.242 & 0.208 & 0.097 & 0.141 \\
2108691 & 21.52 & 21.42 & 21.66 & 21.83 & 21.64 & 22.15 & 21.6745 & 21.5316 & 21.9262 & 0.171 & 0.198 & 0.085 & 0.179 \\
2493844 & 21.94 & 21.90 & 22.10 & 22.39 & 22.28 & 22.92 & 22.1761 & 22.1008 & 22.5190 & 0.239 & 0.337 & 0.068 & 0.096 \\
\enddata
\tablecomments{$\mathrm{min}_{r}$: minimum magnitude (i.e. brightest magnitude) of the Fourier fit to the light curve in the \rps band, $\mathrm{min}_{i}$: minimum magnitude (i.e. brightest magnitude) of the Fourier fit to the light curve in the \ips band, $\mathrm{min}_{g}$: minimum magnitude (i.e. brightest magnitude) of the Fourier fit to the light curve in the \gps band, $\mathrm{max}_{r}$: maximum magnitude (i.e. faintest magnitude) of the Fourier fit to the light curve in the \rps band, $\mathrm{max}_{i}$: maximum magnitude (i.e. faintest magnitude) of the Fourier fit to the light curve in the \ips band, $\mathrm{max}_{g}$: maximum magnitude (i.e. faintest magnitude) of the Fourier fit to the light curve in the \gps band,
\rps: \rps band magnitude (i.e. not extinction corrected as the \nrps magnitude but also determined with with the mean flux of the light curve converted to magnitudes), \ips: \ips band magnitude, \gps: \gps band magnitude, $\mathrm{A}_{21,\ips}$: amplitude ratio in the \ips band (analog to $\mathrm{A}_{21}$ but for the \ips band), $\mathrm{A}_{21,\gps}$: amplitude ratio in the \gps band, $\mathrm{A}_{e,21,\ips}$: error of the amplitude ratio in the \ips band (analog to $\mathrm{A}_{e,21}$, i.e. error from the Fourier decomposition), $\mathrm{A}_{e,21,\gps}$: error of the amplitude ratio in the \gps band 
}
\end{deluxetable}

\addtocounter{table}{-1}

\begin{deluxetable}{ccccccccccccccc}
\tabletypesize{\footnotesize}
\rotate
\tablecaption{lc-info.dat (continued)}
\tablewidth{0pt}
\tablehead{
\colhead{id} & \colhead{$\mathrm{\varphi}_{21,\ips}$} & \colhead{$\mathrm{\varphi}_{21,\gps}$} & \colhead{$\mathrm{\varphi}_{e,21,\ips}$} & \colhead{$\mathrm{\varphi}_{e,21,\gps}$} & \colhead{$a_{i,\rps}$} & \colhead{$a_{i,\ips}$} & \colhead{$a_{i,\gps}$} & \colhead{$a_{i,e,\rps}$} & \colhead{$a_{i,e,\ips}$} & \colhead{$a_{i,e,\gps}$} & \colhead{$\mathrm{gap}_{\rps}$} & \colhead{$\mathrm{gap}_{\ips}$} & \colhead{$\mathrm{gap}_{\gps}$}
}
\startdata
704730  & 4.467 & 1.253 & 0.057 & 0.729 & ... & ... & ... & ... & ... & ... & 0.030 & 0.037 & 0.177 \\
2933467 & 4.470 & 4.706 & 0.712 & 0.629 & ... & ... & ... & ... & ... & ... & 0.028 & 0.043 & 0.168 \\ 
3012073 & 4.235 & 6.248 & 0.439 & 0.607 & ... & ... & ... & ... & ... & ... & 0.034 & 0.054 & 0.231 \\
2108691 & 4.620 & 1.859 & 0.512 & 0.713 & ... & ... & ... & ... & ... & ... & 0.021 & 0.028 & 0.146 \\
2493844 & 4.190 & 4.391 & 0.307 & 0.333 & ... & ... & ... & ... & ... & ... & 0.021 & 0.030 & 0.156 \\
\enddata
\tablecomments{$\mathrm{\varphi}_{21,\ips}$: phase difference of the first two Fourier coefficients in the \ips band (analog to $\mathrm{\varphi}_{21,\gps}$), $\mathrm{\varphi}_{21,\gps}$: phase difference in the \gps band, $\mathrm{\varphi}_{e,21,\ips}$: error of the phase difference in the \ips band (analog to $\mathrm{\varphi}_{e,21}$ the method described in section \ref{detection} is used to determine the error), $\mathrm{\varphi}_{e,21,\gps}$: error of the phase difference in the \gps band, $a_{i,\rps}$: Fourier coefficients in the \rps band as defined in equation 3 in K13 (the order of the fit is $N=5$ so there are 11 Fourier parameters which are included in the electronically available tables but which we do not display here), $a_{i,\ips}$: Fourier coefficients in the \ips band (the order of the fit is $N=5$, therefore there are 11 Fourier parameters), $a_{i,\gps}$:  Fourier coefficients in the \gps band (the order of the fit is $N=3$, but there are still 11 columns where the coulmns for the higher orders are set to a default value of -1), $a_{i,e,\rps}$: fit error of the Fourier coefficients in the \rps band, $a_{i,e,\ips}$: fit error of the Fourier coefficients in the \ips band, $a_{i,e,\gps}$: fit error of the Fourier coefficients in the \gps band, $\mathrm{gap}_{\rps}$: largest phase ($\Theta$) gap in the folded \rps band lightcurve ($\Theta \in [0,1]$), $\mathrm{gap}_{\ips}$: largest phase gap in the folded \ips band light curve, $\mathrm{gap}_{\gps}$: largest phase gap in the folded \gps band light curve
}
\end{deluxetable}

\begin{deluxetable}{ccccccccccccc}
\tabletypesize{\footnotesize}
\rotate
\tablecaption{other.dat\label{table_other}}
\tablewidth{0pt}
\tablehead{
\colhead{id} & \colhead{pretype} & \colhead{$\mathrm{mult}_{\rps}$} & \colhead{$\mathrm{mult}_{\ips}$} & \colhead{$\mathrm{mult}_{\gps}$} & \colhead{$\mathrm{decrise}_{\rps}$} & \colhead{$\mathrm{decrise}_{\ips}$} & \colhead{$\mathrm{decrise}_{\gps}$} & \colhead{$\mathrm{\Rmnum{4}}_{\rps}$} & \colhead{$\mathrm{\Rmnum{4}}_{\ips}$} & \colhead{$\mathrm{\Rmnum{4}}_{\gps}$} & \colhead{E(B-V)}
}
\startdata
704730  & FO & 1 & 1 & 1 & 1.65507 & 1.30657 & 0.69590 & 0.19687 & 0.12460 & 0.25758 & 0.000000 \\
2933467 & FO & 1 & 1 & 1 & 1.63201 & 1.96267 & 1.14940 & 0.15451 & 0.21594 & 0.13723 & 0.000000 \\ 
3012073 & FO & 1 & 1 & 1 & 1.70609 & 1.99371 & 1.15403 & 0.12867 & 0.13750 & 0.06240 & 0.000000 \\
2108691 & FO & 1 & 1 & 1 & 1.72378 & 1.43226 & 0.91662 & 0.24642 & 0.27868 & 0.22139 & 0.069950 \\
2493844 & FO & 1 & 1 & 1 & 1.34169 & 1.76063 & 1.69879 & 0.19212 & 0.23818 & 0.12833 & 0.000000 \\
\enddata
\tablecomments{pretype: Cepheid type before the outlier clipping (see section \ref{clipping}), $\mathrm{mult}_{\rps}$: number of different skycells in which the Cepheid is in the \rps band, $\mathrm{mult}_{\ips}$: number of different skycells in which the Cepheid is in the \ips band, $\mathrm{mult}_{\gps}$: number of different skycells in which the Cepheid is in the \gps band, $\mathrm{decrise}_{\rps}$: decline/rise factor in the \rps band as defined in section 5 in K13, $\mathrm{decrise}_{\ips}$: decline/rise factor in the \ips band, $\mathrm{decrise}_{\gps}$: decline/rise factor in the \gps band, $\mathrm{\Rmnum{4}}_{\rps}$: noise factor in the \rps band as defined by the selection criterion IV in table \ref{tabcuts-ri}, $\mathrm{\Rmnum{4}}_{\ips}$: noise factor in the \ips band as defined by the selection criterion IV in table \ref{tabcuts-ri}, $\mathrm{\Rmnum{4}}_{\gps}$: noise factor in the \gps band as defined by the selection criterion IV in table \ref{tabcuts-g}, E(B-V): color excess from the \citet{2009A&A...507..283M} map but improved where possible by the instability strip as discussed in section \ref{insta}
}
\end{deluxetable}

\addtocounter{table}{-1}

\begin{deluxetable}{ccccccccccccc}
\tabletypesize{\footnotesize}
\rotate
\tablecaption{other.dat (continued)}
\tablewidth{0pt}
\tablehead{
\colhead{id} & \colhead{$\mathrm{err}_{\rps}$} & \colhead{$\mathrm{err}_{\ips}$} & \colhead{$\mathrm{err}_{\gps}$} & \colhead{$\mathrm{err}_{ok,\rps}$} & \colhead{$\mathrm{err}_{ok,\ips}$} & \colhead{$\mathrm{err}_{ok,\gps}$} & \colhead{$\mathrm{err}_{m,\rps}$} & \colhead{$\mathrm{err}_{m,\ips}$} & \colhead{$\mathrm{err}_{m,\gps}$} & \colhead{$\mathrm{err}_{sd,\rps}$} & \colhead{$\mathrm{err}_{sd,\ips}$} & \colhead{$\mathrm{err}_{sd,\gps}$} 
}
\startdata
704730  & 0.0010 & 0.0013 & 0.0051 & 1.00000 & 1.00000 & 1.00000 & 18.4443 & 18.3302 & 18.7396 & 0.0010 & 0.0013 & 0.0051 \\
2933467 & 0.0168 & 0.0202 & 0.0336 & 0.99979 & 1.00000 & 1.00000 & 21.8318 & 21.7569 & 22.1082 & 0.0168 & 0.0201 & 0.0336 \\
3012073 & 0.0147 & 0.0220 & 0.0293 & 1.00000 & 1.00000 & 1.00000 & 21.7767 & 21.7338 & 22.0380 & 0.0147 & 0.0219 & 0.0293 \\
2108691 & 0.0139 & 0.0160 & 0.0250 & 1.00000 & 1.00000 & 1.00000 & 21.6747 & 21.5319 & 21.9267 & 0.0139 & 0.0160 & 0.0249 \\
2493844 & 0.0218 & 0.0228 & 0.0332 & 0.99979 & 0.99792 & 1.00000 & 22.1765 & 22.1012 & 22.5198 & 0.0217 & 0.0227 & 0.0331 \\
\enddata
\tablecomments{$\mathrm{err}_{\rps}$: magnitude error in the \rps band determined from the $1\sigma$ range method, $\mathrm{err}_{\ips}$: magnitude error in the \ips band determined from the $1\sigma$ range method, $\mathrm{err}_{\gps}$: magnitude error in the \gps band determined from the $1\sigma$ range method, $\mathrm{err}_{ok,\rps}$: percentage ($\in[0,1]$) of \rps band epochs used in all the 10000 light curve realizations (see section \ref{detection}; epochs can be cut if the sum of the drawn reference frame flux and difference frame flux is negative or the signal to noise ratio is smaller than two),  $\mathrm{err}_{ok,\ips}$: percentage of \ips band epochs (analog to $\mathrm{err}_{ok,\rps}$),  $\mathrm{err}_{ok,\gps}$: percentage of \gps band epochs (analog to $\mathrm{err}_{ok,\rps}$), $\mathrm{err}_{m,\rps}$: mean \rps band magnitude of the distribution used to calculate the error, $\mathrm{err}_{m,\ips}$: mean \ips band magnitude of the distribution used to calculate the error, $\mathrm{err}_{m,\gps}$: mean \gps band magnitude of the distribution used to calculate the error, $\mathrm{err}_{sd,\rps}$: standard deviation of the \rps band distribution used to calculate the error, $\mathrm{err}_{sd,\ips}$: standard deviation of the \ips band distribution used to calculate the error, $\mathrm{err}_{sd,\gps}$: standard deviation of the \gps band distribution used to calculate the error
}
\end{deluxetable}

\addtocounter{table}{-1}

\begin{deluxetable}{ccccccccccccc}
\tabletypesize{\footnotesize}
\rotate
\tablecaption{other.dat (continued)}
\tablewidth{0pt}
\tablehead{
\colhead{id} & \colhead{$\mathrm{A}_{e,21,m,r}$} & \colhead{$\mathrm{A}_{e,m,21,i}$} & \colhead{$\mathrm{A}_{e,m,21,g}$} & \colhead{$\mathrm{A}_{e,21,c,r}$} & \colhead{$\mathrm{A}_{e,c,21,i}$} & \colhead{$\mathrm{A}_{e,c,21,g}$} & \colhead{$\mathrm{\varphi}_{e,m,21,r}$} & \colhead{$\mathrm{\varphi}_{e,m,21,i}$} & \colhead{$\mathrm{\varphi}_{e,m,21,g}$} & \colhead{$\mathrm{\varphi}_{e,c,21,r}$} 
}
\startdata
704730  & 0.00000 & 0.00000 & 0.00008 & 0.00000 & 0.00000 & 0.00008 & 0.00011 & 0.00010 & 0.00009 & 0.00010 \\
2933467 & 0.00006 & 0.00044 & 0.00144 & 0.00006 & 0.00044 & 0.00144 & 0.00418 & 0.00492 & 0.01575 & 0.00417 \\ 
3012073 & 0.00031 & 0.00076 & 0.00243 & 0.00031 & 0.00076 & 0.00243 & 0.00178 & 0.00302 & 0.01289 & 0.00178 \\
2108691 & 0.00002 & 0.00036 & 0.00165 & 0.00002 & 0.00036 & 0.00165 & 0.00365 & 0.00157 & 0.00049 & 0.00365 \\
2493844 & 0.00008 & 0.00095 & 0.00069 & 0.00008 & 0.00095 & 0.00069 & 0.00634 & 0.00518 & 0.00966 & 0.00634 \\
\enddata
\tablecomments{$\mathrm{A}_{e,21,m,r}$: amplitude ratio error in the \rps band determined from the $1\sigma$ range method, $\mathrm{A}_{e,21,m,i}$: amplitude ratio error in the \ips band determined from the $1\sigma$ range method, $\mathrm{A}_{e,21,m,g}$: amplitude ratio error in the \gps band determined from the $1\sigma$ range method, $\mathrm{A}_{e,21,c,r}$: amplitude ratio error in the \rps band determined from the $\pm \sigma/2$ range method, $\mathrm{A}_{e,21,c,i}$: amplitude ratio error in the \ips band determined from the $\pm \sigma/2$ range method, $\mathrm{A}_{e,21,c,g}$: amplitude ratio error in the \gps band determined from the $\pm \sigma/2$ range method, $\mathrm{\varphi}_{e,m,21,r}$: phase difference error in the \rps band determined from the $1\sigma$ range method, $\mathrm{\varphi}_{e,m,21,i}$: phase difference error in the \ips band determined from the $1\sigma$ range method, $\mathrm{\varphi}_{e,m,21,g}$: phase difference error in the \gps band determined from the $1\sigma$ range method, $\mathrm{\varphi}_{e,c,21,r}$: phase difference error in the \rps band determined from the $\pm \sigma/2$ range method
}
\end{deluxetable}

\addtocounter{table}{-1}

\begin{deluxetable}{ccccccccccccc}
\tabletypesize{\footnotesize}
\rotate
\tablecaption{other.dat (continued)}
\tablewidth{0pt}
\tablehead{
\colhead{id} & \colhead{$\mathrm{\varphi}_{e,c,21,i}$} & \colhead{$\mathrm{\varphi}_{e,c,21,g}$} & \colhead{$\mathrm{P}_{e,\ips}$} &  \colhead{$\mathrm{P}_{e,\gps}$} & \colhead{$\mathrm{P}_{e,c,\rps}$} & \colhead{$\mathrm{P}_{e,c,\ips}$} & \colhead{$\mathrm{P}_{e,c,\gps}$} & \colhead{$\mathrm{P}_{e,ok,\rps}$} & \colhead{$\mathrm{P}_{e,ok,\ips}$} & \colhead{$\mathrm{P}_{e,ok,\gps}$}

}
\startdata	
704730  & 0.00010 & 0.00009 & 0.000223 & -1.000000 & 0.000201 & 0.000226 & -1.000000 & 1.00000 & 1.00000 & -1.00000 \\
2933467 & 0.00492 & 0.01573 & 0.000148 & 0.020554  & 0.000074 & 0.000166 & -1.000000 & 1.00000 & 1.00000 & 0.99540  \\ 
3012073 & 0.00302 & 0.01283 & 0.568224 & 0.002856  & 0.000265 & 0.568463 & -1.000000 & 1.00000 & 1.00000 & 1.00000  \\
2108691 & 0.00157 & 0.00049 & 0.003929 & -1.000000 & 0.000113 & 0.003789 & -1.000000 & 1.00000 & 1.00000 & -1.00000 \\
2493844 & 0.00517 & 0.00965 & 0.000149 & 0.364768  & 0.000077 & 0.000150 & -1.000000 & 1.00000 & 1.00000 & 0.99630  \\
\enddata
\tablecomments{$\mathrm{\varphi}_{e,c,21,i}$: phase difference error in the \ips band determined from the $\pm \sigma/2$ range method, $\mathrm{\varphi}_{e,c,21,g}$: phase difference error in the \gps band determined from the $\pm \sigma/2$ range method, $\mathrm{P}_{e,\ips}$: period error in the \ips band determined with the $1\sigma$ range method, $\mathrm{P}_{e,\gps}$: period error in the \gps band determined with the $1\sigma$ range method, $\mathrm{P}_{e,c,\rps}$: period error in the \rps band determined with the $\pm \sigma/2$ range method, $\mathrm{P}_{e,c,\ips}$: period error in the \ips band determined with the $\pm \sigma/2$ range method, $\mathrm{P}_{e,c,\gps}$: period error in the \gps band determined with the $\pm \sigma/2$ range method, $\mathrm{P}_{e,ok,\rps}$: percentage ($\in[0,1]$) of the 10000 bootstrapped light curves in the \rps band for which SigSpec was able to determine a period, $\mathrm{P}_{e,ok,\ips}$: percentage ($\in[0,1]$) of the 10000 bootstrapped light curves in the \ips band for which SigSpec was able to determine a period, $\mathrm{P}_{e,ok,\gps}$: percentage ($\in[0,1]$) of the 10000 bootstrapped light curves in the \gps band for which SigSpec was able to determine a period
}
\end{deluxetable}

\addtocounter{table}{-1}

\begin{deluxetable}{ccccccccccccc}
\tabletypesize{\footnotesize}
\rotate
\tablecaption{other.dat (continued)}
\tablewidth{0pt}
\tablehead{
\colhead{id} & \colhead{$\mathrm{A}_{e,21,m,b}$} & \colhead{$\mathrm{A}_{e,21,c,b}$}  & \colhead{$\mathrm{\varphi}_{e,21,m,b}$} & \colhead{$\mathrm{\varphi}_{e,21,c,b}$} & \colhead{$\mathrm{b}_{FM}$} & \colhead{$\mathrm{b}_{FO}$} & \colhead{$\mathrm{b}_{T2}$}  & \colhead{$\mathrm{b}_{UN}$} & \colhead{$\mathrm{E(B-V)}_{min}$}	& \colhead{$\mathrm{E(B-V)}_{max}$}	
}
\startdata	
704730  & 0.07236 & -1.00000 & 0.36157 & 0.36095 & 0.003 & 0.993 & 0.000 & 0.004 & 0.000000 & 0.000000 \\
2933467 & 0.05093 & 0.05650  & 0.24158 & 0.24993 & 0.000 & 1.000 & 0.000 & 0.000 & 0.000000 & 0.000000 \\ 
3012073 & 0.06951 & 0.08891  & 0.25159 & 0.25679 & 0.001 & 0.971 & 0.000 & 0.028 & 0.000000 & 0.000000 \\
2108691 & 0.07095 & 0.08381  & 0.33562 & 0.34947 & 0.000 & 0.979 & 0.000 & 0.021 & 0.000000 & 0.139900 \\
2493844 & 0.04659 & 0.05073  & 0.21222 & 0.21361 & 0.000 & 1.000 & 0.000 & 0.000 & 0.000000 & 0.000000 \\
\enddata
\tablecomments{$\mathrm{A}_{e,21,m,b}$: amplitude ratio error in the \rps band determined from the $1\sigma$ range method where the distribution is obtained from 10000 bootstrapped light curves from the measured light curve (i.e. not the method described in section \ref{detection} where the reference frame flux is drawn from a distribution), $\mathrm{A}_{e,21,c,b}$: amplitude ratio error in the \rps band determined from the $\pm \sigma/2$ range method with the bootstrapping, $\mathrm{\varphi}_{e,21,m,b}$: phase difference error in the \rps band determined from the$1\sigma$ range method with bootstraping, $\mathrm{\varphi}_{e,21,c,b}$: phase difference error in the \rps band determined from the $\pm \sigma/2$ range method with bootstrapping, $\mathrm{boot}_{FM}$: percentage ($\in[0,1]$) of how often the FM type is assigned in the 10000 realizations (i.e. criterion \Rmnum{6} in section \ref{selection}; the Cepheid type is the pretype i.e. the type before the outlier clipping), $\mathrm{boot}_{FO}$: percentage of how often the FO type is assigned, $\mathrm{boot}_{T2}$: percentage of how often the T2 type is assigned, $\mathrm{boot}_{UN}$: percentage of how often the UN type is assigned, $\mathrm{E(B-V)}_{min}$: smallest possible color excess determined from the \citet{2009A&A...507..283M} map and the instability strip edges as described in section \ref{insta}, $\mathrm{E(B-V)}_{max}$: largest possible color excess
}
\end{deluxetable}

\end{document}